\pdfoutput=1    

\documentclass[pra,onecolumn,showpacs,preprintnumbers,amsmath,amssymb]{revtex4-1} 

\usepackage{amssymb}
\usepackage{amsmath}
\usepackage{graphicx}
\usepackage{siunitx}
\usepackage{mathrsfs}
\usepackage{hyperref}
\usepackage{color}

\newcommand{\R}[2]{\left\la \p R_{#1}(\pvec{#2} | \pvec{k} )/\p\Omega_s\right\ra_\textrm{incoh}}

\renewcommand{\vec}[1]{\mathbf{{#1}}}                         
\newcommand{\pvec}[1]{\mathbf{{#1}}_\parallel }
\newcommand{\vecUnit}[1]{\mathbf{\hat{#1}}}                         
\newcommand{\pvecUnit}[1]{\mathbf{\hat{#1}}_\parallel }

\newcommand{\imu}{\mathrm{i}}
\newcommand{\dint}{\mathrm{d}}
\renewcommand{\Re}{\mathrm{Re}\,}
\renewcommand{\Im}{\mathrm{Im}\,}

\newcommand{\ppol}{p}
\newcommand{\spol}{s}

\newcommand{\etal}{\textit{et al.}}

\newcommand{\hz}{\hat{\zeta}}
\newcommand{\nn}{\nonumber}

\newcommand{\zxp}{\zeta({\textbf x}_{\|})}
\newcommand{\zxpp}{\zeta({\textbf x}\, '\!\!_{\|} )}
\newcommand{\bkp}{{\textbf k}_{\|}}
\newcommand{\bqp}{{\textbf q}_{\|}}
\newcommand{\bpp}{{\textbf p}_{\|}}
\newcommand{\qp}{q_{\|}}
\newcommand{\kp}{k_{\|}}
\newcommand{\pp}{p_{\|}}
\newcommand{\bxp}{{\textbf x}_{\|}}
\newcommand{\bx}{{\textbf x}}
\renewcommand{\a}{\alpha}
\newcommand{\p}{\partial }
\newcommand{\fr}{\frac{1}{2}}
\newcommand{\ve}{\varepsilon}
\newcommand{\sfr}{^{\frac{1}{2}}}

\newcommand{\w}{\omega }

\newcommand{\la}{\langle}
\newcommand{\ra}{\rangle}
\newcommand{\xp}{x_{\|}}

\begin{document}

\title{Numerical studies of the scattering of light from a two-dimensional randomly rough interface between two dielectric media}
\author{\O{}. S. Hetland$^1$}
\email{Oyvind.Hetland@ntnu.no}
\author{A. A. Maradudin$^2$}
\author{T. Nordam$^1$}
\author{I. Simonsen$^{1}$}
\affiliation{$^1$Department of Physics, NTNU Norwegian University of Science and Technology, NO-7491 Trondheim, Norway}
\affiliation{$^2$Department of Physics and Astronomy, University of California, Irvine CA 92697, U.S.A.}

\date{\today}


\begin{abstract}
The scattering of polarized light incident from one dielectric medium on its two-dimensional randomly rough interface with a second dielectric medium is studied. A reduced Rayleigh equation for the scattering amplitudes is derived for the case where p- or s-polarized light is incident on this interface, with no assumptions being made regarding the dielectric functions of the media. Rigorous, purely numerical, nonperturbative solutions of this equation are obtained. 
They are used to calculate the reflectivity and reflectance of the interface, the mean differential reflection coefficient, and the full angular distribution of the intensity of the scattered light. 
These results are obtained for both the case where the medium of incidence is the optically less dense medium, and in the case where it is the optically more dense medium. 
Optical analogues of the Yoneda peaks observed in the scattering of x-rays from metal surfaces are present in the results obtained in the latter case. 
Brewster scattering angles for diffuse scattering are investigated, reminiscent of the Brewster angle for flat-interface reflection, but strongly dependent on the angle of incidence.
When the contribution from the transmitted field is added to that from the scattered field it is found that the results of these calculations satisfy unitarity with an error smaller than $10^{-4}$.
\end{abstract}

\maketitle


\section{Introduction}
\label{sec:introduction}
In the great majority of the theoretical studies of the scattering of light from a two-dimensional randomly rough surface of a dielectric medium, the medium of incidence has been vacuum. Recent reviews of such studies can be found in Refs. \citenum{Leskova2011} and \citenum{Nordam2013a}.
As a result of this restriction, effects associated with total internal reflection, which requires that the medium of incidence be optically more dense than the scattering medium, were not considered in  these studies. There have been exceptions to this general practice, however.

By the use of the stochastic functional approach \cite{Nakayama1980}, Kawanishi~\etal~\cite{Kawanishi1997} studied the coherent and incoherent scattering of an electromagnetic wave from a two-dimensional randomly rough interface separating two different dielectric media. The light could be incident on the interface from either medium. The theoretical approach used in this work \cite{Kawanishi1997} is perturbative in nature, and applicable only to weakly rough interfaces. Nevertheless its use yielded interesting results, including the presence of Yoneda peaks in the angular dependence of the intensity of the light scattered back into the medium of incidence when the latter was the optically more dense medium. 
These are sharp, asymmetric peaks occurring at the critical angle for total internal reflection for a fixed angle of incidence for both p- and s-polarization of the incident light. These peaks were first observed experimentally in the scattering of x-rays incident from air on a metal surface \cite{Yoneda1963} and have subsequently been studied theoretically in the context of the scattering of x-rays~\cite{Vineyard1982,Sinha1988,Leskova1997} and neutrons \cite{Sinha1988} from rough surfaces.

In Ref.~\citenum{Kawanishi1997}, Kawanishi~\etal~also observed angles of zero scattering intensity, to first order in their approach, in the distributions of the intensity of the incoherently scattered light, when the incident light was p-polarized. Due to their resemblance to the Brewster angle in the reflectivity from a flat interface, they dubbed these angles the ``Brewster scattering angles''. These were observed, in both reflection and transmission, for light incident from either medium.

Both the Yoneda peaks and the Brewster scattering angles seem to have had their first appearance in optics in the paper by Kawanishi~\etal~\cite{Kawanishi1997}. They have yet to be observed experimentally in this context. 
It should be mentioned that in an earlier numerical investigation of light scattering from one-dimensional dielectric rough surfaces, Nieto-Vesperinas and S\'{a}nchez-Gil~\cite{Nieto-Vesperinas1992} observed ``sidelobes'' in the angular intensity distributions. However, these authors did not associate these features with the Yoneda peak phenomenon, even though we believe doing so would have been correct.

In a subsequent paper Soubret~\etal~\cite{Soubret2001a} derived a reduced Rayleigh equation for the scattering amplitudes when an electromagnetic wave is incident from one dielectric medium on its two-dimensional randomly rough interface with a second dielectric medium. The solution of this equation was obtained in the form of expansions of the scattering amplitudes in powers of the surface profile function through terms of third order. However, in obtaining the numerical results presented in this paper~\cite{Soubret2001a}, the medium of incidence was assumed to be vacuum.

In this paper we present a study of this problem free from some of the restrictive assumptions and approximations present in the earlier studies of scattering of polarized light from two-dimensional randomly rough dielectric surfaces. We first derive a reduced Rayleigh equation for the scattering amplitudes when p- or s-polarized light is incident from a dielectric medium whose dielectric constant is $\ve_1$, on its two-dimensional randomly rough interface with a dielectric medium whose dielectric constant is $\ve_2$.
The dielectric constant $\ve_1$ can be smaller or larger than $\ve_2$. This equation is then solved by a rigorous, purely numerical, nonperturbative approach. The scattering amplitudes obtained in this way are then used to calculate the reflectivity and reflectance of the interface as a function of the angle of incidence, and also the effect of surface roughness on the contribution to the mean differential reflection coefficient from the light scattered incoherently (diffusely), and the full angular dependence of the intensity of the incoherently scattered light. It is hoped that the presentation of these results will stimulate and motivate experimental studies of such scattering systems.

\section{The Scattering System}
\label{sec:tss}
The system we study in this paper consists of a dielectric medium (medium 1), whose dielectric constant is $\ve_1$, in the region $x_3 > \zxp$, and a dielectric medium (medium 2), whose dielectric constant is $\ve_2$, in the region $x_3 < \zxp$ [Fig. 1].  Here $\bxp = (x_1, x_2, 0)$ is an arbitrary vector in the plane $x_3 = 0$, and we assume that both $\ve_1$ and $\ve_2$ are real and positive.
The surface profile function $\zxp$ is assumed to be a single-valued function of $\bxp$ that is differentiable with respect to $x_1$ and $x_2$, and constitutes a stationary, zero-mean, isotropic, Gaussian random process defined by
\begin{align}
  \la \zxp \zxpp \ra = \delta^2W(|\bxp - \bxp ' |) , \label{eq:2.1}
\end{align}
where $W(x_\parallel)$ is the \textit{normalized surface height autocorrelation function}, with the property that $W(0)=1$.
The angle brackets here and in all that follows denote an average over the ensemble of realizations of the surface profile function.  The root-mean-square height of the surface is given by
\begin{align}
  \delta = \la \zeta^2(\bxp )\ra ^{\fr} .\label{eq:2.2}
\end{align}
\noindent The power spectrum of the surface roughness $g(k_\parallel )$ is defined by
\begin{align}
  g(k_\parallel ) = \int \dint^2\xp W(x_\parallel )\exp (-\imu\bkp \cdot \bxp ) . \label{eq:2.3}
\end{align}
\noindent For $W(x_\parallel )$ we assume the Gaussian function $W(x_\parallel )=\exp\left(-\xp^2/a^2 \right),$
where the characteristic length $a$ is the transverse correlation length of the surface roughness.  The corresponding power spectrum is given by
\begin{align}
  g(k_\parallel ) = \pi a^2 \exp \left( -\frac{a^2\kp^2} {4}\right) . \label{eq:2.5}
\end{align}

\begin{figure}
  \centering
  \includegraphics{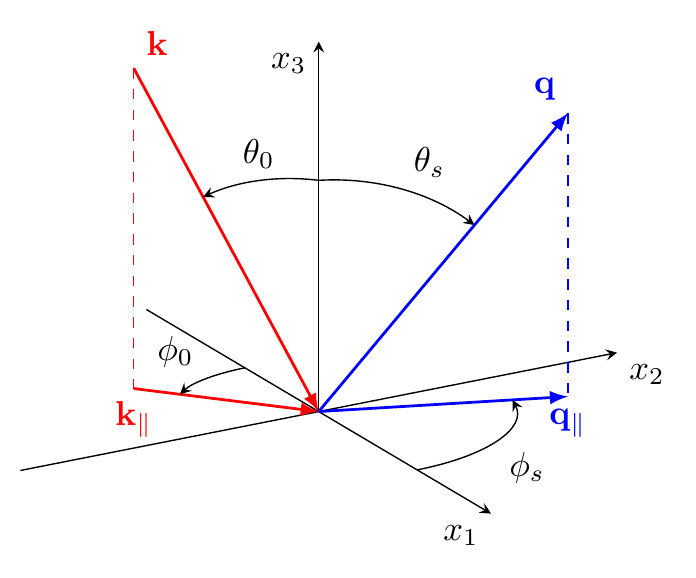} 
  \caption{A sketch of the scattering geometry assumed in this work. The figure also shows the coordinate system used, angles of incidence($\theta_0,\phi_0$) and scattering ($\theta_s,\phi_s$), and the corresponding lateral wave vectors $\pvec{k}$ and $\pvec{q}$, respectively.}
\end{figure}

\section{The Reduced Rayleigh Equation}

The interface $x_3 = \zxp$ is illuminated from the region $x_3 > \zxp$ (medium 1) by an electromagnetic wave of frequency $\w$. The total electric field in this region is the sum of an incoming incident field and an outgoing scattered field,
\begin{align}
  \vec{E}^>(\vec{x} |\omega ) &= \vec{E}_0(\pvec{k}) \exp [\imu\vec{Q}_0(\pvec{k})\cdot\vec{x} ]
  + \int \!\frac{\dint^2q_\parallel}{(2\pi )^2}\; \vec{A}(\pvec{q} ) \exp [\imu\vec{Q}_1(\pvec{q} )\cdot \vec{x} ], 
  \label{eq:3.1}
\end{align}
while the electric field in the region $x_3 < \zxp$ is an outgoing transmitted field,
\begin{align}
  \vec{E}^<(\vec{x} |\omega ) = \int \!\frac{\dint^2q_\parallel}{(2\pi )^2}\; \vec{B}(\pvec{q} ) \exp [\imu \vec{Q}^-_2(\pvec{q} )\cdot \vec{x} ]. \label{eq:3.2}
\end{align}
In writing these equations we have introduced the functions
\begin{subequations}
  \label{eq:3.3}
  \begin{align}
    \vec{Q}_0(\bkp ) &= \pvec{k} - \alpha_1(\kp ) \vecUnit{x}_3     \label{eq:3.3a}\\
    \vec{Q}_1(\bqp ) &= \pvec{q} + \alpha_1(\qp )  \vecUnit{x}_3    \label{eq:3.3b}\\
    \vec{Q}^{\pm}_2(\bqp ) &= \pvec{q} \pm \alpha_2(\qp )  \vecUnit{x}_3 ,  \label{eq:3.3c}
  \end{align}
\end{subequations}
where $(i = 1, 2)$
\begin{align}
  \alpha_i(q_\parallel ) = \left[\ve_i\left(\frac{\w}{c}\right)^2 - \qp^2 \right]\sfr
  \qquad \Re \alpha_i(q_\parallel ) > 0, \Im \alpha_i(q_\parallel ) > 0
  . \label{eq:3.4}
\end{align}
Here $\bkp = (k_1,k_2,0)$, and a caret over a vector indicates that it is a unit vector.  A frequency dependence of the field of the form $\exp (-\imu \w t)$ has been assumed, but not indicated explicitly.

The boundary conditions satisfied  by these fields at the interface $x_3 = \zxp$ are the continuity of the tangential components of the electric field:
\begin{align}
  \vec{n} \times  \vec{E}_0(\bkp ) & \exp [ \imu \pvec{k} \cdot \bxp -  \imu \alpha_1(\kp ) \zxp ]
  + \int \frac{\dint^2\qp}{(2\pi )^2}\; \vec{n} \times \vec{A}(\bqp ) \exp [\imu\bqp \cdot\bxp + \imu\alpha_1(\qp )\zxp ]\nn\\
        &= \int \frac{\dint^2\qp}{(2\pi )^2}\; \vec{n}\times \vec{B} (\bqp ) \exp [\imu\bqp \cdot\bxp - \imu\alpha_2(\qp )\zxp ] ; 
   \label{eq:3.5}
\end{align}
the continuity of the tangential components of the magnetic field:
\begin{align}
\begin{aligned}
  \vec{n} \times   [\imu\vec{Q}_0(\bkp )  \times \vec{E}_0(\bkp ) ] & \exp [\imu\bkp \cdot\bxp - \imu\alpha_1(\kp ) \zxp ]
   + \int \frac{\dint^2\qp}{(2\pi )^2}\; \vec{n} \times [\imu\vec{Q}_1(\bqp ) \times \vec{A}(\bqp )]
                \exp [\imu\bqp\cdot\bxp + \imu\alpha_1(\qp )\zxp ] 
    \\
    &= \int \frac{\dint^2\qp}{(2\pi )^2}\; \vec{n} \times [\imu\vec{Q}_2^-(\bqp ) \times \vec{B}(\bqp )]
                \exp [\imu\bqp\cdot\bxp - \imu\alpha_2(\qp )\zxp ] ; 
    \label{eq:3.6}
\end{aligned}
\end{align}
and the continuity of the normal component of the electric displacement:
\begin{align}
\begin{aligned}
 \ve_1 \vec{n} \cdot  \vec{E}_0(\bkp ) & \exp [\imu\bkp \cdot\bxp - \imu\alpha_1(\kp )\zxp ] 
  + \ve_1\int \!\frac{\dint^2\qp}{(2\pi )^2}\; \vec{n} \cdot \vec{A}(\bqp ) \exp [\imu\bqp \cdot \bxp + \imu\alpha_1(\qp ) \zxp ]
 \\  
  & = \ve_2 \int \!\frac{\dint^2 \qp}{(2\pi )^2}\; \vec{n} \cdot \vec{B} (\bqp ) \exp [\imu\bqp \cdot \bxp - \imu\alpha_2 (\qp ) \zxp ] . \label{eq:3.7}
\end{aligned}
\end{align}
The vector $\vec{n}\equiv \vec{n}(\pvec{x})$ entering these equations is a vector normal to the surface $x_3 = \zxp$ at each point of it, directed into medium $1$:
\begin{align}
\begin{aligned}  
  \label{eq:3.8}
    \vec{n}(\pvec{x}) = \left(-\frac{\p \zxp}{\p x_1}, - \frac{\p\zxp}{\p x_2}, 1\right).
  \end{aligned}
  \end{align}
%
Equation~(\ref{eq:3.7}) is redundant, but its inclusion simplifies the subsequent analysis.
We now proceed to eliminate the transmission amplitude ${\bf B}(\bqp )$ from this set of equations to obtain an equation that relates the scattering amplitude $\vec{A}(\bqp )$ to the amplitude of the incident field ${\bf E}_0(\bkp )$.

We begin by taking the vector cross product of Eq.~(\ref{eq:3.5}) with $ \ve_2\vec{Q}_2^+(\bpp ) \exp [-\imu\vec{Q}_2^+(\bpp) \cdot\{\bxp + \vecUnit{x}_3\zxp\} ] $; we next multiply Eq.~(\ref{eq:3.6}) by $-\imu\ve_2\exp [-\imu\vec{Q}_2^+(\pvec{p})\cdot\{\bxp +\vecUnit{x}_3 \zxp\} ]$; and finally multiply Eq.~(\ref{eq:3.7}) by $-\vec{Q}_2^+(\bpp ) \exp [-\imu\vec{Q}_2^+(\bpp) \cdot\{\bxp + \vecUnit{x}_3\zxp\} ] $, where $\bpp =(p_1,p_2,0)$ is an arbitrary wave vector in the plane $x_3 = 0$.  
When we add the three equations obtained in this way, and integrate the sum over $\bxp$ we obtain an equation that can be written in the form
\begin{align} 
\begin{aligned}  
   \ve_2 &
           \vec{Q}_2^+(\bpp )            \times \left[ \vec{V}_{E}(\bpp |\bkp ) \times \vec{E}_0(\bkp ) \right]
         + \ve_2 \vec{V}_{E}(\bpp |\bkp ) \times \left[ \vec{Q}_0(\bkp ) \times \vec{E}_0(\bkp )        \right]
         - \ve_1 \vec{Q}_2^+(\bpp )              \left[ \vec{V}_{E}(\bpp |\bkp )\cdot \vec{E}_0(\bkp )  \right] 
  \\ &
  \quad
  + \int \!\frac{\dint^2\qp}{(2\pi )^2}\; 
       \left\{ 
       \ve_2 \vec{Q}_2^+(\bpp )     \times \left[ \vec{V}_{A}(\bpp |\bqp )\times \vec{A}(\bqp ) \right]
      +\ve_2 \vec{V}_A(\bpp |\bqp ) \times \left[ \vec{Q}_1(\bqp )        \times \vec{A}(\bqp ) \right]
      -\ve_1 \vec{Q}_2^+(\bpp )            \left[ \vec{V}_{A}(\bpp |\bqp )\cdot \vec{A}(\bqp )  \right]   
       \right\} 
  \\&
   = \ve_2 \int \frac{\dint^2\qp}{(2\pi )^2}\, 
     \left\{ 
           \vec{Q}_2^+(\bpp )       \times \left[ \vec{V}_B(\bpp |\bqp )   \times \vec{B}(\bqp )   \right]
         + \vec{V}_B(\bpp |\bqp ) \times \left[ \vec{Q}_2^-(\bqp )         \times \vec{B}(\bqp )   \right]
         - \vec{Q}_2^+ (\bpp )             \left[ \vec{V}_{B}(\bpp |\bqp ) \cdot \vec{B}(\bqp )    \right]
      \right\} ,
\label{eq:3.9}
\end{aligned}
\end{align}
where
\begin{subequations}
\label{eq:3.10}
\begin{align}
  \vec{V}_E(\bpp |\bkp ) 
  &= \int \dint^2\xp\; \vec{n}(\pvec{x})   \exp \left\{ -\imu(\bpp - \bkp ) \cdot \bxp - \imu [\alpha_2(\pp ) + \alpha_1(\kp ) ] \zxp \right\}
  \label{eq:3.10a}
  \\
  \vec{V}_A(\bpp |\bqp ) 
  &= \int \dint^2\xp\; \vec{n}(\pvec{x})   \exp \left\{ -\imu(\bpp - \bqp ) \cdot \bxp - \imu [\alpha_2(\pp ) - \alpha_1(\qp ) ] \zxp \right\} 
   \label{eq:3.10b}
   \\
   \vec{V}_B(\bpp |\bqp ) 
   &= \int \dint^2\xp\; \vec{n}(\pvec{x})  \exp \left\{ -\imu (\bpp -\bqp ) \cdot \bxp - \imu [\alpha_2(\pp ) + \alpha_2(\qp ) ] \zxp \right\} .
   \label{eq:3.10c}
\end{align}
\end{subequations}
At this point it is convenient to introduce the representation
\begin{align}
  \exp [-\imu\gamma\zxp ] = \int \!\frac{\dint^2Q_{\|}}{(2\pi )^2}\; I(\gamma |\vec{Q}_{\|}) \exp (\imu\vec{Q}_{\|} \cdot \bxp ).
  \label{eq:3.11}
\end{align}
On differentiating both sides of Eq.~(\ref{eq:3.11}) with respect to $x_{j}(j = 1, 2)$ we obtain the result
\begin{align}
  -\frac{\p \zxp}{\p x_{j}} \exp [-\imu\gamma\zxp ] 
  = \int\!\frac{\dint^2Q_{\|}}{(2\pi )^2}\; \frac{Q_{j}}{\gamma} I(\gamma |\vec{Q}_{\|})\exp (\imu\vec{Q}_{\|}\cdot\bxp ) .
  \label{eq:3.12}
\end{align}
Finally, to be able to evaluate the function $I(\gamma |{\bf Q}_{\|})$ we need the inverse of Eq.~(\ref{eq:3.11}), namely
\begin{align}
  I(\gamma |\vec{Q}_{\|}) &= \int \!\dint^2\xp\; \exp (-\imu\vec{Q}_{\|}\cdot\bxp )
  \exp [-\imu\gamma \zxp ]
  = \sum^{\infty}_{n=0} \frac{(-\imu\gamma )^n}{n!} \hz^{(n)}(\vec{Q}_{\|}) , 
  \label{eq:3.13}
\end{align}
 where
\begin{subequations}
  \label{eq:3.14}
  \begin{align}
    \hz^{(0)}(\vec{Q}_{\|}) 
    &= (2\pi )^2 \delta (\vec{Q}_{\|}) 
    \label{eq:3.14a}
    \\
    \hz^{(n)}(\vec{Q}_{\|}) 
    &= \int \dint^2\xp \; \zeta^n(\bxp ) \exp (-\imu\vec{Q}_{\|}\cdot \bxp ),\qquad n \geq 1 . 
    \label{eq:3.14b}
  \end{align}
\end{subequations}
On combining Eqs.~(\ref{eq:3.10})--(\ref{eq:3.13}) with Eqs.~(\ref{eq:3.3}) and (\ref{eq:3.8}) we obtain the results
\begin{subequations}
  \label{eq:3.15}
 \begin{align}
   \vec{V}_E(\bpp |\bkp ) 
   &= \left[ \vec{Q}_2^+(\bpp) - \vec{Q}_0(\bkp) \right]
      \frac{I(\alpha_2(\pp ) + \alpha_1(\kp )|\bpp - \bkp )}{\alpha_2(\pp )+\alpha_1(\kp )} 
      \label{eq:3.15a} 
      \\
   \vec{V}_A(\bpp |\bqp ) 
   &= \left[ \vec{Q}_2^+(\bpp) - \vec{Q}_1(\bqp) \right]
      \frac{I(\alpha_2(\pp ) - \alpha_1(\qp )|\bpp - \bqp )}{\alpha_2(\pp )-\alpha_1(\qp )} 
      \label{eq:3.15b}
      \\
  \vec{V}_B(\bpp |\bqp ) 
   &= \left[ \vec{Q}_2^+(\bpp) - \vec{Q}_2^-(\bqp) \right]
      \frac{I(\alpha_2(\pp ) + \alpha_2(\qp )|\bpp - \bqp )}{\alpha_2(\pp )+\alpha_2(\qp )} .
\label{eq:3.15c} 
\end{align}
\end{subequations}
When the results given by Eq.~(\ref{eq:3.15}) are substituted into Eq.~(\ref{eq:3.9}), the latter becomes
\begin{align}
  \vec{Q}_2^+(\bpp) \times \left[ \vec{Q}_2^+(\bpp) \cdot \vec{E}_0(\bkp) \right]
  & \frac{I( \alpha_2(\pp ) + \alpha_1(\kp )|\bpp - \bkp )}{ \alpha_2(\pp )+\alpha_1(\kp )}
  \nn\\
  +  \int \frac{\dint^2\qp}{(2\pi )^2}\; \vec{Q}_2^+(\bpp) \times &\left[ \vec{Q}_2^+(\bpp)\cdot \vec{A}(\bqp)\right]
    \frac{I(-\alpha_1(\qp ) + \alpha_2(\pp )|\bpp - \bqp )}{-\alpha_1(\qp )+\alpha_2(\pp )}
  = 0 .
\label{eq:3.16}
\end{align}
Thus, the amplitude of the transmitted field ${\bf B}(\bqp)$ has been eliminated from the problem and we have obtained an equation for the scattering amplitude ${\bf A}(\bqp )$ alone.

To transform Eq.~(\ref{eq:3.16}) into a more useful form we first introduce three mutually perpendicular unit vectors:
\begin{subequations}
  \label{eq:3.21}
  \begin{align}
    \vecUnit{a}_0(\bpp ) &= \frac{c}{\sqrt{\ve_2}\w} \left[ \pvec{p} + \vecUnit{x}_3\alpha_2(\pp) \right] = \frac{c}{\sqrt{\ve_2}\w} \vec{Q}^+_2(\bpp)
    \label{eq:3.21a}\\
    \vecUnit{a}_1(\bpp ) &= \frac{c}{\sqrt{\ve_2}\w} \left[ \pvecUnit{p}\alpha_2(\pp ) - \vecUnit{x}_3\pp \right]
    \label{eq:3.21b}\\
    \vecUnit{a}_2(\bpp ) &= \vecUnit{x}_3 \times \pvecUnit{p} . 
    \label{eq:3.21c}
  \end{align}   
\end{subequations}
In terms of these vectors Eq.~(\ref{eq:3.16}) becomes:
\begin{align}
  \left\{ \left[ \vecUnit{a}_0(\bpp) \cdot \vec{E}_0(\bkp) \right] \vecUnit{a}_0(\bpp) - \vec{E}_0(\bkp) \right\}
  \frac{I(\alpha_2(\pp ) + \alpha_1(\kp )|\bpp - \bkp )}{\alpha_2(\pp )+\alpha_1(\kp )}
  \qquad\qquad\qquad\qquad\qquad\nn\\
  + \int \frac{\dint^2\qp}{(2\pi )^2}\; \left\{\left[ \vecUnit{a}_0(\bpp) \cdot \vec{A}(\bqp) \right] \vecUnit{a}_0(\bpp) - \vec{A}(\bqp)\right\}
  \frac{I(\alpha_2(\pp )-\alpha_1(\qp )|\bpp - \bqp )}{\alpha_2(\pp )-\alpha_1(\qp )}
  = 0 . 
\label{eq:3.18}
\end{align}
We now write the vectors $\vec{E}_0(\bkp )$ and $\vec{A}(\bqp )$ in the forms
\begin{subequations}
  \label{eq:3.19}
  \begin{align}
    \vec{E}_0(\bkp ) = \vec{\hat{e}}_p^{(i)}(\bkp ) E_{0p}(\bkp ) + \vecUnit{e}_s^{(i)}(\bkp ) E_{0s}(\bkp ) ,
    \label{eq:3.19a}
  \end{align}
with
  \begin{align}
    \vecUnit{e}^{(i)}_p(\bkp ) &= \frac{c}{\sqrt{\ve_1}\w}\left[ \pvecUnit{k}\alpha_1(\kp ) + \vecUnit{x}_3\kp \right]
    \label{eq:3.19b}\\
    \vecUnit{e}^{(i)}_s(\bkp ) &= \pvecUnit{k} \times \vecUnit{x}_3 , 
    \label{eq:3.19c}
  \end{align}
\end{subequations}
and
\begin{subequations}
  \label{eq:3.20}
  \begin{align}
    \vec{A}(\bqp ) = \vecUnit{e}^{(s)}_p(\bqp ) A_p(\bqp ) + \vecUnit{e}^{(s)}_{s}(\bqp )A_s(\bqp ) ,
    \label{eq:3.20a}
  \end{align}
with
  \begin{align}
    \vecUnit{e}^{(s)}_p(\bqp ) &= \frac{c}{\sqrt{\ve_1}\w}\left[ - \pvecUnit{q} \alpha_1(\qp ) + \vecUnit{x}_3\qp \right]
  \label{eq:3.20b}\\
    \vecUnit{e}^{(s)}_s(\bqp ) &= \pvecUnit{q} \times \vecUnit{x}_3 .
    \label{eq:3.20c}
  \end{align}
\end{subequations}
In these expressions $E_{0p}(\bkp )$ and $E_{0s}(\bkp )$ are the amplitudes of the p-and s-polarized components of the incident field with respect to the plane of incidence, defined by the vectors $\pvecUnit{k}$ and $\vecUnit{x}_3$.  Similarly, $A_p(\bqp )$ and $A_s(\bqp )$ are the amplitudes of the p- and s-polarized components of the scattered field with respect to the plane of scattering defined by the vectors $\pvecUnit{q}$ and $\vecUnit{x}_3$.

Equation~\eqref{eq:3.18} is a vector equation: it is a set of three coupled equations. However, there are only two unknowns, namely $A_p(\bqp )$ and $A_s(\bqp )$. Consequently, one of these equations is redundant. To obtain $A_p(\bqp )$ and $A_s(\bqp )$ in terms of $E_{0p}(\bkp)$ and $E_{0s}(\bkp )$ we proceed as follows.
We take the scalar product of Eq.~(\ref{eq:3.18}) with each of the three unit vectors given by Eq.~\eqref{eq:3.21} in turn. The results are:
\begin{subequations}
  \label{eq:3.22}
  \begin{align}
    & \vecUnit{a}_0(\bpp ) \cdot \left[ \mbox{\rm Eq.~\eqref{eq:3.18}}\right]: \qquad 0 = 0 ; 
    \label{eq:3.22a}\\
    %
    & \vecUnit{a}_1(\bpp ) \cdot \left[ \mbox{\rm Eq.~\eqref{eq:3.18}}\right]: \qquad 
    \nn\\
    & \quad - \vecUnit{a}_1(\bpp ) \cdot \vec{E}_0(\bkp)
    \frac{I(\alpha_2(\pp )+\alpha_1(\kp )|\bpp - \bkp )}{\alpha_2(\pp )+\alpha_1(\kp )}
    = \int \frac{\dint^2\qp}{(2\pi )^2}\; \vecUnit{a}_1(\bpp ) \cdot \vec{A}(\bqp)
    \frac{I(\alpha_2(\pp )-\alpha_1(\qp )|\bpp - \bqp )}{\alpha_2(\pp )-\alpha_1(\qp )} ;
    \label{eq:3.22b}\\
    %
    & \vecUnit{a}_2(\bpp ) \cdot \left[ \mbox{\rm Eq.~\eqref{eq:3.18}}\right]:  
    \nn\\
    & \quad - \vecUnit{a}_2(\bpp ) \cdot \vec{E}_0(\bkp)
    \frac{I(\alpha_2(\pp )+\alpha_1(\kp )|\bpp - \bkp )}{\alpha_2(\pp )+\alpha_1(\kp )}
    = \int \frac{\dint^2\qp}{(2\pi )^2}\; \vecUnit{a}_2(\bpp ) \cdot \vec{A}(\bqp)
    \frac{I(\alpha_2(\pp )-\alpha_1(\qp )|\bpp - \bqp )}{\alpha_2(\pp )-\alpha_1(\qp )} .
    \label{eq:3.22c}
  \end{align}  
\end{subequations}
Equations~\eqref{eq:3.22b} and \eqref{eq:3.22c} are the two equations we seek.

With the use of Eqs.~\eqref{eq:3.21} and \eqref{eq:3.19}--\eqref{eq:3.20}, Eqs.~\eqref{eq:3.22b} and \eqref{eq:3.22c} can be rewritten in the form $(\alpha = p, s, \, \beta = p, s)$
\begin{align}
  A_{\alpha}(\bqp ) = \sum_{\beta} R_{\alpha\beta}(\bqp |\bkp ) E_{0\beta}(\bkp ) . 
  \label{eq:3.23}
\end{align}
On combining Eqs.~\eqref{eq:3.22b} and \eqref{eq:3.22c} with Eq.~\eqref{eq:3.23} we find that the scattering amplitudes $\{ R_{\alpha\beta}(\bqp |\bkp )\}$ are the solutions of the equation
\begin{align}
  \int \frac{\dint^2\qp}{(2\pi )^2} \; \frac{I(\alpha_2(\pp )-\alpha_1(\qp )|\bpp -\bqp )}{\alpha_2(\pp )-\alpha_1(\qp )}
  \vec{M}^{+} (\bpp |\bqp ) \vec{R}(\bqp |\bkp )
  = -\frac{I(\alpha_2(\pp )+\alpha_1(\kp )|\bpp -\bkp )}{\alpha_2(\pp )+\alpha_1(\kp )}
  \vec{M}^{-} (\bpp |\bkp ),
  \label{eq:3.24}
\end{align}
where
\begin{subequations}
\label{eq:3.25}
  \begin{align}
    \vec{M}^{\pm}(\bpp |\bqp ) &= 
    \left( 
      \begin{array}{cc}
        \frac{1}{\sqrt{\ve_1\ve_2}} [\pp\qp \pm \alpha_2(\pp )\, \pvecUnit{p} \cdot \pvecUnit{q} \,\alpha_1(\qp ) ] 
        & \quad 
        - \frac{1}{\sqrt{\ve_2}}\frac{\w}{c}\alpha_2(\pp )\, [\vec{\hat{p}}_{\|} \times \vec{\hat{q}}_{\|}]_3
        \\ 
        \pm\frac{1}{\sqrt{\ve_1}}\frac{\w}{c} [\vec{\hat{p}}_{\|}\times \vec{\hat{q}}_{\|}]_3 \,\alpha_1(\qp ) 
        & \quad 
        \frac{\w^2}{c^2}\, \pvecUnit{p} \cdot \pvecUnit{q} 
      \end{array} 
    \right) \label{eq:3.25a}
  \end{align}
and
  \begin{align}
    \vec{R} (\bqp |\bkp ) 
    = 
    \left( 
      \begin{array}{cc}
        R_{pp}(\bqp |\bkp ) & R_{ps}(\bqp |\bkp )\\
        R_{sp}(\bqp |\bkp ) & R_{ss}(\bqp |\bkp ) 
      \end{array}
    \right) .
    \label{eq:3.25b}
  \end{align}
\end{subequations}
Equation~\eqref{eq:3.24} is the reduced Rayleigh equation for the scattering amplitudes.

\section{The Mean Differential Reflection Coefficient}

From the knowledge of the scattering amplitudes the mean differential reflection coefficient, the reflectivity, and the reflectance can be calculated. The differential reflection coefficient $\p R/\p\Omega_s$ is defined such that $(\p R/\p\Omega_s)d\Omega_s$ is the fraction of the total time-average flux incident on the interface that is scattered into the element of solid angle $d\Omega_s$ about the scattering direction defined by the polar and azimuthal scattering angles $(\theta_s,\phi_s)$.  To obtain the mean differential reflection coefficient we first note that the magnitude of the total time-averaged flux incident on the interface is given by
\begin{align}
  P_\textrm{inc} &= - \Re \frac{c}{8\pi} \int \!\dint^2\xp \left\{ \vec{E}^*_0(\bkp ) \times \left[\frac{c}{\w} \vec{Q}_0(\pvec{k} )\times \vec{E}_0 (\bkp ) \right] \right\}_3
  \exp \left\{ [-\imu \vec{Q}_0^*(\bkp ) + \imu \vec{Q}_0(\bkp ) ] \cdot \vec{x}  \right\}
  \nn\\
  &= -\Re \frac{c^2}{8\pi \w} \int \!\dint^2\xp\; \left\{ \left| \vec{E}_0 (\bkp ) \right|^2 \vec{Q}_0(\bkp ) - \left[ \vec{E}_0^*(\bkp ) \cdot \vec{Q}_0(\bkp ) \right] \vec{E}_0(\bkp )\right\}_3
  \nn\\
  &= \Re \frac{c^2}{8\pi\w} \int \!\dint^2\xp \; \alpha_1(\kp ) \left|\vec{E}_0(\bkp )\right|^2
  \nn\\
  &= S \frac{c^2}{8\pi\w} \alpha_1(\kp ) \left| \vec{E}_0(\bkp )\right|^2 . 
  \label{eq:4.1}
\end{align}
In this result $S$ is the area of the $x_1x_2$ plane covered by the randomly rough surface.  The minus sign on the right-hand side of the first equation compensates for the fact that the 3-component of the incident flux is negative, and we have used the fact that $\alpha_1(\kp )$ is real, so that $\vec{Q}_0(\bkp )$ is real, and $\vec{E}^*_0(\bkp ) \cdot \vec{Q}_0 (\bkp ) = 0$.

In a similar fashion we note that the total time-averaged scattered flux is given by
\begin{align}
  P_\textrm{sc} =& \,\Re\, \frac{c}{8\pi} \int \!\dint^2\xp\! \!\int\frac{\dint^2\qp}{(2\pi )^2}\! \int\frac{\dint^2\qp '}{(2\pi )^2}
    \left\{ \vec{A}^*(\bqp ) \times \left[\frac{c}{\w} \vec{Q}_1(\bqp) \times \vec{A}(\bqp ') \right]\right\}_3
    \exp \left[ -\imu(\vec{Q}_1^*(\bqp) - \vec{Q}_1(\bqp '))\cdot \bx\right] \nn\\
  =&  \,\Re \frac{c^2}{8\pi\w} \int \!\frac{\dint^2\qp}{(2\pi )^2}\,  
        \left \{ \left| \vec{A}(\bqp )\right|^2 \vec{Q}_1 (\bqp )  - \left[ \vec{A}^*(\bqp )\cdot \vec{Q}_1(\bqp ) \right] \vec{A} (\bqp ) \right\}_3     
       \exp \left[ - 2 \Im \alpha_1(\qp ) x_3 \right] \nn\\
  =&  \,\Re \frac{c^2}{8\pi\w} \int \!\frac{\dint^2\qp}{(2\pi )^2}\,  
       \left \{  \alpha_1(\qp) \left| \vec{A}(\bqp )\right|^2  - \frac{c}{\sqrt{\ve_1}\omega}\qp \left[ \vec{A}^*(\bqp )\cdot \vec{Q}_1(\bqp ) \right] A_p (\qp ) \right\}
       \exp \left[- 2 \Im \alpha_1(\qp ) x_3\right]\nn\\
  =&  \,\Re \frac{c^2}{8\pi\w} \int \!\frac{\dint^2\qp}{(2\pi )^2}\; \alpha_1(\qp) \left|{\bf A}(\bqp )\right|^2   
        \exp \left[- 2 \Im\alpha_1(\qp ) x_3 \right]   
    \nn\\ & 
        - \,\Re \frac{c^2}{8\pi\w} \int \!\frac{\dint^2\qp}{(2\pi )^2}\; [\alpha_1(\qp ) - \alpha_1^*(\qp)]   
      \frac{c^2 }{\ve_1\w^2} \qp^2 \left|A_p(\qp )\right|^2 \exp \left[-2\Im\alpha_1(\qp )x_3 \right] .\label{eq:4.3}
\end{align}
The integral in the second term is purely imaginary.  Thus we have
\begin{align}
  P_\textrm{sc} = \frac{c^2}{32\pi^3\w} \int\limits_{\qp < \sqrt{\ve_1}\frac{\w}{c}} \dint^2\qp \; \alpha_1(\qp ) \left|\vec{A}(\bqp )\right|^2 . 
  \label{eq:4.4}
\end{align}
The wave vectors $\bkp$ and $\bqp$ can be expressed in terms of the polar and azimuthal angles of incidence $(\theta_0,\phi_0)$ and scattering $(\theta_s,\phi_s)$, respectively, by
\begin{subequations}
  \label{eq:4.5}
  \begin{align}
    \bkp &= \sqrt{\ve_1} \frac{\w}{c} \sin\theta_0 (\cos\phi_0,\sin\phi_0,0)
    \label{eq:4.5a}\\
    \bqp &= \sqrt{\ve_1} \frac{\w}{c} \sin \theta_s (\cos\phi_s,\sin\phi_s, 0). 
    \label{eq:4.5b}
  \end{align}
\end{subequations}
From these results it follows that
\begin{align}
  \dint^2\qp = \ve_1 \left( \frac{\w}{c}\right)^2 \cos\theta_s \, \dint\Omega_s ,\label{eq:4.6}
\end{align}
where $\dint\Omega_s = \sin\theta_s \, \dint\theta_s \, \dint\phi_s$.  The total time-averaged scattered flux therefore becomes
\begin{align}
  P_\textrm{sc} &= \frac{\ve^{3/2}_1\w^2}{32\pi^3c} \int \!\dint\Omega_s\; \cos^2\theta_s \left[ \left|A_p(\bqp )\right|^2 + \left|A_s(\bqp )\right|^2 \right].  
\label{eq:4.7}
\end{align}
Similarly, the total time averaged incident flux, Eq.~(\ref{eq:4.1}), becomes
\begin{align}
  P_\textrm{inc} &= S\frac{\sqrt{\ve_1}c}{8\pi} \cos\theta_0 \left[ \left|E_{0p}(\bkp )\right|^2 + \left|E_{0s}(\bkp )\right|^2\right] .\label{eq:4.8}
\end{align}
Thus by definition, the differential reflection coefficient is given by
\begin{align}
  \frac{\p R}{\p \Omega_s} &= \frac{1}{S} \ve_1 \left(\frac{\w}{2\pi c}\right)^2 \frac{\cos^2\theta_s}{\cos\theta_0} 
  \frac{\left|A_p(\bqp )\right|^2 + \left|A_s(\bqp )\right|^2}{ \left|E_{0p}(\bkp )\right|^2 + \left|E_{0s}(\bkp )\right|^2}. 
\label{eq:4.9}
\end{align}
From this result and Eq.~\eqref{eq:3.23} we find that the contribution to the differential reflection coefficient when an incident plane wave of polarization  $\beta$, the projection of whose wave vector on the mean scattering plane is $\bkp$, is reflected into a plane wave of polarization $\alpha$, the projection of whose wave vector on the mean scattering plane is $\bqp$, is given by
\begin{align}
  \frac{\p R_{\alpha\beta}(\bqp |\bkp )}{\p\Omega_s} 
  &= 
  \frac{1}{S} \ve_1 \left( \frac{\w}{2\pi c}\right)^2 \frac{\cos^2\theta_s}{\cos\theta_0}  
  \left| R_{\alpha\beta}(\bqp |\bkp )\right|^2 .
  \label{eq:4.10}
\end{align}
As we are dealing with scattering from a randomly rough interface, it is the average of this function over the ensemble of realizations of the surface profile function that we need to calculate.  This is the mean differential reflection coefficient, which is defined by
\begin{align}
  \left\la \frac{\p R_{\alpha\beta}(\bqp |\bkp )}{\p \Omega_s}\right\ra 
  &= 
  \frac{1}{S} \ve_1 \left( \frac{\w}{2\pi c}\right)^2 \frac{\cos^2\theta_s}{\cos\theta_0}  
  \left\la \left| R_{\alpha\beta} (\bqp |\bkp ) \right|^2\right\ra . 
  \label{eq:4.11}
\end{align}
If we write the scattering amplitude $R_{\alpha\beta}(\bqp |\bkp )$ as the sum of its mean value and the fluctuation from this mean, 
\begin{align}
  R_{\alpha\beta}(\bqp |\bkp ) 
  &= 
  \left\la R_{\alpha\beta}(\bqp |\bkp )\right\ra  + \left[R_{\alpha\beta}(\bqp |\bkp ) - \left\la R_{\alpha\beta}(\bqp |\bkp )\right\ra \right] ,     
  \label{eq:4.12}
\end{align}
then each of these two terms contributes separately to the mean differential reflection coefficient,
\begin{align}
  \left\la \frac{\p R_{\alpha\beta}(\bqp |\bkp )}{\p\Omega_s}\right\ra 
  &= 
  \left\la \frac{\p R_{\alpha\beta}(\bqp |\bkp )}{\p\Omega_s}\right\ra_\textrm{coh} 
  + \left\la \frac{\p R_{\alpha\beta}(\bqp |\bkp )}{\p\Omega_s}\right\ra_\textrm{incoh} , 
  \label{eq:4.12-mislabeled}
\end{align}
where
\begin{align}
  \left\la \frac{\p R_{\alpha\beta}(\bqp |\bkp )}{\p \Omega_s}\right\ra_\textrm{coh} 
  &= 
  \frac{1}{S} \ve_1 \left( \frac{\w}{2\pi c}\right)^2\frac{\cos^2\theta_s}{\cos\theta_0}
 \left|\left\la R_{\alpha\beta} (\bqp |\bkp )\right\ra \right|^2 
  \label{eq:4.13}
\end{align}
and
\begin{align}
  \left\la \frac{\p R_{\alpha\beta}(\bqp |\bkp )}{\p \Omega_s}\right\ra_\textrm{incoh} 
  &= 
  \frac{1}{S} \ve_1 \left( \frac{\w}{2\pi c}\right)^2\frac{\cos^2\theta_s}{\cos\theta_0} 
  \left[ \left\la  \left| R_{\alpha\beta} (\bqp |\bkp )-\left\la R_{\alpha\beta}(\bqp |\bkp )\right\ra \right|^2\right\ra \right] \nn\\
  &= \frac{1}{S} \ve_1 \left( \frac{\w}{2\pi c}\right)^2 \frac{\cos^2\theta_s}{\cos\theta_0} 
  \left[ \left\la  \left| R_{\alpha\beta} (\bqp |\bkp )\right|^2\right\ra -\left|\Big\la R_{\alpha\beta}(\bqp |\bkp )\Big\ra \right|^2 \right]. \quad \label{eq:4.14}
\end{align}
The former contribution describes the coherent (specular) reflection of the incident field from a randomly rough surface, while the latter contribution describes the incoherent (diffuse) component of the scattered light.

\subsection{Reflectivity and Reflectance}
%
\noindent Equation \eqref{eq:4.13} is the starting point for obtaining the reflectivity of the two-dimensional randomly rough interface. We begin with the result that 
\begin{align}
  \la R_{\alpha\beta}(\bqp |\bkp )\ra = (2\pi )^2 \delta (\bqp - \bkp )\delta_{\alpha\beta}R_{\alpha}(\kp ) .\label{eq:4.15}
\end{align}
The presence of the delta function is due to the stationarity of the randomly rough surface; the Kronecker symbol $ \delta_{\alpha\beta}$ arises from the conservation of angular momentum in the scattering process; and the result that $R_{\alpha}(\kp )$ depends on $\bkp$ only through its magnitude is due to the isotropy of the random roughness.

With the result given by Eq.~(\ref{eq:4.15}), the expression for $\la \p R_{\alpha\beta}(\bqp |\bkp )/\p\Omega_s \ra_\textrm{coh}$ given by Eq.~(\ref{eq:4.13}), becomes
\begin{align}
  \left\la \frac{\p R_{\alpha\alpha} (\pvec{q} |\pvec{k} )}{\p\Omega_s}\right\ra_\textrm{coh} 
  &= 
  \ve_1 \left( \frac{\w}{c}\right)^2\frac{\cos^2\theta_s}{\cos\theta_0}
  \left| R_{\alpha}(\kp ) \right|^2 \, \delta (\pvec{q} - \pvec{k}), 
  \label{eq:4.16}
\end{align}
where we have used the result 
\begin{align}
  \left[ (2\pi )^2\delta (\bqp - \bkp) \right]^2 
  &=
  (2\pi )^2 \delta (\vec{0}) \,
  (2\pi )^2\delta (\pvec{q} - \pvec{k}  ) 
  = S (2\pi )^2 \delta (\bqp - \bkp ) \label{eq:4.17}
\end{align}
in obtaining this expression. We next use the relation
\begin{align}
  \delta (\bqp - \bkp ) 
  =& 
  \frac{1}{\kp } \delta ( q_\parallel  - \kp )\, \delta (\phi_s - \phi_0) 
\label{eq:4.18-new}
\end{align}
together with the relations
\begin{align}
  \kp = \sqrt{\ve_1}\frac{\omega}{c}\sin\theta_0, \qquad \qp = \sqrt{\ve_1}\frac{\omega}{c}\sin\theta_s,
  \label{eq:k_and_q_as_functions_of_angles} 
\end{align}
to obtain
\begin{align}
  \left\la \frac{\p R_{\alpha\alpha} (\bqp |\bkp )}{\p\Omega_s}\right\ra_\textrm{coh} 
  &= 
  \sqrt{\ve_1} \left( \frac{\w}{c}\right)\frac{\cos^2\theta_s}{\cos\theta_0}\frac{1}{\kp}
  \left| R_{\alpha}(\kp ) \right|^2 \, \delta (\sin\theta_s - \sin\theta_0) \delta (\phi_s - \phi_0 )
  \nn\\
  &=
  \frac{\cos^2\theta_s}{\cos^2\theta_0}
  \left| R_{\alpha}(\kp ) \right|^2 \, \frac{\delta (\theta_s - \theta_0) \delta (\phi_s - \phi_0 )}{\sin\theta_0}
  \nn\\
  &=
  \left| R_{\alpha}(\kp ) \right|^2 \, \frac{\delta (\theta_s - \theta_0) \delta (\phi_s - \phi_0 )}{\sin\theta_0}.
  \label{eq:4.20-new}
\end{align}
The reflectivity, ${\mathcal R}_{\alpha}(\theta_0)$, for light of $\alpha$ polarization is defined by
\begin{align}
  {\mathcal R}_{\alpha}(\theta_0) 
  = \int^{\frac{\pi}{2}}_{0} \dint\theta_s\, \sin\theta_s \int^{\pi}_{-\pi} \dint\phi_s\, 
  \left\la \frac{\p R_{\alpha\alpha}(\bqp |\bkp )}{\p\Omega_s}\right\ra_\textrm{coh}
  = \left| R_{\alpha}(\kp) \right|^2.
  \label{eq:4.21-new}
\end{align}
The function $R_{\alpha}(\kp )$ is obtained from Eq.~\eqref{eq:4.15}, with the aid of the result that $(2\pi)^2\delta(\vec{0})=S$, in the form
\begin{align}
  \label{eq:4.22}
  R_{\alpha}(\kp ) =   R_{\alpha}\left( \sqrt{ \ve_1 }\frac{\w}{c}\sin\theta_0 \right) 
                   = \frac{1}{S} \left< R_{\alpha\alpha}(\pvec{k} | \pvec{k}) \right>.
\end{align}

In addition to the reflectivity~\eqref{eq:4.21-new} that depends only on the co-polarized light reflected coherently by the rough interface, it is also of interest to introduce the \textit{reflectance} for $\beta$-polarized light defined as 
\begin{subequations}
  \label{eq:reflectance}
\begin{align}
  \label{eq:reflectance-A}
  {\mathscr R}_{\beta}(\theta_0) 
  =&  
  \sum_{\alpha=p,s}  {\mathscr R}_{\alpha\beta}(\theta_0),  
\end{align}
where
\begin{align}
  \label{eq:reflectance-B}
  {\mathscr R}_{\alpha\beta}(\theta_0)   
  &=
  \int^{\frac{\pi}{2}}_{0} \dint\theta_s\, \sin\theta_s \int^{\pi}_{-\pi} \dint\phi_s\,
  \left< \frac{\p R_{\alpha\beta}(\bqp |\bkp )}{\p\Omega_s} \right>. 
\end{align}
\end{subequations}
In short, the reflectance measures the fraction of the power flux incident on the rough surface that was reflected by it, taking both specularly and diffusely reflected light into account: In view of Eq.~\eqref{eq:4.12-mislabeled}, the reflectance is the sum of a contribution from light that has been reflected coherently and a contribution from light that has been reflected incoherently by the rough interface, ${\mathscr R}_{\beta}(\theta_0) = {\mathscr R}_{\beta}(\theta_0)_{\mathrm{coh}} + {\mathscr R}_{\beta}(\theta_0)_{\mathrm{incoh}}$, and both co- and cross-polarized reflected light contribute. 
Since cross-polarized coherently reflected light is not allowed [see Eq.~\eqref{eq:4.15}], the coherent contribution to the reflectance for $\beta$-polarized light equals the reflectivity for $\beta$-polarized light; ${\mathscr R}_{\beta}(\theta_0)_{\mathrm{coh}} = {\mathcal R}_{\beta}(\theta_0)$. Equation~\eqref{eq:reflectance-A} can therefore also be written in the form
\begin{align}
  {\mathscr R}_{\beta}(\theta_0)  &=    {\mathcal R}_{\beta}(\theta_0) 
    +  \sum_{\alpha=p,s} {\mathscr R}_{\alpha\beta}(\theta_0)_{\mathrm{incoh}}. 
    \label{eq:reflectance_sum}
\end{align}
If the incident light is not purely p- or s-polarized, the reflectance and the reflectivity of the rough surface will have to be calculated on the basis of weighted sums of the expressions in Eqs.~\eqref{eq:4.21-new} and  \eqref{eq:reflectance_sum}, where the weights reflect the fractions of the different polarizations contained within the incident light.

\section{Numerical solution of the reduced Rayleigh equation}

The simulation results to be presented in this work were obtained by a nonperturbative numerical solution of the reduced Rayleigh equation \eqref{eq:3.24}, which was carried out in the following manner; A realization of the surface profile function was generated on a grid of $N_x\times N_x$ points within a square region of the $x_1 x_2$-plane of edges $L$. This surface profile enters Eq.~\eqref{eq:3.24} through the function $I(\gamma | \pvec{Q})$, given by Eq.~\eqref{eq:3.13}. Utilizing the Taylor expansion detailed in Eq.~\eqref{eq:3.13}, the Fourier transform of $\zeta^n(\pvec{x})$ was calculated by use of the fast Fourier transform~\cite{Nordam2013a}, and the Taylor series was truncated at the finite order $N_T$.
In evaluating the $\pvec{q}$ integral in Eq.~\eqref{eq:3.24}, the infinite limits of integration were replaced by finite limits $|\pvec{q} |< {\mathcal Q}/2$, and the integration was carried out by a two-dimensional version of the extended midpoint rule \cite[p.~135]{Book:Press1996} applied to a circular subsection of a grid of $N_q\times N_q$ points in the $q_1q_2$-plane, whose size and discretization was determined by the Nyquist sampling theorem \cite[p.~494]{Book:Press1996} and the properties of the discrete Fourier transform~\cite{Nordam2013a}. In momentum space, these limits lead to discretization intervals of $\Delta q=2\pi/L$ along the orthogonal axes of the $q_1q_2$-plane, and upper limits on the magnitude of resolved wave vectors are given by ${\mathcal Q}=\Delta q \lfloor N_x/2 \rfloor $, where $\lfloor \cdot \rfloor$ denotes the floor function~\cite[p.~948]{Book:Press1996}. The resulting linear system of equations was solved by LU factorization and back substitution.

These calculations were performed simultaneously for incident light of both p- and s-polarization, and they were performed for a large number $N_p$ of realizations of the surface profile function $\zxp$. The resulting scattering amplitude $R_{\alpha\beta}(\pvec{q} |\pvec{k} )$ and its squared modulus $|R_{\alpha\beta}(\pvec{q} |\pvec{k} )|^2$ were obtained for each realization. An arithmetic average of the $N_p$ results for these quantities yielded the mean values $\la R_{\alpha\beta}(\bqp \bkp )\ra $ and $\la |R_{\alpha\beta}(\bqp |\bkp )|^2\ra$ that enter Eqs.~\eqref{eq:4.22}, \eqref{eq:reflectance} and \eqref{eq:4.14} for the reflectivity, reflectance and the mean differential reflection coefficient, respectively.
A more detailed description of the numerical method can be found in Ref.~\citenum{Nordam2013a}.

\section{Results and discussions}

The two-dimensional randomly rough dielectric interfaces we study in this work were defined by an isotropic Gaussian height distribution of rms height $\delta=\lambda/40$, and an isotropic Gaussian correlation function of transverse correlation length $a=\lambda/4$ [Eq.~\eqref{eq:2.5}]. They covered a square region of the $x_1 x_2$-plane of edge $L=25\lambda$, giving an area $S=L^2 = 25\lambda \times 25\lambda$. The incident light was assumed to be a p- or s-polarized plane wave of wavelength $\lambda$ in vacuum. One of the two media in our configuration was assumed to be vacuum with a dielectric constant $\ve=1.0$, and the other medium was assumed to be a photoresist defined by the dielectric constant $\ve=2.64$. Since the dielectric constants entering the calculations are independent of the wavelength, all lengths appearing in them can be scaled with respect to $\lambda$. The angles of incidence were $(\theta_0,\phi_0)$, where the azimuthal angle of incidence was set to $\phi_0=\ang{0}$, without loss of generality. We remark that this value of $\phi_0$ was chosen since it coincided with one of the two axes of the numerical grid, but that it is, due to the isotropy of the roughness, an arbitrary choice in the sense that results for any other value of $\phi_0$ can be obtained from the results presented through a trivial rotation. 
%
The surface profiles were generated by the Fourier filtering method (see Refs.~\citenum{Maradudin1990} and \citenum{Simonsen2011}) on a grid of $N_x\times N_x =321\times321$ points. The values used for $N_x$ and $L$ correspond to $\mathcal{Q}=6.4\,\omega/c$, where $\mathcal{Q}$ is the limit in the $I(\gamma|{\mathbf Q}_\parallel)$-integrals, Eq.~\eqref{eq:3.13}, and we used the first $N_T = 18$ terms of the Taylor expansion in the calculation of these integrals.

\smallskip
Investigating the energy conservation of our simulation results can be a useful test of their accuracy. In combining simulation results from the current work with corresponding results obtained for the mean differential transmission coefficient $\left< \partial T_{\alpha\beta}/\partial \Omega_t\right>$ through the use of computationally similar methods~\cite{Simonsen2010,Hetland2016b}, we may add the total reflected and transmitted power for any lossless system. When the reflectance is added to the transmittance for any of the systems investigated in the current work, it is found that the results of these calculations satisfy unitarity~\cite{Simonsen2010}, a measure of energy conservation, with an error smaller than $10^{-4}$. This testifies to the accuracy of the approach used, and it is also a good indicator of satisfactory discretization. It should be noted, however, that unitarity is a necessary, but not sufficient, condition for the correctness of the presented results. In a separate investigation~\cite{Simonsen2012-04}, unitarity was found to be satisfied to a satisfactory degree for surfaces with a root mean square roughness up to about three times larger than the roughness used in obtaining the results presented in this paper.

\subsection{Normal incidence}

\begin{figure}[tbh]
  \centering
  \includegraphics[width=0.47\columnwidth]{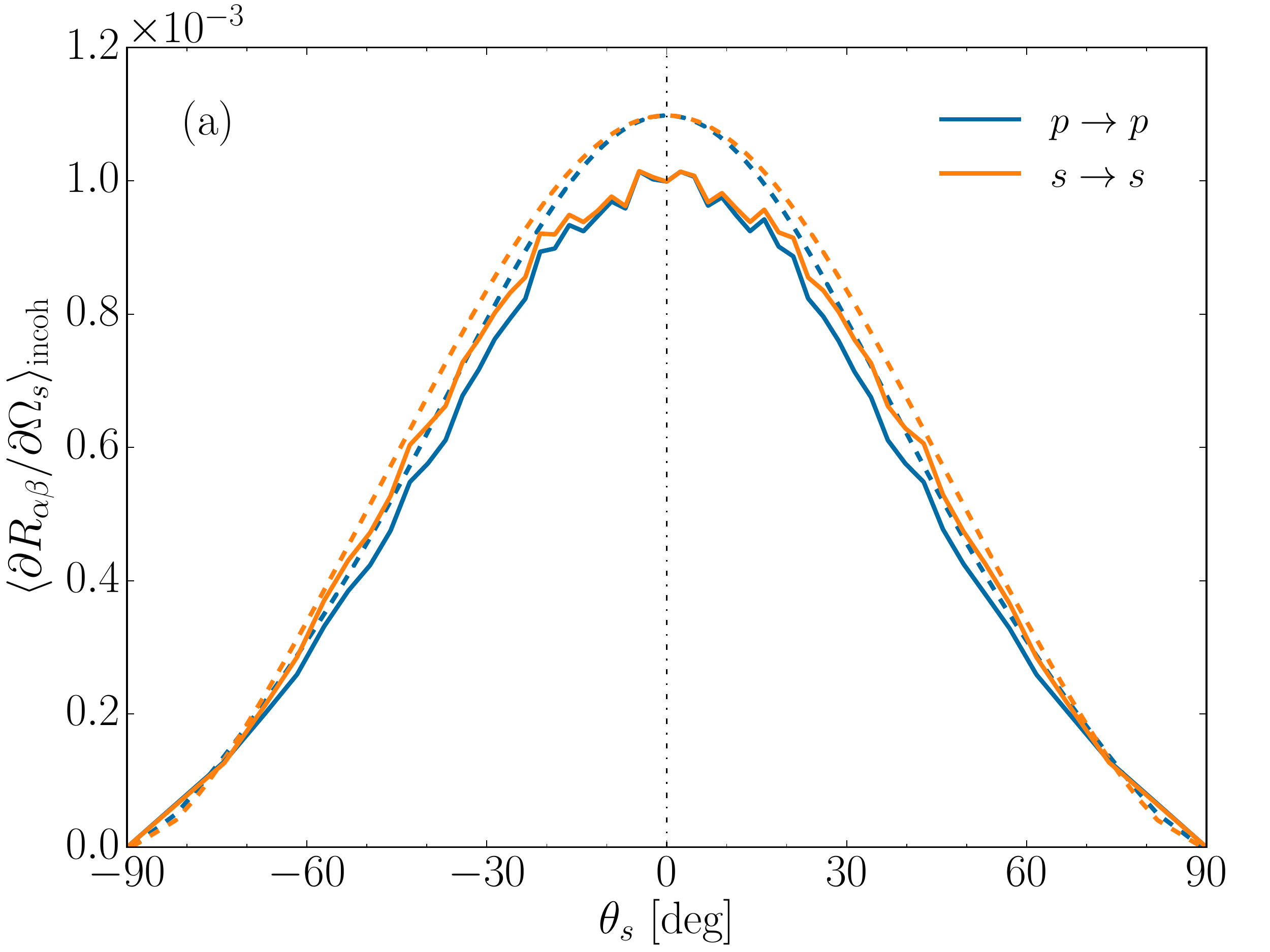}
  \includegraphics[width=0.47\columnwidth]{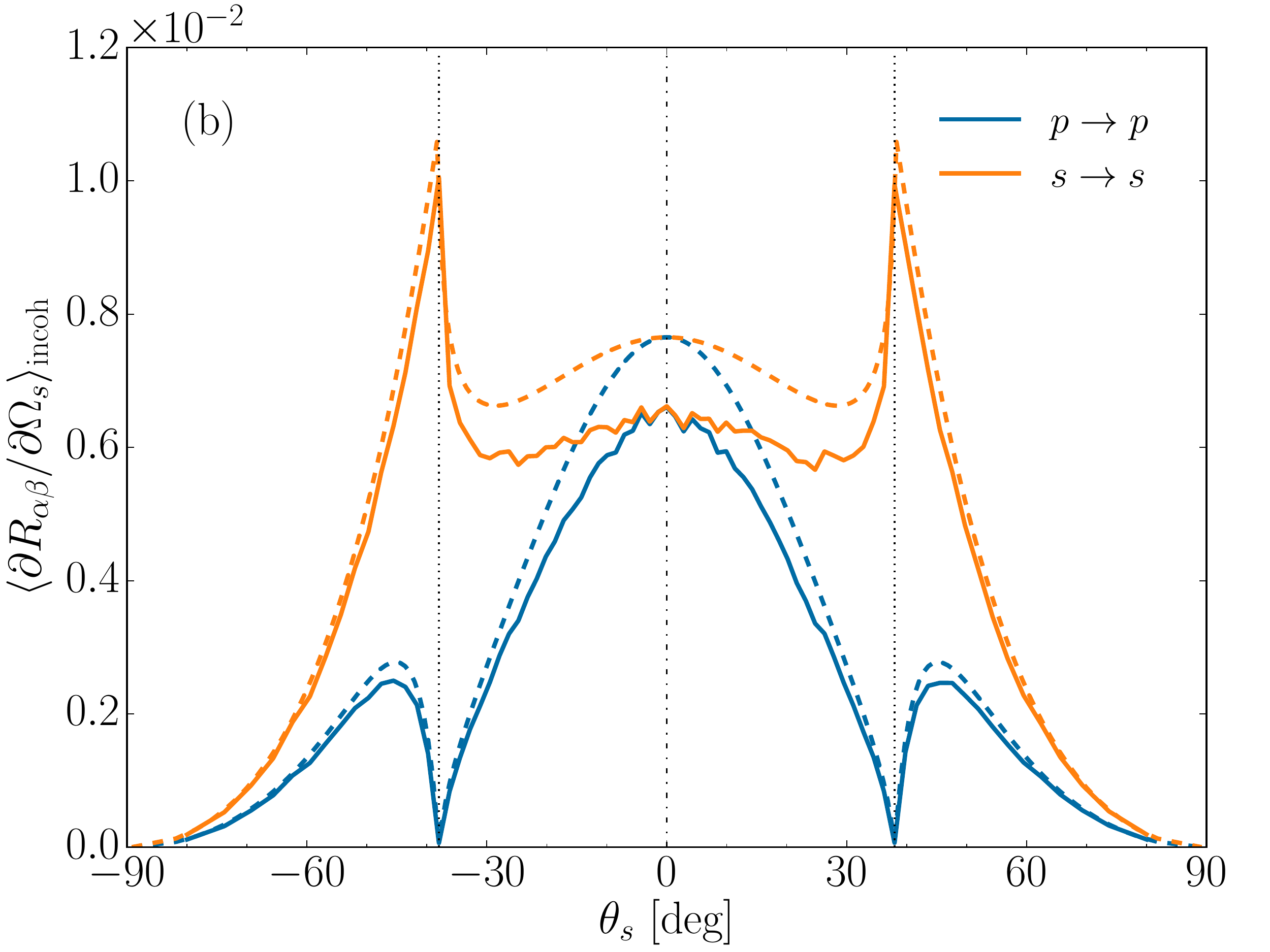}
  \caption{The contribution to the incoherent component of the mean differential reflection coefficient from the in-plane, co-polarized scattering of p- and s-polarized light incident normally [$\theta_0=\ang{0}$] on (a) a random vacuum-dielectric interface [$\ve_1=1.0$, $\ve_2=2.64$] and (b) a dielectric-vacuum interface [$\ve_1=2.64$, $\ve_2=1.0$] as a function of the polar angle of scattering $\theta_s$. The solid curves were obtained on the basis of numerically solving the reduced Rayleigh equations, Eq.~\eqref{eq:3.24}, for an ensemble of \num{4500} surface realizations. The dashed curves are results from small amplitude perturbation theory, Eq.~\eqref{eq:5.20_R}, included for comparison.The specular direction of reflection is indicated by the vertical dash-dotted line at $\theta_s=\ang{0}$, and in Fig.~\protect\ref{fig:inplane_mdrc_theta_0}(b), the dotted lines at $|\theta_s|=\theta_c=\sin^{-1}{\sqrt{\ve_2/\ve_1}}\approx\ang{38.0}$ indicate the positions of the critical angle for total internal reflection (as expected for a flat surface system). Results for cross-polarized scattering have not been indicated since they are generally suppressed in the plane-of-incidence. The wavelength of the incident light in vacuum was $\lambda$.  The rough interface was assumed to have a root-mean-square roughness of $\delta = \lambda/40$, and it was characterized by an isotropic Gaussian power spectrum, Eq.~\eqref{eq:2.5}, of transverse correlation length $a=\lambda/4$. In the numerical calculations it was assumed that the surface covered an area $L\times L$, with $L=25 \lambda$, and the surface was discretized on a grid of $321 \times 321$ points.
 }
\label{fig:inplane_mdrc_theta_0}
\end{figure}

In Fig.~\ref{fig:inplane_mdrc_theta_0} we display the contribution to the in-plane~($\pvecUnit{q}\parallel\pvecUnit{k}$) incoherent components of the mean differential reflection coefficient (DRC) as a function of the polar angle of scattering when the random surface is illuminated from the vacuum side at normal incidence by p- and s-polarized light, Fig.~\ref{fig:inplane_mdrc_theta_0}(a), and when it is illuminated from the dielectric medium side, Fig.~\ref{fig:inplane_mdrc_theta_0}(b). Notice that the unit vectors $\pvecUnit{q}=\pvec{q}/q_\parallel$ and $\pvecUnit{k}=\pvec{k}/k_\parallel$ are well defined also for  $\theta_s=\ang{0}$ and $\theta_0=\ang{0}$, respectively, as follows from  Eqs.~\eqref{eq:4.5} and \eqref{eq:k_and_q_as_functions_of_angles}. Only results for in-plane co-polarized scattering are presented, since in-plane cross-polarized scattering is suppressed due to the absence of contribution from single-scattering processes. 
An ensemble of \num{4500} realizations of the surface profile function was used to produce the numerical results that this figure is based on. This ensemble size is more than adequate in terms of the interpretation of the results and their features, but we note that a larger ensemble size would have reduced the jaggedness that can be observed in all the (solid line) results presented in this work. 

From Fig.~\ref{fig:inplane_mdrc_theta_0}(a) it is observed that the curves corresponding to the two polarizations are featureless, and are nearly identical. In contrast, the curves presented in Fig.~\ref{fig:inplane_mdrc_theta_0}(b) are rather different for the two polarizations; they display both peaks and dips in $\ppol\to\ppol$ scattering, and peaks in $\spol\to\spol$ scattering.
The origins of these features can be understood through small amplitude perturbation theory (SAPT). The contribution to the mean differential reflection coefficients from light scattered incoherently can to the lowest nonzero order in the surface profile function $\zeta(\bxp)$ be expressed as (see the appendix for details):
\begin{subequations}
  \label{eq:5.20_R}
\begin{align}
\left\la \frac{\p R_{pp}(\pvec{q} | \pvec{k} )}{\p\Omega_s}\right\ra_\textrm{incoh} 
   &= 
      \frac{\delta^2}{\pi^2} \ve_1 (\ve_2-\ve_1)^2  
      \left( \frac{\w}{c} \right)^2 \frac{\cos^2\theta_s}{\cos\theta_0} g(|\bqp -\bkp |) 
      \nn\\ & \quad\times 
      \frac{1}{|d_p(\qp )|^2} \left|\ve_2\qp\kp-\ve_1\alpha_2(\qp )(\pvecUnit{q}\cdot\pvecUnit{k}) \alpha_2(\kp ) \right|^2 \frac{\alpha^2_1(\kp )}{|d_p(\kp )|^2} 
     \label{eq:5.20a}
    \\
 \left\la \frac{\p R_{sp}(\bqp |\bkp )}{\p\Omega_s}\right\ra_\textrm{incoh} 
   &= 
    \frac{\delta^2}{\pi^2} \ve_1^2(\ve_2-\ve_1)^2  
    \left( \frac{\w}{c}\right)^4 \frac{\cos^2\theta_s}{\cos\theta_0} g(|\bqp - \bkp |) 
    \frac{\a^2_2(\kp )}{|d_s(\qp )|^2} \left([\pvecUnit{q} \times \pvecUnit{k}]_3\right)^2 \frac{\a^2_1(\kp)}{|d_p(\kp )|^2} \qquad
    \label{eq:5.20sp}
    \\
 \left\la \frac{\p R_{ps}(\bqp |\bkp )}{\p\Omega_s}\right\ra_\textrm{incoh} 
   &= 
    \frac{\delta^2}{\pi^2} \ve_1^2(\ve_2-\ve_1)^2  
    \left( \frac{\w}{c}\right)^4 \frac{\cos^2\theta_s}{\cos\theta_0} g(|\bqp - \bkp |) 
    \frac{\a^2_2(\qp )}{|d_p(\qp )|^2} \left([\pvecUnit{q} \times \pvecUnit{k}]_3\right)^2 \frac{\a^2_1(\kp)}{|d_s(\kp )|^2} \qquad
    \label{eq:5.20ps}
    \\
 \left\la \frac{\p R_{ss}(\bqp |\bkp )}{\p\Omega_s}\right\ra_\textrm{incoh} 
   &= 
    \frac{\delta^2}{\pi^2} \ve_1(\ve_2-\ve_1)^2  
    \left( \frac{\w}{c}\right)^6 \frac{\cos^2\theta_s}{\cos\theta_0} g(|\bqp - \bkp |) 
    \frac{1}{|d_s(\qp )|^2} (\pvecUnit{q}\cdot\pvecUnit{k})^2 \frac{\a^2_1(\kp )}{|d_s(\kp )|^2}, \qquad
    \label{eq:5.20b}
\end{align}
\end{subequations}
where the functions $d_\alpha(\qp)$ and $d_\alpha(\kp)$ for $\alpha=p,s$ are presented in Eq.~\eqref{app:eq:A11} as $d^+_\alpha(\qp)$ and $d^+_\alpha(\kp)$. 
%
\begin{figure}[tbh]
  \centering
  \includegraphics[width=0.47\columnwidth]{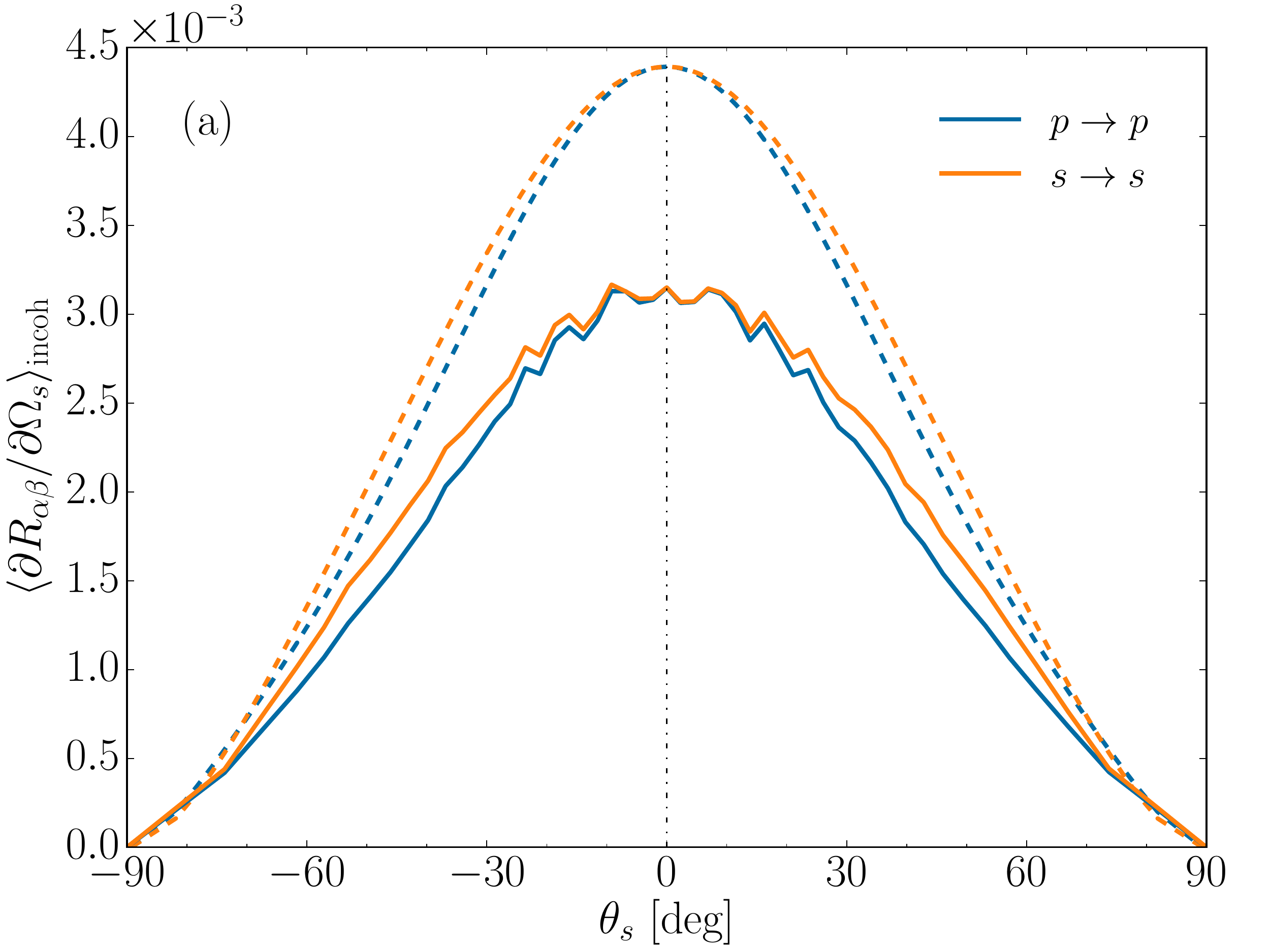}
  \includegraphics[width=0.47\columnwidth]{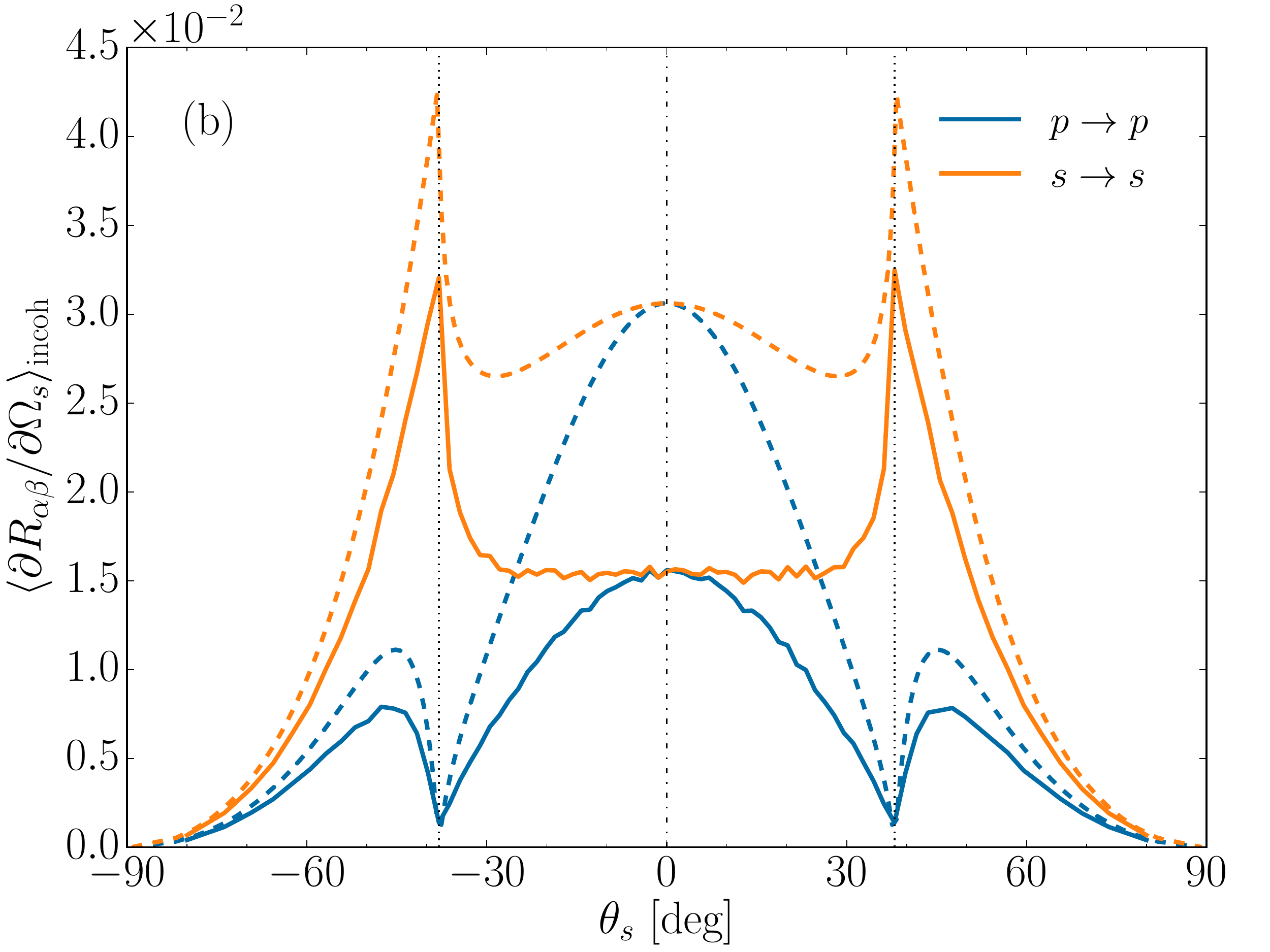}
  \caption{Same as Fig.~\ref{fig:inplane_mdrc_theta_0} but for the root-mean-square roughness $\delta=\lambda/20$. 
 }
\label{fig:inplane_mdrc_theta_0_div20}
\end{figure}
The results of a numerical evaluation of Eq.~\eqref{eq:5.20_R} for normal incidence and in-plane scattering [$\pvecUnit{q}\parallel\pvecUnit{k}$], are displayed as dashed curves in Fig.~\ref{fig:inplane_mdrc_theta_0}. By comparing the curves obtained from small amplitude perturbation theory to the results obtained from a purely numerical solution of the reduced Rayleigh equation, Eq.~\eqref{eq:3.24}, we conclude that SAPT for the considered level of roughness, even to lowest non-zero order in the surface profile function as in Eq.~\eqref{eq:5.20_R}, reproduces all the important features found in the mean differential reflection coefficients fairly well, but with a discrepancy in the amplitudes. This discrepancy decreases with decreasing surface roughness (results not shown). For example, similar comparisons for surfaces with an rms-roughness of $\delta=\lambda/80$, but with the same correlation length $a=\lambda/4$, show that the ability of Eq.~\eqref{eq:5.20_R} to reproduce the results based on the RRE is excellent for such weakly rough surfaces. However, for surfaces of rms-roughness $\delta=\lambda/20$ and the same correlation length, significant discrepancies are observed both in intensity (or amplitude) and angular dependence between the curves obtained on the basis of SAPT and the corresponding curves resulting from a numerical solution of the RRE [Fig.~\ref{fig:inplane_mdrc_theta_0_div20}].  For instance, from Fig.~\ref{fig:inplane_mdrc_theta_0_div20}(b) it is observed that the angular dependence of the $\spol\to \spol$ scattered intensity around the normal scattering direction is not correctly reproduced by SAPT; in this angular interval the numerical simulation results are almost constant and therefore essentially independent of $\theta_s$. These results illustrate the importance and necessity to go beyond lowest order SAPT or to do numerical simulations. We therefore stress the point that even if we in the following often turn to SAPT for interpretation of the nonperturbative solutions to the RRE, any conclusion drawn on the basis of Eq.~\eqref{eq:5.20_R} is correct only to the lowest non-zero order in the surface profile function. 

Results similar to those presented in Figs.~\ref{fig:inplane_mdrc_theta_0} and \ref{fig:inplane_mdrc_theta_0_div20}  but for scattering \textit{out-of-plane} [$\pvecUnit{q}\cdot\pvecUnit{k}=0$] are not presented, since, for normal incidence, the results for co-polarized in-plane scattering are the same as the results for cross-polarized out-of-plane scattering when the polarization of the scattered light is the same in the two cases.  This symmetry is expected for isotropic surfaces like the ones we are investigating when the lateral momentum of the incident light is zero, supported by the observation that Eq.~\eqref{eq:5.20a} evaluated in-plane equals Eq.~\eqref{eq:5.20ps} evaluated out-of-plane, and correspondingly for Eqs.~\eqref{eq:5.20b} and \eqref{eq:5.20sp}, when $\kp=0$ and $\theta_0=0$.

In order to simplify the subsequent discussion we here express $d_\alpha(\qp)$ and $d_\alpha(\kp)$ in terms of the polar angles of incidence and scattering using the relations in Eq.~\eqref{eq:k_and_q_as_functions_of_angles}:
\begin{subequations}
\label{eq:5.21}
\begin{align}
d_p(\qp ) &= \sqrt{\ve_1}\frac{\w}{c} \bigg\{ \ve_2\cos\theta_s +  \ve_1 \bigg[ \frac{\ve_2}{\ve_1}-\sin^2\theta_s \bigg]\sfr \bigg\} \label{eq:5.21a}\\
d_s(\qp ) &= \sqrt{\ve_1}\frac{\w}{c} \bigg\{ \cos\theta_s + \bigg[ \frac{\ve_2}{\ve_1}-\sin^2\theta_s \bigg]\sfr \bigg\} \label{eq:5.21b}\\
d_p(\kp ) &= \sqrt{\ve_1}\frac{\w}{c} \bigg\{ \ve_2\cos\theta_0 +  \ve_1 \bigg[ \frac{\ve_2}{\ve_1}-\sin^2\theta_0 \bigg]\sfr \bigg\} \label{eq:5.21c}\\
d_s(\kp ) &= \sqrt{\ve_1}\frac{\w}{c} \bigg\{ \cos\theta_0 + \bigg[ \frac{\ve_2}{\ve_1}-\sin^2\theta_0 \bigg]\sfr \bigg\} . \label{eq:5.21d}
\end{align}
\end{subequations}
%
We see from Eq.~(\ref{eq:5.21}) that when $\ve_2$ is greater than $\ve_1$, both $d_p(\qp )$ and $d_s(\qp )$ are real continuous functions of $\theta_s$ in the interval $0<|\theta_s|<\pi/2$, and so therefore are $|d_p(\qp )|^2$ and $|d_s(\qp )|^2$. Hence, no features are introduced into the corresponding mean differential reflection coefficients by these functions.  
However, when $\ve_1$ is greater than $\ve_2$, the function $\left[(\ve_2/\ve_1)-\sin^2\theta_s\right]\sfr$ present in $d_p(\qp )$ and $d_s (\qp )$ vanishes when $|\theta_s|$ equals the critical angle $\theta_c = \sin^{-1}\sqrt{\ve_2/\ve_1}$ for total internal reflection for the corresponding flat-surface system, and becomes purely imaginary as $|\theta_s|$ increases beyond this angle. The functions $|d_p(\qp )|^{2}$ and $|d_s(\qp)|^{2}$ therefore both have minima at $|\theta_s|=\theta_c$.
In the case of $\spol\to\spol$ in-plane scattering at normal incidence, the minima in the function $|d_s(\qp)|^2$ for $\ve_1>\ve_2$ lead to sharp peaks at $|\theta_s|=\theta_c$ in $\R{ss}{q}$, as displayed in Fig.~\ref{fig:inplane_mdrc_theta_0}(b). These same peaks will then also be present for $\ppol\to\spol$ out-of-plane scattering at normal incidence. However, for $\ppol\to\ppol$ in-plane scattering, while there are still minima in the function $|d_p(\qp)|^2$, we see from Eq.~\eqref{eq:5.20a} that $\R{pp}{q}$, to lowest order in the surface profile function, vanishes when the function 
\begin{align}
  F(\bqp|\bkp) = \left| \ve_2\qp\kp - \ve_1\alpha_2(\qp )(\pvecUnit{q}\cdot\pvecUnit{k})\alpha_2(\kp ) \right|^2 
  \label{eq:pp_zero}
\end{align}
vanishes.  For normal incidence ($\kp=0$) and in-plane scattering ($\pvecUnit{q}\parallel\pvecUnit{k}$), we see from this expression that $\R{pp}{q}$ will vanish when $\a_2(\qp)=0$, which is when $\qp=\sqrt{\ve_2}\w/c$ [see Eq.~\eqref{eq:k_and_q_as_functions_of_angles}]. This will be the case for $\theta_s<\ang{90}$ only when $\ve_1>\ve_2$, and the expression for $\R{pp}{q}$ in Eq.~\eqref{eq:5.20a} will in this case be zero for $|\theta_s|=\theta_c$, explaining the dips shown in Fig.~\ref{fig:inplane_mdrc_theta_0}(b) for $\ppol\to\ppol$ scattering. We note in passing that the out-of-plane distribution of $\R{ps}{q}$ also shows dips at the same polar angles, due to the factor $\a_2(\qp)$ in Eq.~\eqref{eq:5.20ps}, but these dips will be present regardless of the angle of incidence.

The peaks observed for $\theta_s\geq\theta_c$ in Figs.~\ref{fig:inplane_mdrc_theta_0} and \ref{fig:inplane_mdrc_theta_0_div20} for $\ve_1>\ve_2$ are optical analogues of the \textit{Yoneda peaks} observed by Y.~Yoneda in the scattering of x-rays from a metal surface~\cite{Yoneda1963}
and described as ``quasi-anomalous scattering peaks'' in the two-dimensional work by Kawanishi~\etal~\cite{Kawanishi1997}. The Yoneda peaks were originally observed as sharp peaks for incidence close to the grazing angle, as the difference in the dielectric constants of the two scattering media is very small at x-ray frequencies. In the following, by Yoneda peaks we will mean well-defined maxima in the angular distribution of the intensity of the scattered light at, or slightly above, the critical polar angle for total internal reflection, when $\varepsilon_1 > \varepsilon_2$. 

Although the mathematical origin of the Yoneda peaks is clear from Eqs.~\eqref{eq:5.20_R} and \eqref{eq:5.21}, namely, they are associated with the minimum of the functions $|d_{p,s}(q_\parallel)|^2$ and $|d_{p,s}(k_\parallel)|^2$, a physical interpretation of them is still under discussion. Thus, Warren and Clarke~\cite{Warren1965} in a study of the reflection of x-rays from a polished surface (mirror), proposed that these peaks arise when the incident beam at a grazing angle of incidence $\theta$ that is greater then the grazing critical angle for total internal reflection $\theta_c$ is scattered through a small angle $\beta$ by something just above the mirror surface. The scattered field falls upon the mirror at a grazing angle $\alpha$, and strong reflection occurs when $\alpha<\theta_c$. This reflection is cutoff sharply for $\alpha>\theta_c$ and less sharply for $\alpha<\theta_c$ by the rapidly decreasing intensity of small-angle scattering. This produces an asymmetric peak in the intensity of the scattered field, whose maximum occurs at the critical angle. It was suggested that the scatterer could be a projection on an irregular surface. 

In a subsequent study of the grazing-angle reflection of x-rays from rough metal surfaces with the use of the distorted-wave Born approximation~\cite{Book:Schiff1968}, Vineyard~\cite{Vineyard1982} noted from an examination of the angular dependence of the magnitude of the Fresnel coefficient for transmission through a planar vacuum-metal interface, that it produces a transmitted field on the surface whose angular dependence has the form of an asymmetric peak. The peak maximum occurs at the critical angle, and has a magnitude that is twice that of the incident electric field on the surface, leading to enhanced diffuse scattering at this angle. This ``Vineyard effect'' was invoked by Sinha~\etal~\cite{Sinha1988} as the origin of the Yoneda peaks. This result is mathematically similar to the explanation provided by Eqs.~\eqref{eq:5.20_R} and \eqref{eq:5.21}, but it is not a physical explanation for these peaks.


Such an interpretation was offered by Kawanishi~\etal~\cite{Kawanishi1997}, who suggested that the Yoneda peaks may be associated with the presence of lateral waves~\cite{Book:Tamir1982} propagating along the interface in the optically less dense medium. This wave satisfies the condition for refraction back into the optically more dense medium, and it therefore leaks energy at every position along the interface, along rays whose scattering angle $\theta_s$ equals $\theta_c$~\cite{Tamir1969}. This radiation is restricted to the range $\theta_c<\theta_s<\pi/2$. This is an attractive explanation, but it needs to be explored more through additional calculations.

\begin{figure}[tbh]
  \centering
  \includegraphics[width=0.8\columnwidth]{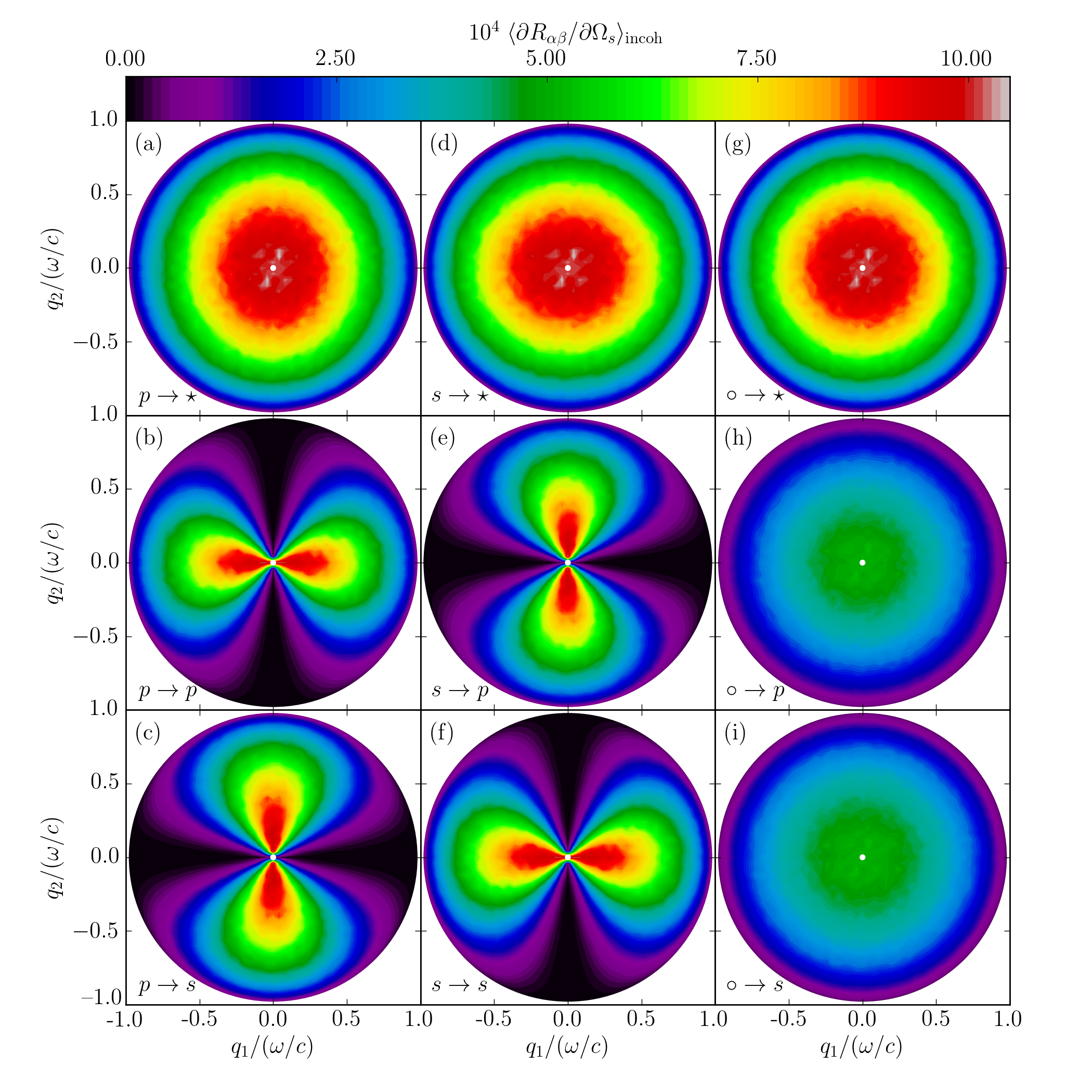}
  \caption{The incoherent component of the mean differential reflection coefficient, showing the full angular intensity distribution as a function of the lateral wave vector of the light scattered by a rough interface between vacuum and a dielectric. The light was incident on the surface from the vacuum, [$\ve_1=1.0$, $\ve_2=2.64$]. The angles of incidence were $(\theta_0,\phi_0)=(\ang{0},\ang{0})$. The position of the specular direction in reflection is indicated by white dots. The parameters assumed for the scattering geometry and used in performing the numerical simulations had values that are identical to those assumed in obtaining the results of Fig.~\protect\ref{fig:inplane_mdrc_theta_0}(a). The in-plane intensity variations in Figs.~\protect\ref{fig:2Dmdrc_vtp_theta_0}(b) and ~\protect\ref{fig:2Dmdrc_vtp_theta_0}(f) are the curves depicted in Fig.~\protect\ref{fig:inplane_mdrc_theta_0}(a). The star notation, \textit{e.g.} $p\rightarrow\star$,
  indicates that the polarization of the scattered light was not recorded, hence Fig.~\ref{fig:2Dmdrc_vtp_theta_0}(a) is the sum of Figs.~\ref{fig:2Dmdrc_vtp_theta_0}(b) and \ref{fig:2Dmdrc_vtp_theta_0}(c), and Fig.~\ref{fig:2Dmdrc_vtp_theta_0}(d) is the sum of Figs.~\ref{fig:2Dmdrc_vtp_theta_0}(e) and \ref{fig:2Dmdrc_vtp_theta_0}(f). Furthermore, for the subfigures in the third column the open circle in \textit{e.g.} $\circ\rightarrow\star$ symbolizes that the incident light was unpolarized; these simulation results were obtained by taking the arithmetic average of the other two subfigures in the same row. The roughness parameters assumed in obtaining these results were $\delta=\lambda/40$ and $a=\lambda/4$. }
\label{fig:2Dmdrc_vtp_theta_0}
\end{figure}

\begin{figure}[tbh] 
  \centering
  \includegraphics[width=0.8\columnwidth]{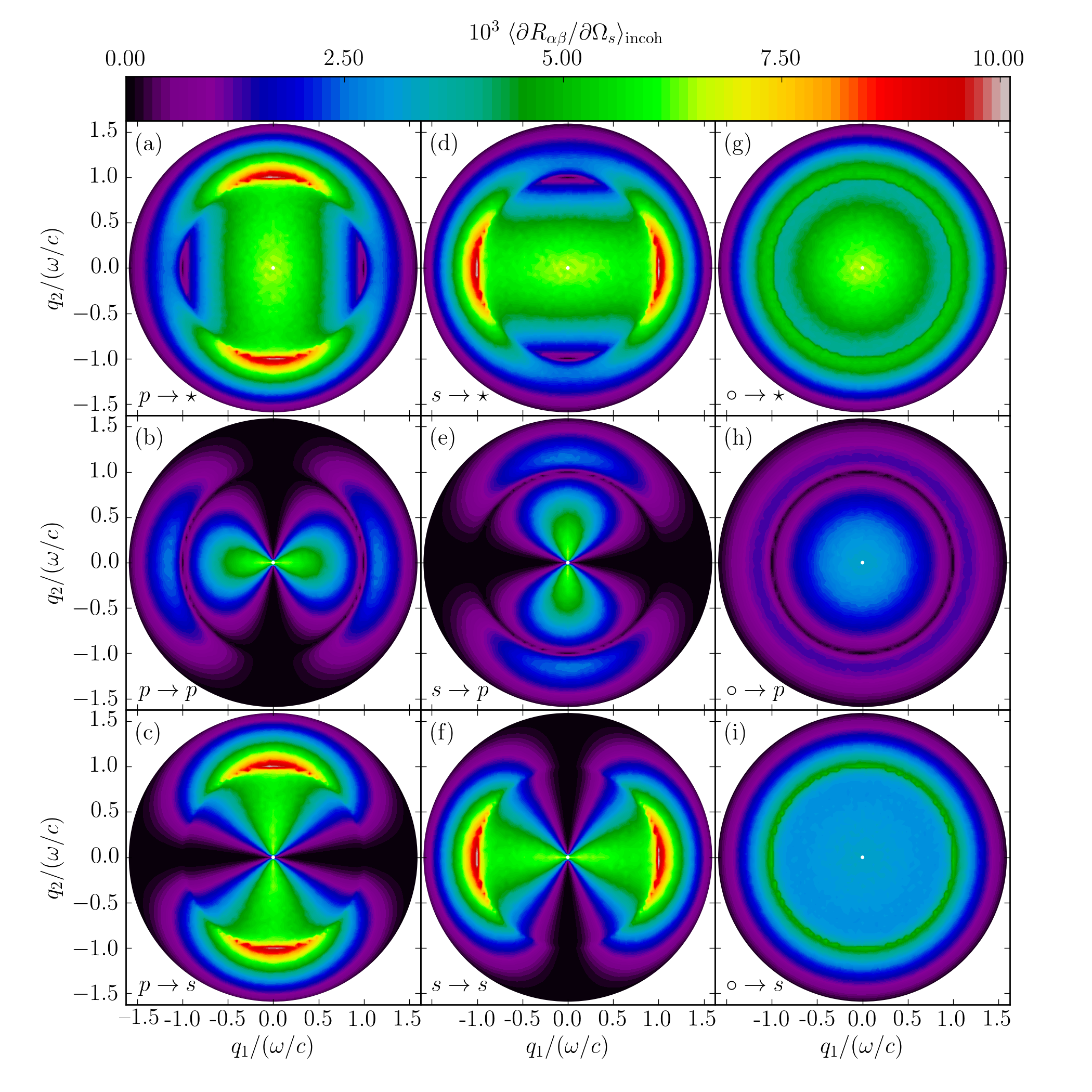}
  \caption{The same as Fig.~\protect\ref{fig:2Dmdrc_vtp_theta_0}, but for light incident from the dielectric side onto the interface with vacuum, [$\ve_1=2.64$; $\ve_2=1.0$]. The in-plane intensity variations in Figs.~\protect\ref{fig:2Dmdrc_ptv_theta_0}(b) and ~\protect\ref{fig:2Dmdrc_ptv_theta_0}(f) are the curves depicted in Fig.~\protect\ref{fig:inplane_mdrc_theta_0}(b).
   Notice the rapid changes in intensity around the polar angle $\theta_s=\theta_c=\sin^{-1}{\sqrt{\ve_1/\ve_2}}$ [or~$\qp=\sqrt{\ve_2}\omega/c$].
  }
 \label{fig:2Dmdrc_ptv_theta_0}
\end{figure}

\bigskip
We have also calculated the full angular intensity distributions of the reflected light. Figures~\ref{fig:2Dmdrc_vtp_theta_0} and \ref{fig:2Dmdrc_ptv_theta_0} present such simulation results for the contribution to the mean differential reflection coefficient from the light that has been scattered incoherently by the randomly rough interface. The angles of incidence were set to $(\theta_0,\phi_0)=(\ang{0},\ang{0})$; it was cross-sectional cuts along the plane of incidence of these angular intensity distributions that resulted in the solid curves presented in Fig.~\ref{fig:inplane_mdrc_theta_0}. The parameters assumed in producing the results of Figs.~\ref{fig:inplane_mdrc_theta_0}(a)  and \ref{fig:2Dmdrc_vtp_theta_0}, and Figs.~\ref{fig:inplane_mdrc_theta_0}(b) and \ref{fig:2Dmdrc_ptv_theta_0} are therefore identical.

Figures~\ref{fig:2Dmdrc_vtp_theta_0} and \ref{fig:2Dmdrc_ptv_theta_0}, and all following full angular intensity distributions, are organized with a similar layout: They are arranged in $3\times 3$ subfigures where each row and column of the array correspond to the angular distribution of the incoherent component of the mean differential reflection coefficient for a given state of polarization of the scattered and incident light, respectively. 
The lower left $2\times 2$ corner of such figures corresponds to the cases where $\beta$-polarized incident light is reflected by the rough interface into $\alpha$-polarized light, denoted $\beta\rightarrow\alpha$ in the lower left corner of each subfigure, where $\alpha=\ppol,\spol$ and the same for $\beta$. Moreover, the first row corresponds to results where the polarization of the reflected light was not recorded (indicated by $\star$); such results are obtained by adding the other two results from the same column. The rightmost column presents results for which the incident light is \textit{unpolarized} (indicated by an open circle, $\circ$); these results are obtained by taking the arithmetic average of the other two results present in the same row.

The lower left $2\times 2$ corners of Figs.~\ref{fig:2Dmdrc_vtp_theta_0} and \ref{fig:2Dmdrc_ptv_theta_0} display dipole-like patterns oriented along the plane of incidence for co-polarization and perpendicular to it for cross-polarization. This is a consequence of the definition used for the polarization vectors in our system. 
Moreover, Figs.~\ref{fig:2Dmdrc_vtp_theta_0}(g)--(i) and \ref{fig:2Dmdrc_ptv_theta_0}(g)--(i) show that for unpolarized incident light at normal incidence, the scattering distributions are independent of the azimuthal angle of scattering $\phi_s$. This rotational symmetry is expected for isotropic surfaces like the ones we are investigating when the lateral momentum of the unpolarized incident light is zero. However, for $\ve_1<\ve_2$ and when the incident light is linearly polarized but the polarization of the reflected light is not recorded, Figs.~\ref{fig:2Dmdrc_vtp_theta_0}(a) and (d), we observe a slight skew in the distributions. This is similar to results presented in other, similar work~\cite{Nordam2013a,Nordam2014}, and is due to the subtle differences between the distributions of $\ppol\to\ppol$ and $\spol\to\spol$ scattered light, as presented for in-plane scattering in Fig.~\ref{fig:inplane_mdrc_theta_0}.

When $\ve_1>\ve_2$ the Yoneda peaks form a circle of equal intensity at the polar angle $\theta_s=\theta_c$ [or $\qp = \sqrt{\ve_2}\omega/c$] in Fig.~\ref{fig:2Dmdrc_ptv_theta_0}(i), where unpolarized incident light is scattered by the surface into s-polarized light. Similarly, a circular groove of close-to-zero intensity [exactly zero according to SAPT, Eq.~\eqref{eq:5.20_R}] can be found at $\theta_s=\theta_c$ in Fig.~\ref{fig:2Dmdrc_ptv_theta_0}(h). The position and circular symmetry of this groove can be understood through the previously mentioned factors of $\a_2(\qp)$ and $\kp$ present in Eq.~\eqref{eq:5.20a} for $\ppol\to\ppol$ polarization and the factor $\a_2(\qp)$ present in Eq.~\eqref{eq:5.20ps} for $\spol\to\ppol$ polarization, since $\a_2(\qp)$ becomes zero when $\qp=\sqrt{\ve_2}\omega/c$ and $\kp$ is zero for normal incidence. It can be of interest to note that we also, as a consequence, observe a $\phi_s$-independent peak in Fig.~\ref{fig:2Dmdrc_ptv_theta_0}(h), at a polar scattering angle significantly larger than $\theta_c$: the same peak as seen for $\ppol\to\ppol$ scattering in Fig.~\ref{fig:inplane_mdrc_theta_0}(b). However, this peak is not as sharp as the peak found at $\theta_c$ in Fig.~\ref{fig:2Dmdrc_ptv_theta_0}(i), and according to our definition it is not a Yoneda peak. 

Equations~\eqref{eq:5.20_R} demonstrate that the angular intensity distributions we are investigating can, to lowest order in the surface profile function, be explained through different factors in these equations with good approximation. 
As an aid in the interpretation of the results presented here and in the following, we notice that the power spectrum of the surface, $g(|\pvec{q}-\pvec{k}|)$ is common for all equations in Eq.~\eqref{eq:5.20_R}. As such, the mean DRC in SAPT to lowest order is essentially a distorted Gaussian on which critical angle effects are superposed.

\subsection{Non-normal incidence}

\begin{figure}[tbh]
  \centering
  \includegraphics[width=0.47\columnwidth]{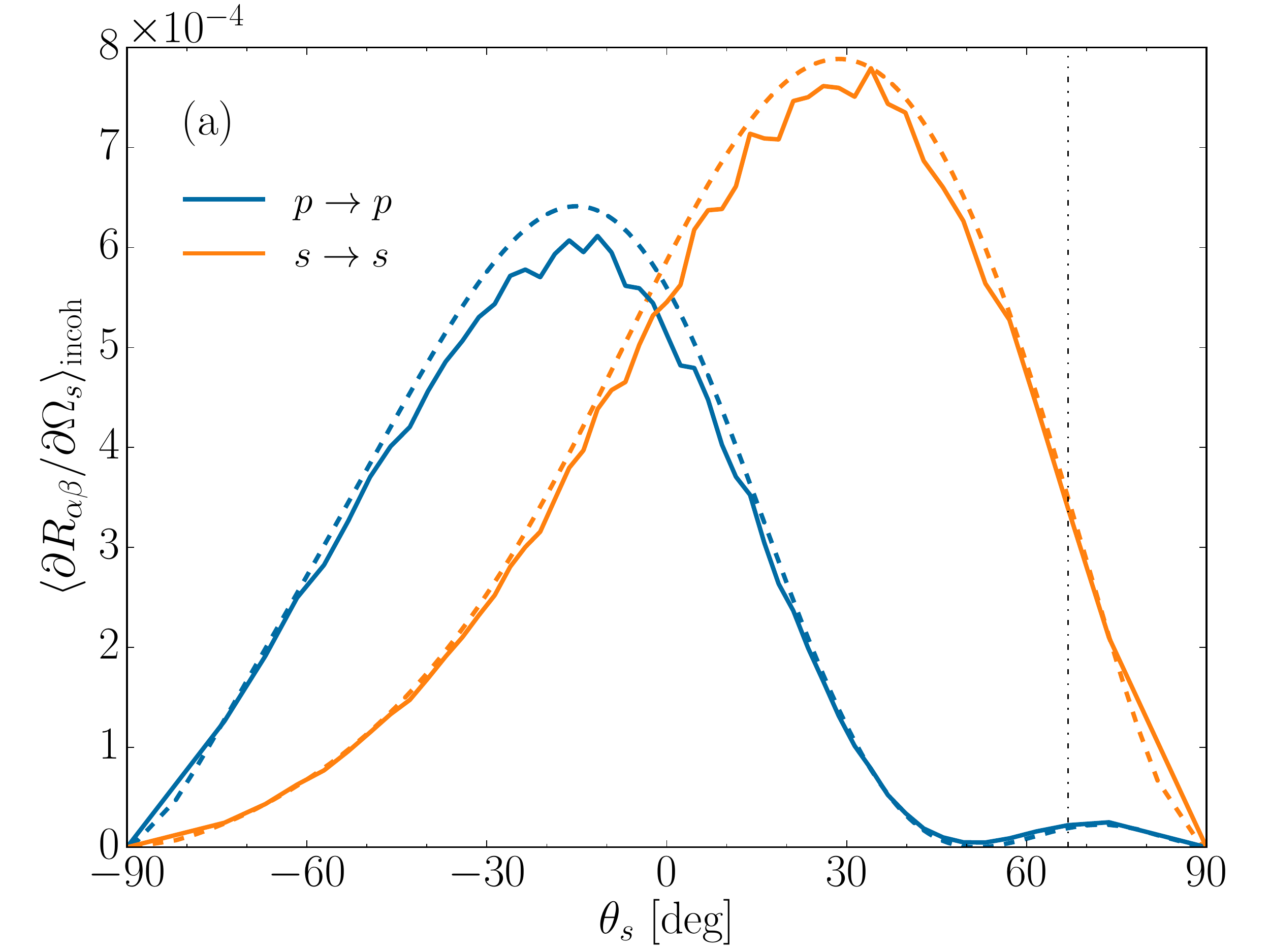}
  \includegraphics[width=0.47\columnwidth]{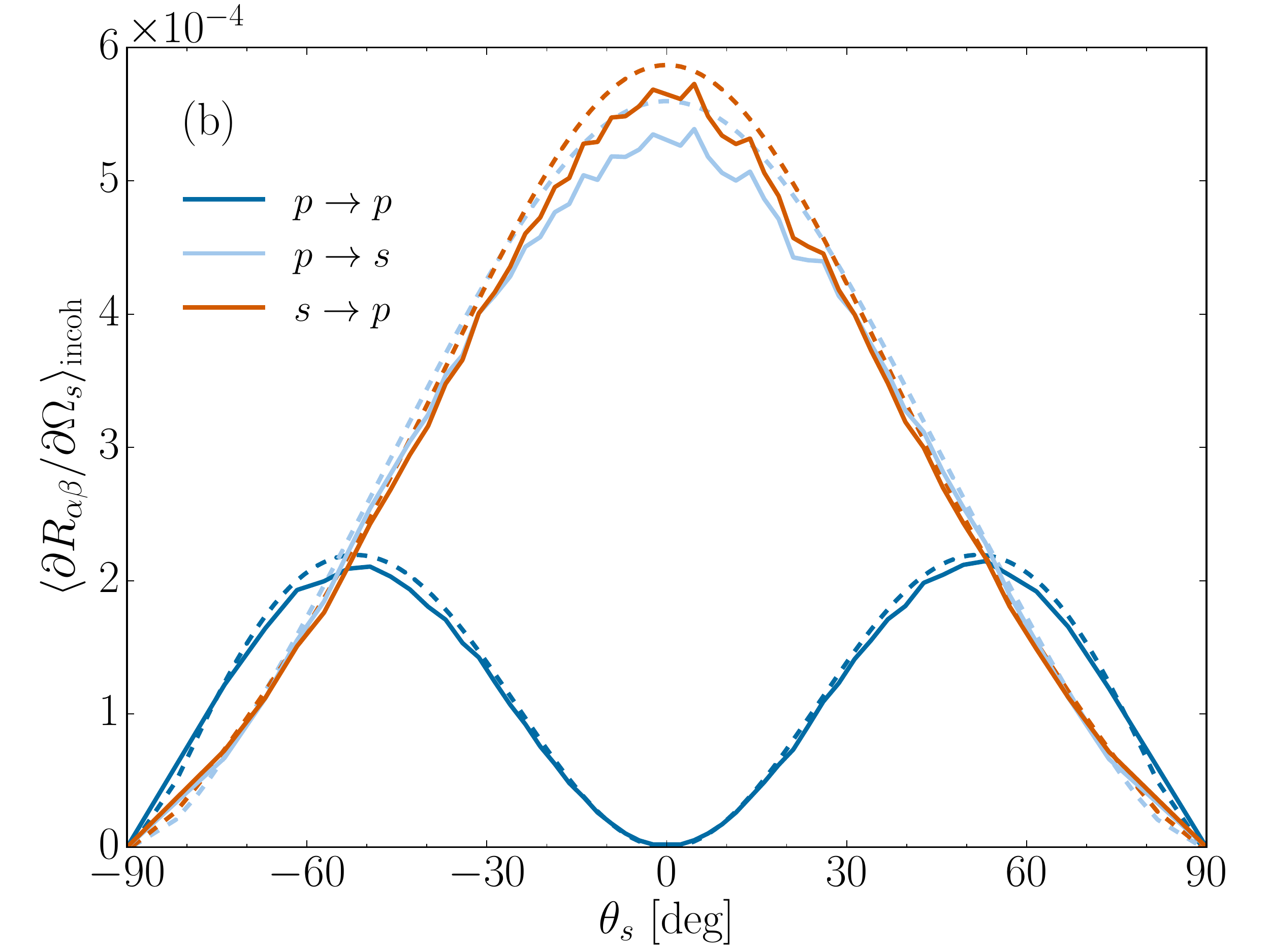}
  \caption{(a) Same as Fig.~\ref{fig:inplane_mdrc_theta_0}(a) but for angles of incidence $(\theta_0,\phi_0)=(\ang{66.9},\ang{0})$.
   (b) Same as Fig.~\ref{fig:cut_mdrc_theta_669}(a) but for out-of-plane scattering [$\phi_s=\pm\ang{90}$]. 
   Results for combinations of the polarizations of the incident and scattered light  for which the scattered intensity was everywhere negligible have been omitted. [Parameters: $\ve_1=1.0$, $\ve_2=2.64$; $\delta=\lambda/40$, $a=\lambda/4$].
 }
\label{fig:cut_mdrc_theta_669}
\end{figure}

\begin{figure}[tbh]
  \centering                               
  \includegraphics[width=0.47\columnwidth]{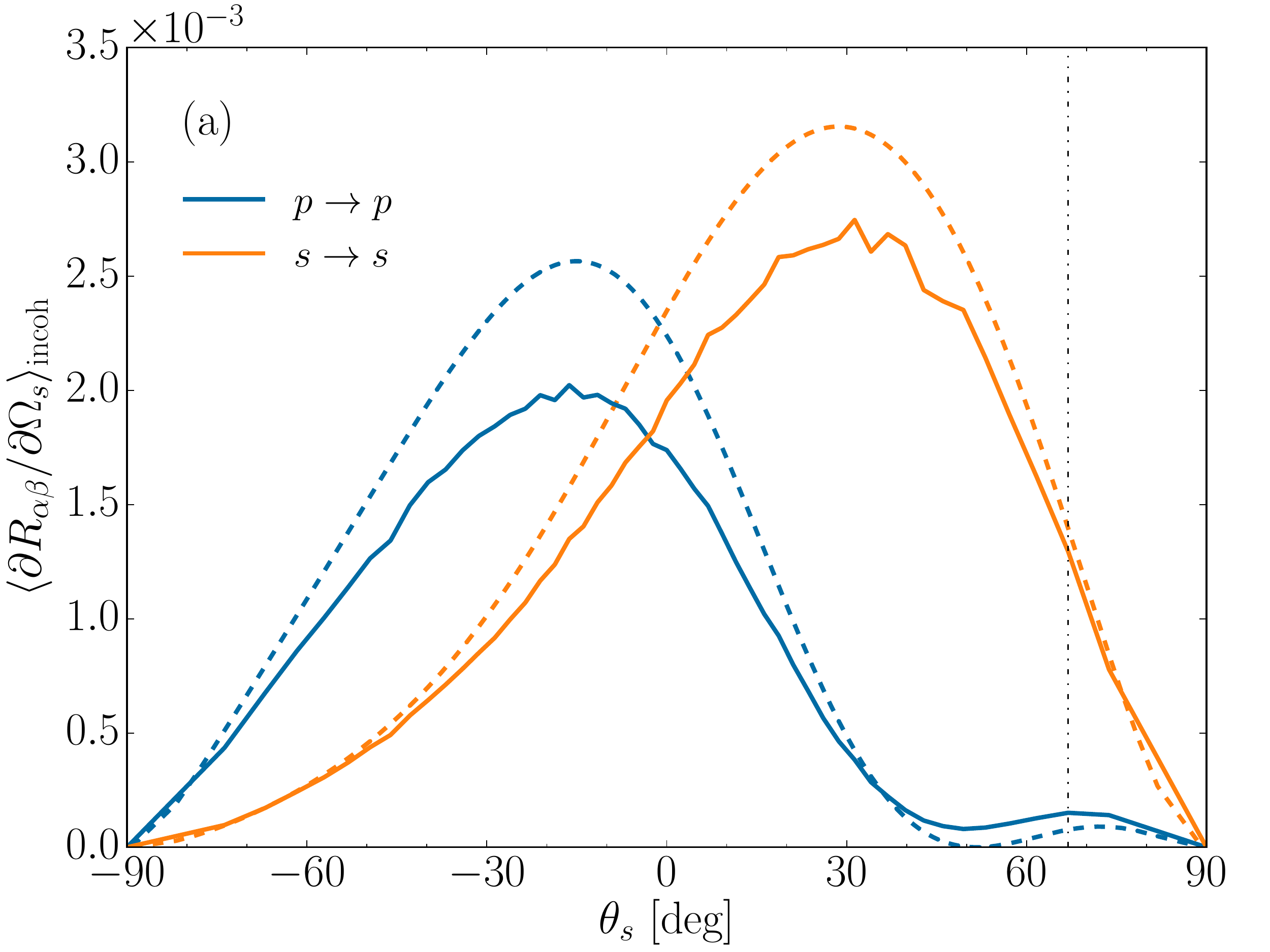}
  \includegraphics[width=0.47\columnwidth]{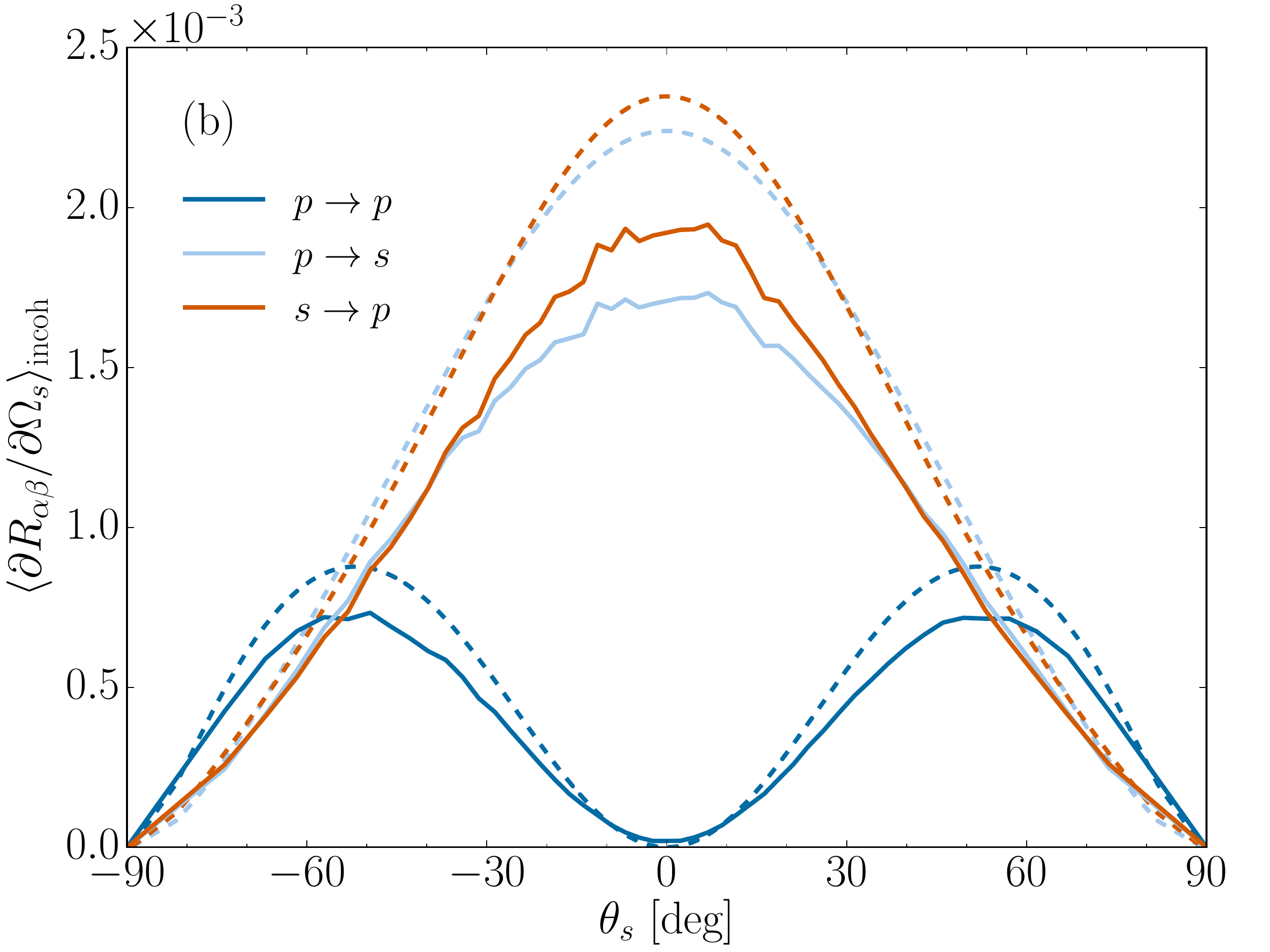}
  \caption{Same as Fig.~\protect\ref{fig:cut_mdrc_theta_669} but for root-mean-square roughness $\delta=\lambda/20$.
  }
\label{fig:cut_mdrc_theta_669_lambda_over_20}
\end{figure}

As a starting point for our discussion of results for non-normal incidence, we in Fig.~\ref{fig:cut_mdrc_theta_669}(a) present the angular dependence of the light scattered incoherently for a grazing angle of incidence from vacuum: $\theta_0=\ang{66.9}$. The scattering distribution for $\spol\to\spol$ scattered light can be seen to have retained its general shape from Fig.~\ref{fig:inplane_mdrc_theta_0}(a), but for $\ppol\to\ppol$ scattering we now observe a new feature: a local minimum at $\theta_s\approx\ang{50}$. In the case of small amplitude perturbation theory, represented in Fig.~\ref{fig:cut_mdrc_theta_669}(a) by dashed curves, $\R{pp}{q}$ goes to zero at the position of this minimum.

In order to explain this minimum for $\ppol\to\ppol$ scattering in Fig.~\ref{fig:cut_mdrc_theta_669}(a), we again turn to Eq.~\eqref{eq:5.20_R}.
For non-normal incidence ($\kp\neq0$), the function $F(\bqp|\bkp)$ in Eq.~\eqref{eq:pp_zero} can only cause $\R{pp}{q}$ to vanish when $\pvecUnit{q}\cdot\pvecUnit{k}$ is positive (forward scattering). Specifically, for in-plane forward scattering [$\pvecUnit{q}\cdot\pvecUnit{k}=1$], $\R{pp}{q}$ will vanish at a polar angle $\Theta_B$ given by
\begin{align}
  \label{eq:zero1}
  \Theta_B(\theta_0) = \sin^{-1}\left( \sqrt{\frac{\ve_2(\ve_2-\ve_1\sin^2\theta_0)}{(\ve_2^2-\ve_1^2)\sin^2\theta_0+\ve_1\ve_2}}\right).
\end{align}
The scattering intensity $\R{pp}{q}$ will therefore, to lowest non-zero order in SAPT, have a zero when $\ve_1<\ve_2$ and $\theta_0$ is in the interval $\sin^{-1}\left[ \ve_2(\ve_2-\ve_1)/(\ve_2^2+\ve_2\ve_1-\ve_1^2)\right]\sfr < \theta_0 < \pi/2$. 
For $\theta_0=\ang{66.9}$, as assumed in producing Fig.~\ref{fig:cut_mdrc_theta_669}(a), we therefore expect a local minimum in $\R{pp}{q}$ at $\theta_s=\Theta_B(\ang{66.9})=\ang{51.7}$, which is in good agreement with the observed value.

The scattering angles defined by $\Theta_B$ were first mentioned in the literature by Kawanishi~\etal~\cite{Kawanishi1997}, where the angular values of $\Theta_B$ were explored through a stochastic functional approach for two-dimensional surfaces. They chose to call the angles at which the first order contribution (according to their approach) to $\R{\alpha p}{q}$ vanishes the \textit{Brewster scattering angles}, as a generalization of the Brewster angle for a flat surface. 
In what follows, following Kawanishi~\etal, we will call the polar angles of scattering in the plane of incidence at which p- and s-polarized light is scattered diffusely (incoherently) into light of any polarization with zero, or nearly zero, intensity, the Brewster scattering angles. 

The Brewster angle $\theta_B$ is defined by the zero in the reflectivity from a flat surface (coherent reflection in the specular direction) for p-polarization at the angle of incidence given by $\theta_0 = \theta_B = \tan^{-1}(\sqrt{\ve_2/\ve_1})$. For one set of $\{\ve_1,\ve_2\}$, there is hence only one Brewster angle for incidence in a given medium. However, in contrast, we would like to stress the fact that the Brewster scattering angles for $\ppol\to\ppol$ scattering are present for a wide range of angles of incidence, given by Eq.~\eqref{eq:zero1} for in-plane scattering. From Eq.~\eqref{eq:zero1} it is also of interest to note that for light \textit{incident} at the Brewster angle (for the corresponding flat-surface system), $\theta_0=\theta_B$, we find that $\Theta_B(\theta_B)=\theta_B$; the scattering intensity for light scattered incoherently vanishes for a scattering angle equal to the Brewster angle.
This attests to the close relation between the Brewster angle for coherent reflection and the Brewster scattering angle $\Theta_B$ for diffuse reflection, and is consistent with the findings of Kawanishi~\etal~\cite{Kawanishi1997}.


Figure~\ref{fig:cut_mdrc_theta_669}(b) presents simulation results for the same configuration as in Fig.~\ref{fig:cut_mdrc_theta_669}(a), but for light scattered out-of-plane [$\pvecUnit{q} \cdot \pvecUnit{k}=0$]. The dot product $\pvecUnit{q} \cdot \pvecUnit{k}$ in Eq.~\eqref{eq:5.20b} indicates that, to lowest non-zero order in SAPT, we should not expect any contribution to the mean DRC from $\spol\to\spol$ out-of-plane incoherently scattered light. However, this is not the case for $\ppol\to\ppol$ scattered light, where, even for $\pvecUnit{q} \cdot \pvecUnit{k}=0$, a closer look at Eq.~\eqref{eq:5.20a} indicates that the out-of-plane scattered intensity is zero only for $\theta_s=0$ [$\qp=0$]. This is precisely what we observe for $\R{pp}{q}$ in Fig.~\ref{fig:cut_mdrc_theta_669}(b).

Figure~\ref{fig:cut_mdrc_theta_669_lambda_over_20} depicts results similar to those presented in Fig.~\ref{fig:cut_mdrc_theta_669} but for an increased surface rms-roughness of $\delta=\lambda/20$ with the remaining parameters unchanged. As for normal incidence, it is found, not surprisingly, that small amplitude perturbation theory is most accurate for the smallest surface roughness. However, the most interesting feature to notice from Fig.~\ref{fig:cut_mdrc_theta_669_lambda_over_20}(a), as compared to  Fig.~\ref{fig:cut_mdrc_theta_669}(a), is the angular position and amplitude of the local minimum of the in-plane $\ppol \to \ppol$ intensity distribution. In the former figure [Fig.~\ref{fig:cut_mdrc_theta_669_lambda_over_20}(a)], the intensity at the position of the minimum is non-zero and it is located at an angle that is smaller than the Brewster scattering angle $\Theta_B(\theta_0)$ predicted by Eq.~\eqref{eq:zero1}. We speculate that this shift in the  Brewster scattering angle is roughness induced in a way similar to how the ``normal'' Brewster angle is shifted by the introduction of surface roughness.

\begin{figure}[tbh]
  \centering
  \includegraphics[width=0.8\columnwidth]{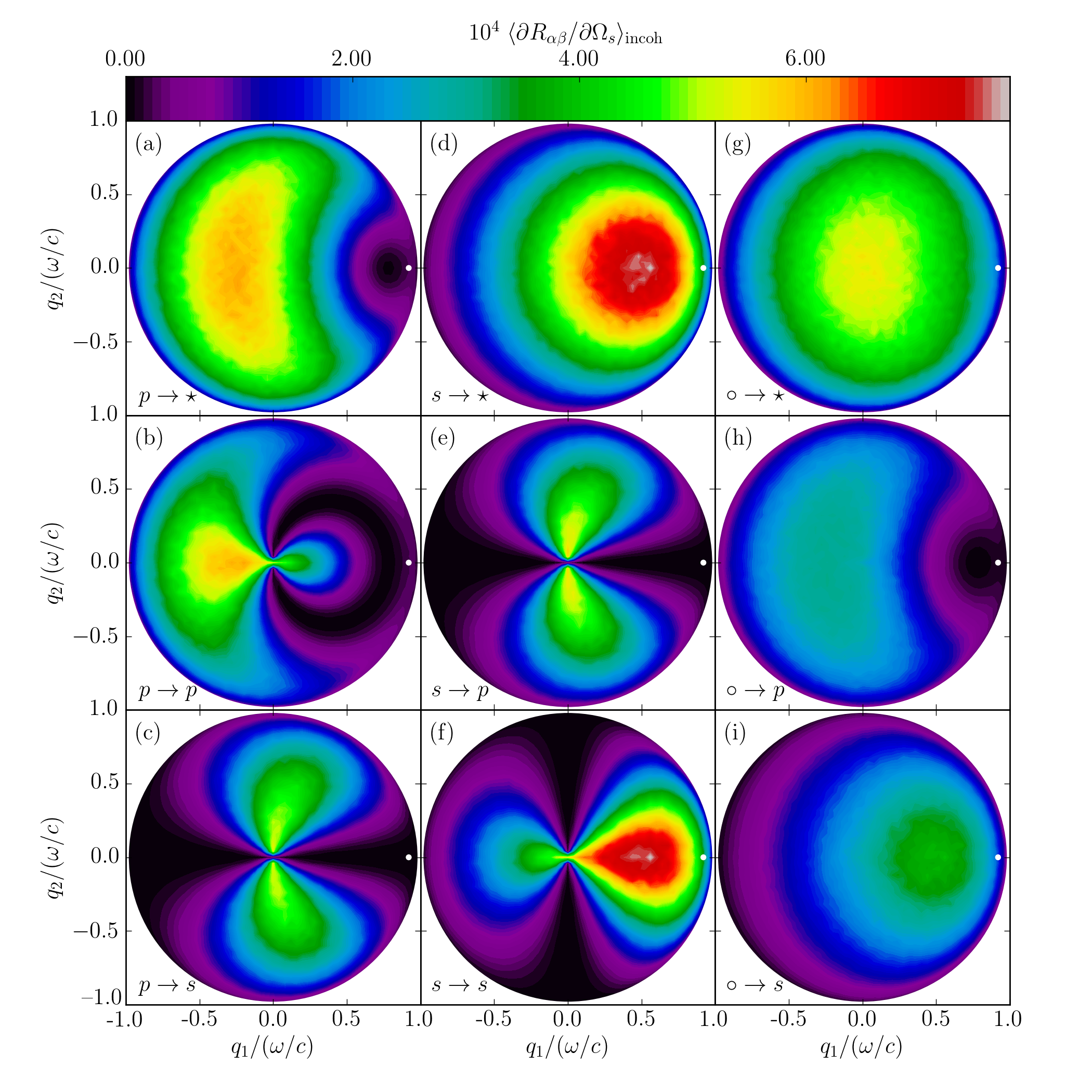}
  \caption{Same as Fig.~\protect\ref{fig:2Dmdrc_vtp_theta_0} but for the angles of incidence $(\theta_0,\phi_0)=(\ang{66.9},\ang{0})$.
    [Parameters: $\ve_1=1.0$, $\ve_2=2.64$; $\delta=\lambda/40$, $a=\lambda/4$].
  }
\label{fig:2Dmdrc_vtp_theta_669}
\end{figure}

\smallskip
The results presented in Fig.~\ref{fig:cut_mdrc_theta_669} were in-plane and out-of-plane cuts from 
Fig.~\ref{fig:2Dmdrc_vtp_theta_669}, which presents the full angular distribution of the contributions to the mean DRC from incoherently scattered light for the angles of incidence $(\theta_0,\phi_0)=(\ang{66.9},\ang{0})$. 
Here the white dots indicate the lateral wavevector of the specular reflection, $\pvec{k}$. Compared to the results presented in Fig.~\ref{fig:2Dmdrc_vtp_theta_0}, Fig.~\ref{fig:2Dmdrc_vtp_theta_669} displays many interesting features that are strongly dependent on both incoming and outgoing polarization, and we are in Fig.~\ref{fig:2Dmdrc_vtp_theta_669} left with symmetry in the distributions only about the plane of incidence.
%
For $\ppol\to \ppol$ polarized reflection, Fig.~\ref{fig:2Dmdrc_vtp_theta_669}(b), we observe that a significant fraction of the incoherently scattered light has shifted into the backscattering portion of the $\qp$-plane as the angle of incidence has increased. 
The opposite is true for $\spol\to \spol$ polarized reflection, Fig.~\ref{fig:2Dmdrc_vtp_theta_669}(f), where the majority of the incoherently scattered light is scattered into the forward portion of the $\qp$-plane. 
This can be understood through small amplitude perturbation theory: In Eq.~\eqref{eq:5.20_R}, the function $F(\bqp|\bkp)$ [Eq.~\eqref{eq:pp_zero}] constitutes the main difference between $\spol\to\spol$ and $\ppol\to \ppol$ polarized scattering, and it is easy to see that this term will enhance the backward scattering and reduce the forward scattering for $\ppol\to\ppol$ polarization. 
%
Additionally, the Brewster scattering angle, which for $\theta_0=\ang{66.9}$ was given by Eq.~\eqref{eq:zero1} and found to be at $\theta_s=\ang{51.7}$ for the parameters assumed, can now be seen to be part of a more general but still localized minimum in both Fig.~\ref{fig:2Dmdrc_vtp_theta_669}(a) and Fig.~\ref{fig:2Dmdrc_vtp_theta_669}(h), \textit{i.e.}, for $\ppol\to\star$ and $\circ\to\ppol$ scattering respectively.
Figure~\ref{fig:2Dmdrc_vtp_theta_669}(b) shows that the Brewster scattering angle for $\ppol\to\ppol$ polarized scattering can be found to be part of an interestingly shaped minimum in the $\qp$-plane. The shape of this minimum can, however, be extracted in a straightforward manner from Eq.~\eqref{eq:pp_zero}.

\begin{figure}[tbh]
  \centering
  \includegraphics[width=0.47\columnwidth]{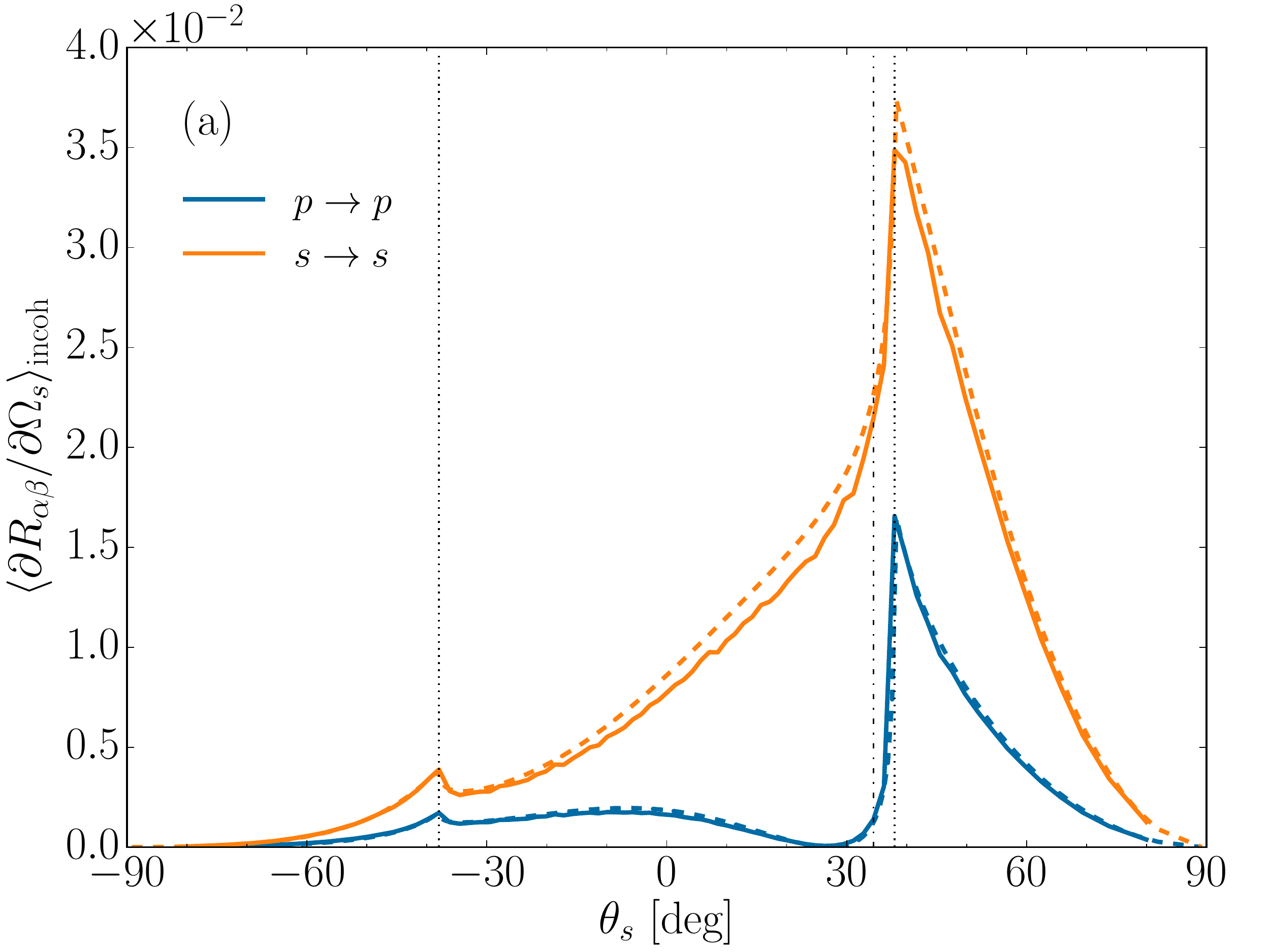}
  \includegraphics[width=0.47\columnwidth]{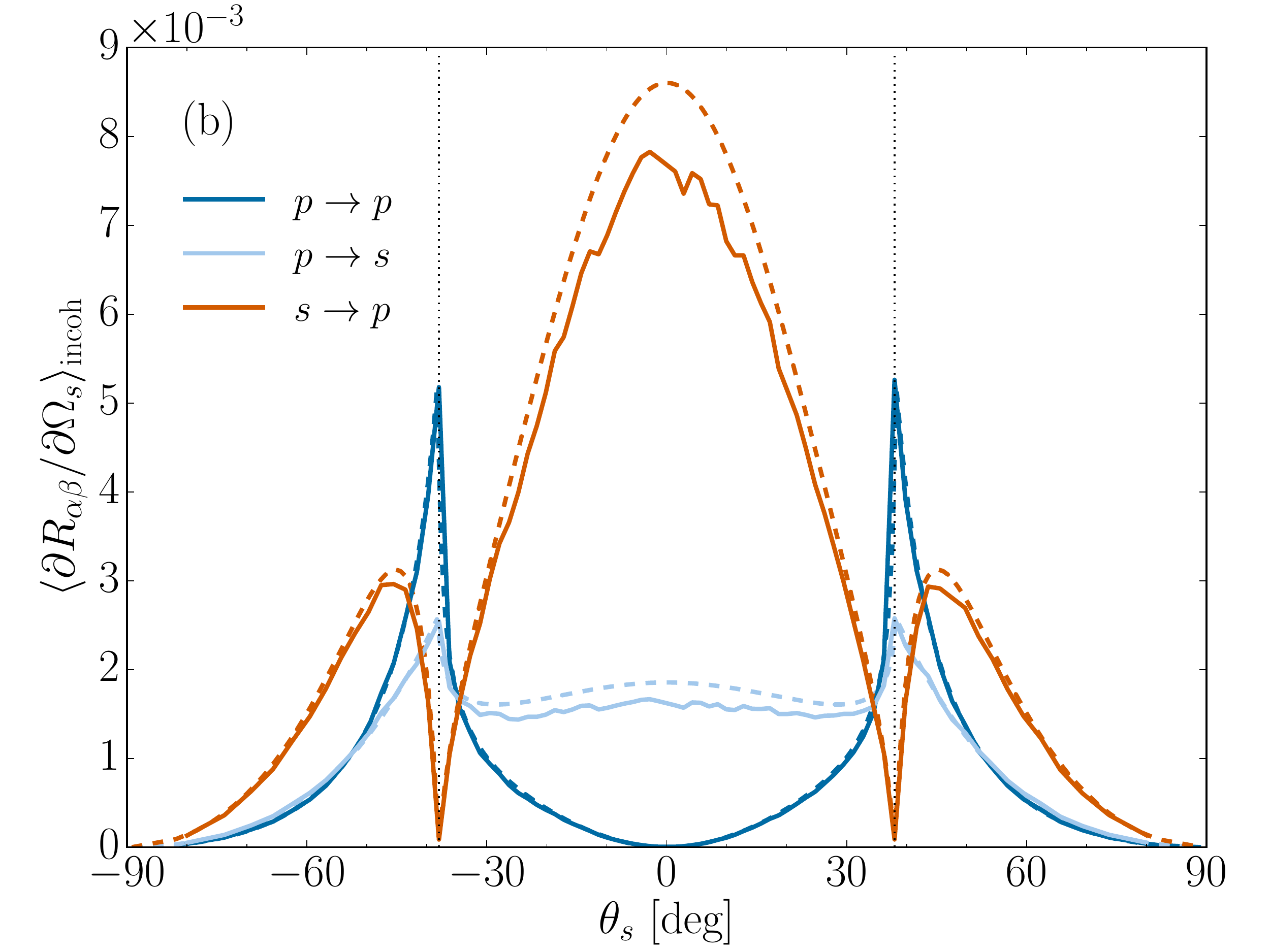}
  \caption{
    (a) Same as Fig.~\ref{fig:inplane_mdrc_theta_0}(b) but for angles of incidence $(\theta_0,\phi_0)=(\ang{34.5},\ang{0})$.
   (b) Same as Fig.~\ref{fig:cut_mdrc_theta_345}(a) but for out-of-plane scattering [$\phi_s=\pm\ang{90}$]. 
   Results for combinations of the polarizations of the incident and scattered light  for which the scattered intensity was everywhere negligible have been omitted. [Parameters: $\ve_1=2.64$, $\ve_2=1.0$; $\delta=\lambda/40$, $a=\lambda/4$].
 }
\label{fig:cut_mdrc_theta_345}
\end{figure}

\begin{figure}[tbh]
  \centering
  \includegraphics[width=0.47\columnwidth]{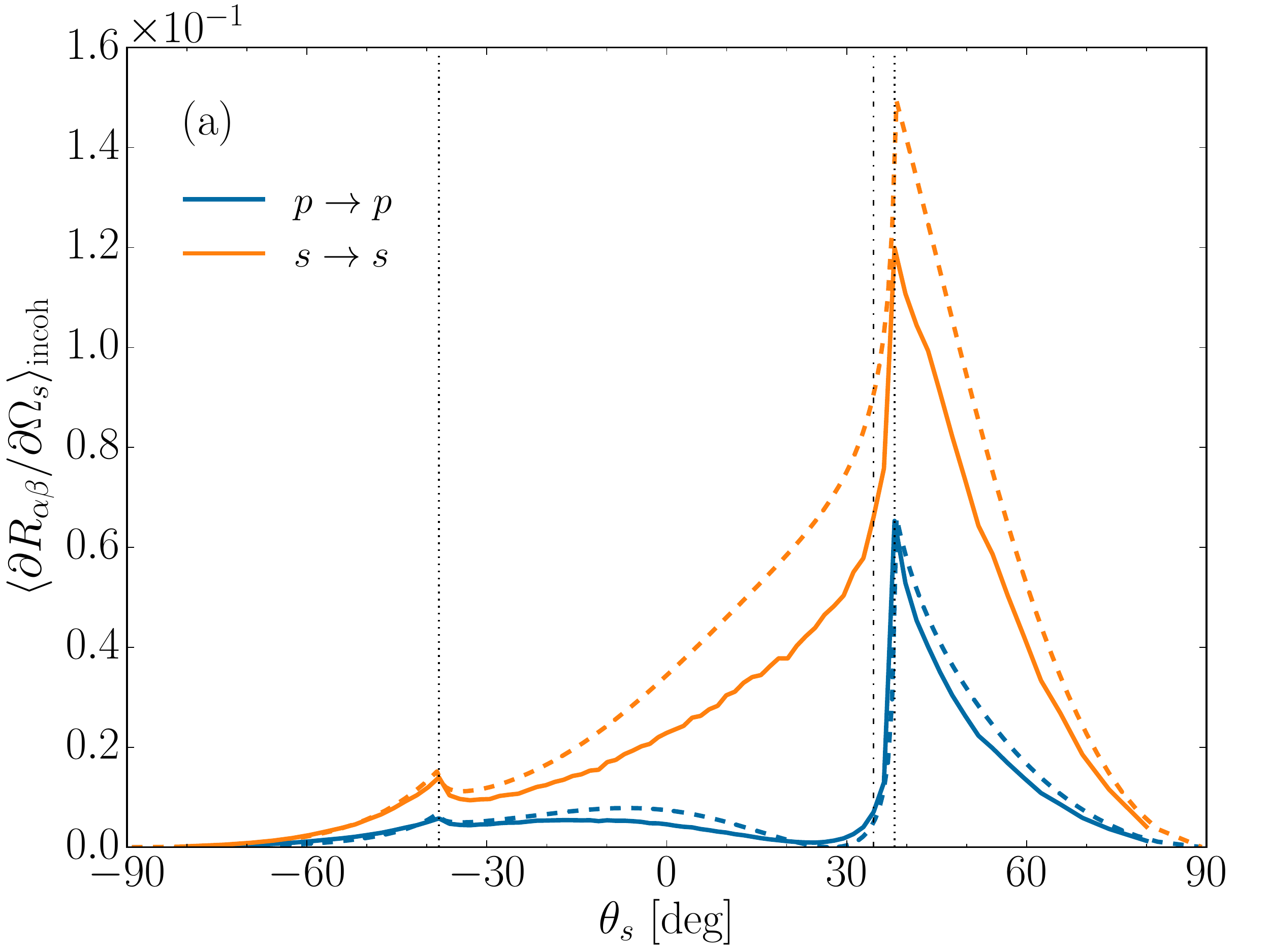}
  \includegraphics[width=0.47\columnwidth]{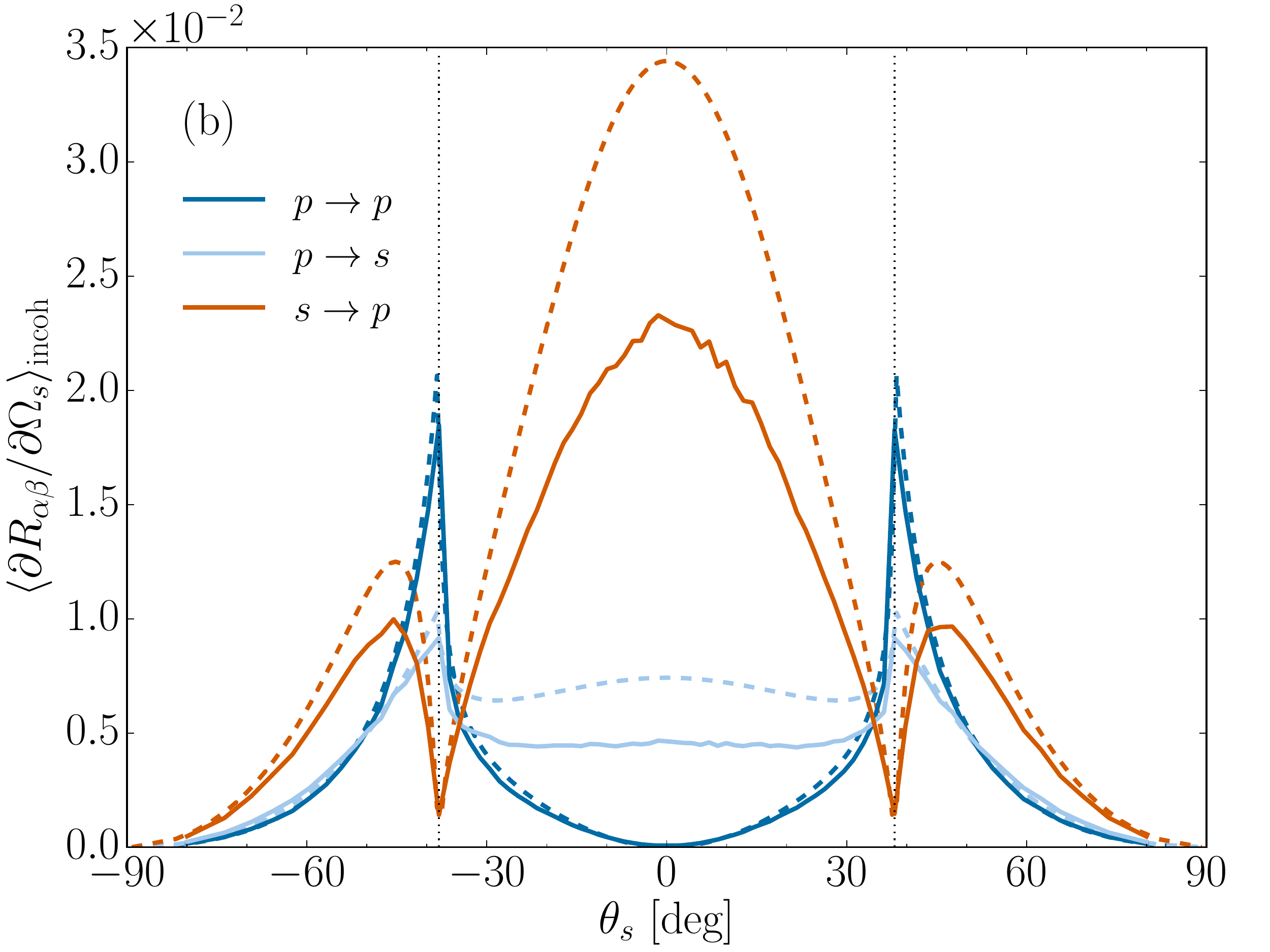}
  \caption{
    Same as Fig.~\protect\ref{fig:cut_mdrc_theta_345} but for root-mean-square roughness $\delta=\lambda/20$.
  }
\label{fig:cut_mdrc_theta_345_lambda_over_20}
\end{figure}

\smallskip
More interesting still is scattering in the inverse configuration, where light is incident from the dielectric side of the rough interface [$\ve_1=2.64$, $\ve_2=1.0$]: solutions of the RRE for this configuration and angles of incidence $(\theta_0,\phi_0)=(\ang{34.5},\ang{0})$, but for otherwise identical parameters as in Figs.~\ref{fig:cut_mdrc_theta_669} and \ref{fig:2Dmdrc_vtp_theta_669}, are presented in Fig.~\ref{fig:cut_mdrc_theta_345}. Analogous with Fig.~\ref{fig:cut_mdrc_theta_669}, Fig.~\ref{fig:cut_mdrc_theta_345}(a) shows the incoherent component of the mean DRC for in-plane scattering, and Fig.~\ref{fig:cut_mdrc_theta_345}(b) shows the corresponding curves for out-of-plane scattering.

In Fig.~\ref{fig:cut_mdrc_theta_345}(a), we now observe that the two dips in $\R{pp}{q}$ at $|\theta_s|=\theta_c$ observed in Fig.~\ref{fig:inplane_mdrc_theta_0}(b) have both turned into Yoneda peaks, albeit with different peak intensities, and that the sharp dip at the same angle for forward scattering have turned into a less sharp local minimum at $\theta_s\approx\ang{27}$. In order to understand these features, we see from Eq.~\eqref{eq:zero1} that, for $\ve_1>\ve_2$, $\R{pp}{q}$ vanishes for $\theta_s=\Theta_B(\theta_0)$ when $\theta_0$ is in the interval $0 < \theta_0 < \sin^{-1}\sqrt{\ve_2/\ve_1}$. This minimum in $\R{pp}{q}$ will shift its polar position towards $\theta_s=\ang{0}$ for increasing $\theta_0$, eventually ``releasing'' the Yoneda peaks in the forward scattering plane originating in the $|d_\a(\qp)|^2$ functions also for $\ppol\to\ppol$ scattering.
In the backscattering plane, we observe through the function given in \eqref{eq:pp_zero} that the negative sign of ($\pvecUnit{q} \cdot \pvecUnit{k}$) will lead to a monotonic increase in the contribution from Eq.~\eqref{eq:pp_zero} to Eq.~\eqref{eq:5.20a} as $\theta_0$ increases, eventually producing a Yoneda peak also for $\pvecUnit{q} \cdot \pvecUnit{k}<0$.
The overall distribution of $\spol\to\spol$ incoherent scattering in Fig.~\ref{fig:cut_mdrc_theta_345}(a) also show a strong forward shift in its scattering intensities, which, as we look to Eq.~\eqref{eq:5.20b}, can be attributed solely to the shifted power spectrum $g(|\pvec{q}-\pvec{k}|)$.

Looking at Fig.~\ref{fig:cut_mdrc_theta_345}(b), we observe several features for out-of-plane scattering that warrant a comment.
Overall, we observe that the scattering distributions are again symmetric about $\theta_s=0$, as is expected for out-of-plane scattering when the surface roughness is isotropic. Moreover, the distribution of $\R{ps}{q}$ appears similar in shape to the distribution of $\R{pp}{q}$ in Fig.~\ref{fig:inplane_mdrc_theta_0}(b). Their similarity can, to lowest non-zero order in SAPT, be attributed to their shared factor of $\a_2(\qp)$ in Eqs.~\eqref{eq:5.20a} and \eqref{eq:5.20ps}, which in both cases vanishes for $\qp=\sqrt{\ve_2}\w/c$, thereby suppressing the Yoneda peaks at this polar angle.
There are no such suppressing factors present in Eq.~\eqref{eq:5.20sp}, and the distribution of $\R{sp}{q}$ therefore displays Yoneda peaks at $|\theta_s|=\theta_c$. 
Similar to what we observed in Fig.~\ref{fig:cut_mdrc_theta_669}(b), we see that the distribution of $\R{pp}{q}$ has a local minimum at $\theta_s=0$; both this minimum and the Yoneda peaks found at $|\theta_s|=\theta_c$ are readily understood through the function in Eq.~\eqref{eq:pp_zero} and the factor $|d_\a(\qp)|^2$, respectively.

We now turn to a scattering system for which the rms-roughness of the surface is increased to $\delta=\lambda/20$, \textit{i.e.}, twice of the roughness assumed in obtaining the results of Fig.\ref{fig:cut_mdrc_theta_345}. Results for the in-plane and out-of-plane scattered intensity distributions for different combinations of the polarizations of the incident and scattered light are presented in Fig.~\ref{fig:cut_mdrc_theta_345_lambda_over_20}. Overall the results in Fig.~\ref{fig:cut_mdrc_theta_345_lambda_over_20} are in qualitative agreement with those of Fig.~\ref{fig:cut_mdrc_theta_345} for the equivalent but less rough scattering system. In general, the increase in surface roughness is again found to result in a poorer agreement between the results obtained on the basis of SAPT and those obtained by a direct numerical solution of the RRE. However, it is interesting to observe that for the case of in-plane as well as out-of-plane $\ppol \to \ppol$ scattering, SAPT seems to give a fair representation of the simulated scattered intensity distributions for both levels of roughness considered in Figs.~\ref{fig:cut_mdrc_theta_345} and \ref{fig:cut_mdrc_theta_345_lambda_over_20}. 
For other combinations of the polarizations of the incident and scattered light this is not the case.

\begin{figure}[tbh] 
  \centering
  \includegraphics[width=0.8\columnwidth]{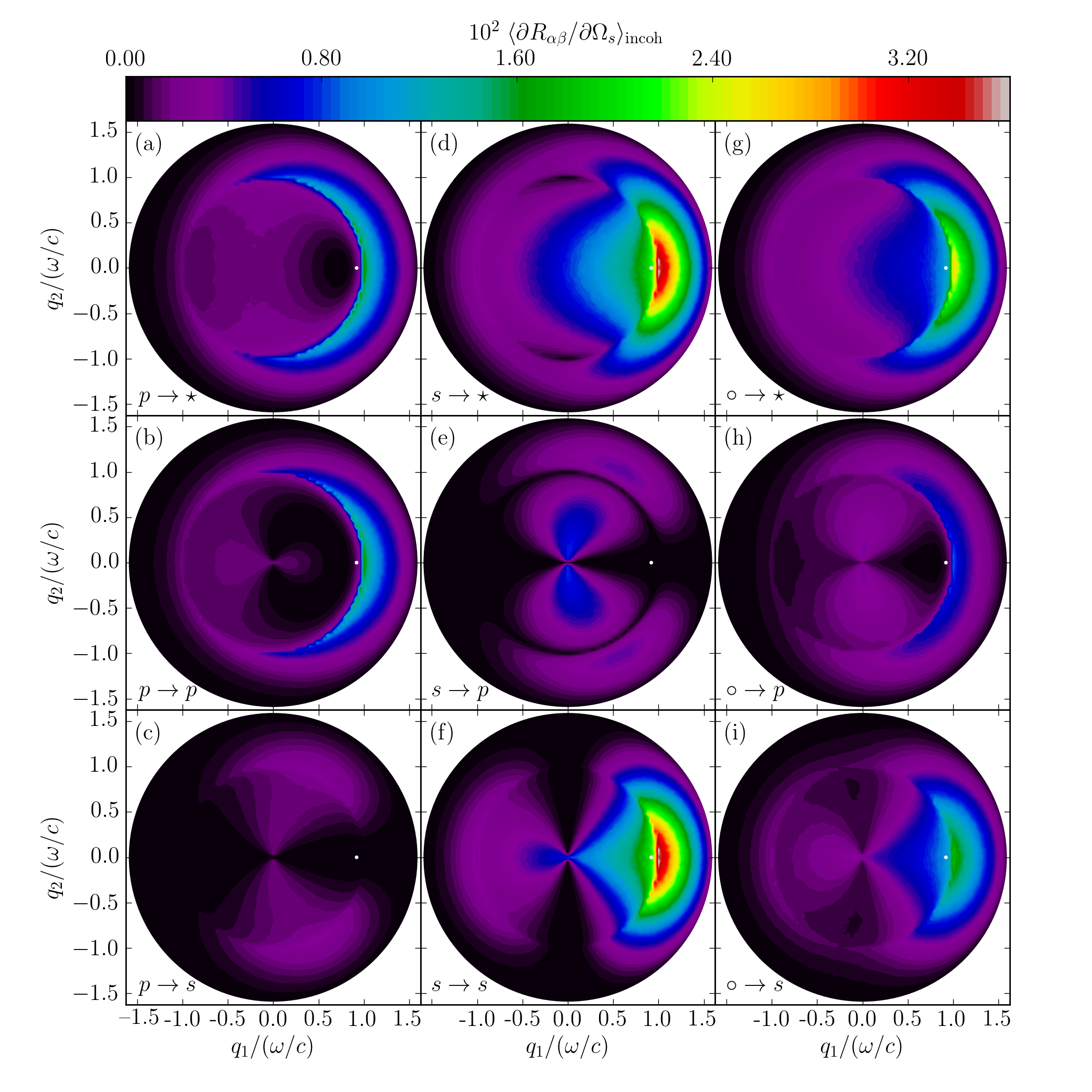}
  \caption{Same as Fig.~\ref{fig:2Dmdrc_ptv_theta_0} 
 but for the angles of incidence $(\theta_0,\phi_0)=(\ang{34.5},\ang{0})$. As can be seen from the position of the white dot, this figure captures the scattering distribution when the polar angle of incidence $\theta_0$ is close to the critical angle $\theta_c=\sin^{-1}{\sqrt{\ve_2/\ve_1}}$ for a corresponding flat-interface system. [Parameters: $\ve_1=2.64$; $\ve_2=1.0$;  $\delta=\lambda/40$, $a=\lambda/4$].
}
 \label{fig:2Dmdrc_ptv_theta_345}
\end{figure}

\smallskip
The results presented in Fig.~\ref{fig:cut_mdrc_theta_345} were, as for Fig.~\ref{fig:cut_mdrc_theta_669}, in-plane and out-of-plane cuts from Fig.~\ref{fig:2Dmdrc_ptv_theta_345} which displays the full angular distribution of the contribution to the mean DRC from the incoherently scattered light for the angles of incidence $(\theta_0,\phi_0)=(\ang{34.5},\ang{0})$. In contrast to what was observed in Fig.~\ref{fig:2Dmdrc_vtp_theta_669}, all four of the lower left $2\times2$ subfigures in Fig.~\ref{fig:2Dmdrc_ptv_theta_345} now have significantly differing appearances.
Similar to our observations in the case of incidence from vacuum, we observe that the Brewster scattering angle described by Eq.~\eqref{eq:zero1} can be seen to be part of a more general but still localized minimum in both Fig.~\ref{fig:2Dmdrc_ptv_theta_345}(a) and Fig.~\ref{fig:2Dmdrc_ptv_theta_345}(h), for $\ppol\to\star$ and $\circ\to\ppol$ scattering, respectively. Further, we still, as in Fig.~\ref{fig:2Dmdrc_ptv_theta_0}, observe Yoneda peaks for all azimuthal angles of scattering. The intensities of these peaks are now, however, significantly stronger in the forward scattering plane, closer to the direction of specular reflection.

\begin{figure}[tbh] 
  \centering
  \includegraphics[width=0.8\columnwidth]{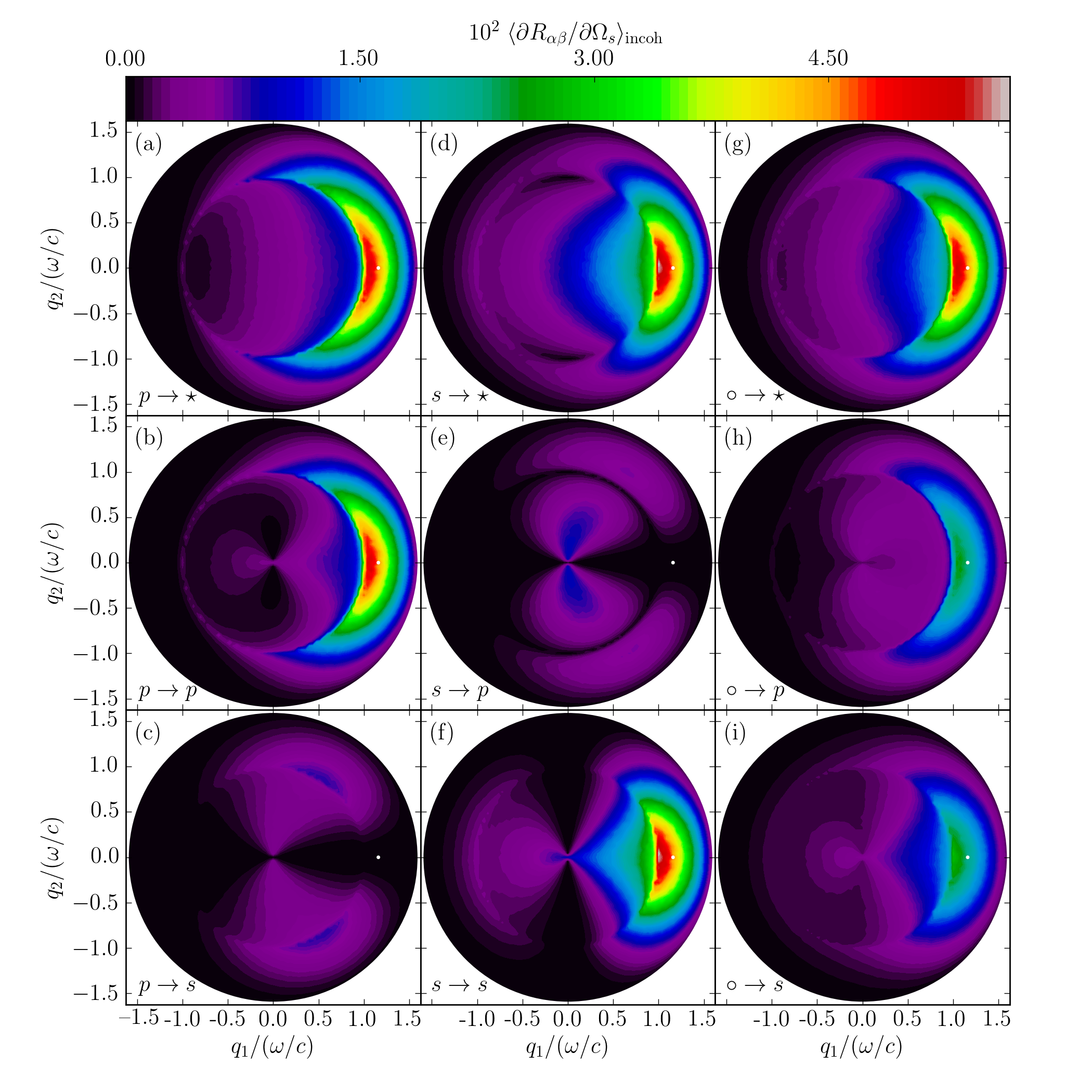}
  \caption{Same as Fig.~\ref{fig:2Dmdrc_ptv_theta_345} 
    but for the angles of incidence $(\theta_0,\phi_0)=(\ang{45.5},\ang{0})$. 
    [Parameters: $\ve_1=2.64$, $\ve_2=1.0$;  $\delta=\lambda/40$, $a=\lambda/4$].
  }
 \label{fig:2Dmdrc_ptv_theta_45}
\end{figure}

We now turn to Fig.~\ref{fig:2Dmdrc_ptv_theta_45}, which is identical to Fig.~\ref{fig:2Dmdrc_ptv_theta_345} but for the angles of incidence $(\theta_0,\phi_0)=(\ang{45.5},\ang{0})$. For these angles of incidence, the light incident on a flat surface would exhibit total internal reflection. Incoherent scattering is, as before, greatly enhanced for $\qp\geq\sqrt{\ve_2}\omega/c$, the part of wavevector-space that is evanescent in the medium of transmission. The intensity of the light scattered diffusely into this region is now comparable for s- and p-polarized light, and we see Yoneda peaks in both forward and backward scattering, for a fairly wide range of azimuthal angles. This can, as before, be understood to lowest non-zero order in SAPT through Eqs.~\eqref{eq:5.20_R} and \eqref{eq:5.21}. The factors $|d_p(\kp)|^{-2}$ and $|d_s(\kp)|^{-2}$ will both have their maxima at $\theta_0=\sin^{-1}{(\sqrt{\ve_2/\ve_1})}$, maxima that coincide with the corresponding maxima for the previously mentioned factors $|d_p(\qp)|^{-2}$ and $|d_s(\qp)|^{-2}$. The contribution from these factors will be the same for all $\phi_s$, but common for all combinations of polarized scattering in Eq.~\eqref{eq:5.20_R} is that the multiplicative factor of the power spectrum will have its principal weight at $|\pvec{q}-\pvec{k}|=0$; explaining the asymmetry about $q_1=0$ and the consequent shift of scattering to the forward scattering portion of the $\qp$-plane.

While there is no Brewster scattering angle for the angle of incidence in Fig.~\ref{fig:2Dmdrc_ptv_theta_45}, we still observe a local minimum in the backward scattering direction close to the critical angle for p-polarized incident light, Fig.~\ref{fig:2Dmdrc_ptv_theta_45}(a).

\begin{figure}[tbh]
  \centering
  \includegraphics[width=0.47\columnwidth]{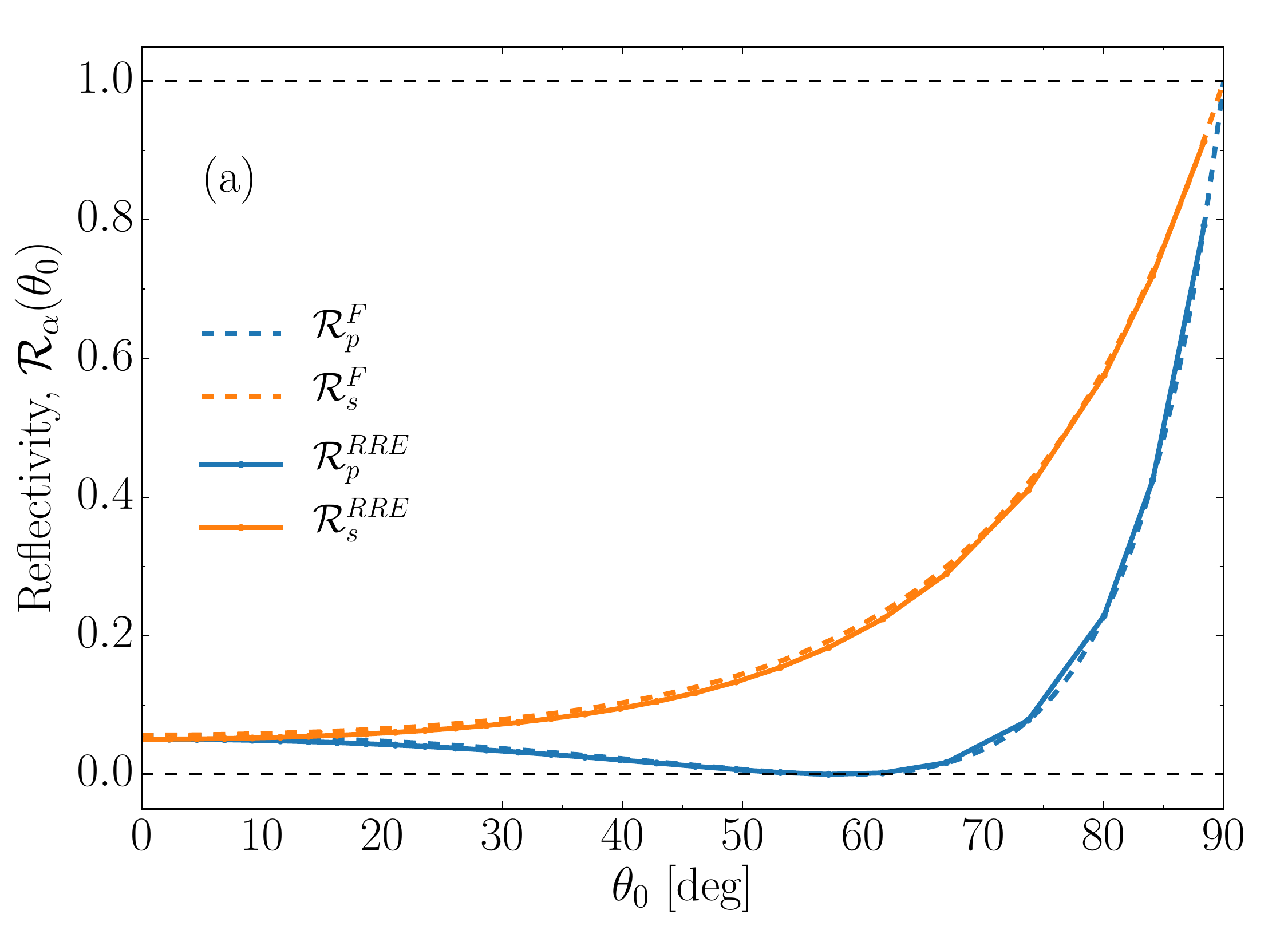}
  \includegraphics[width=0.47\columnwidth]{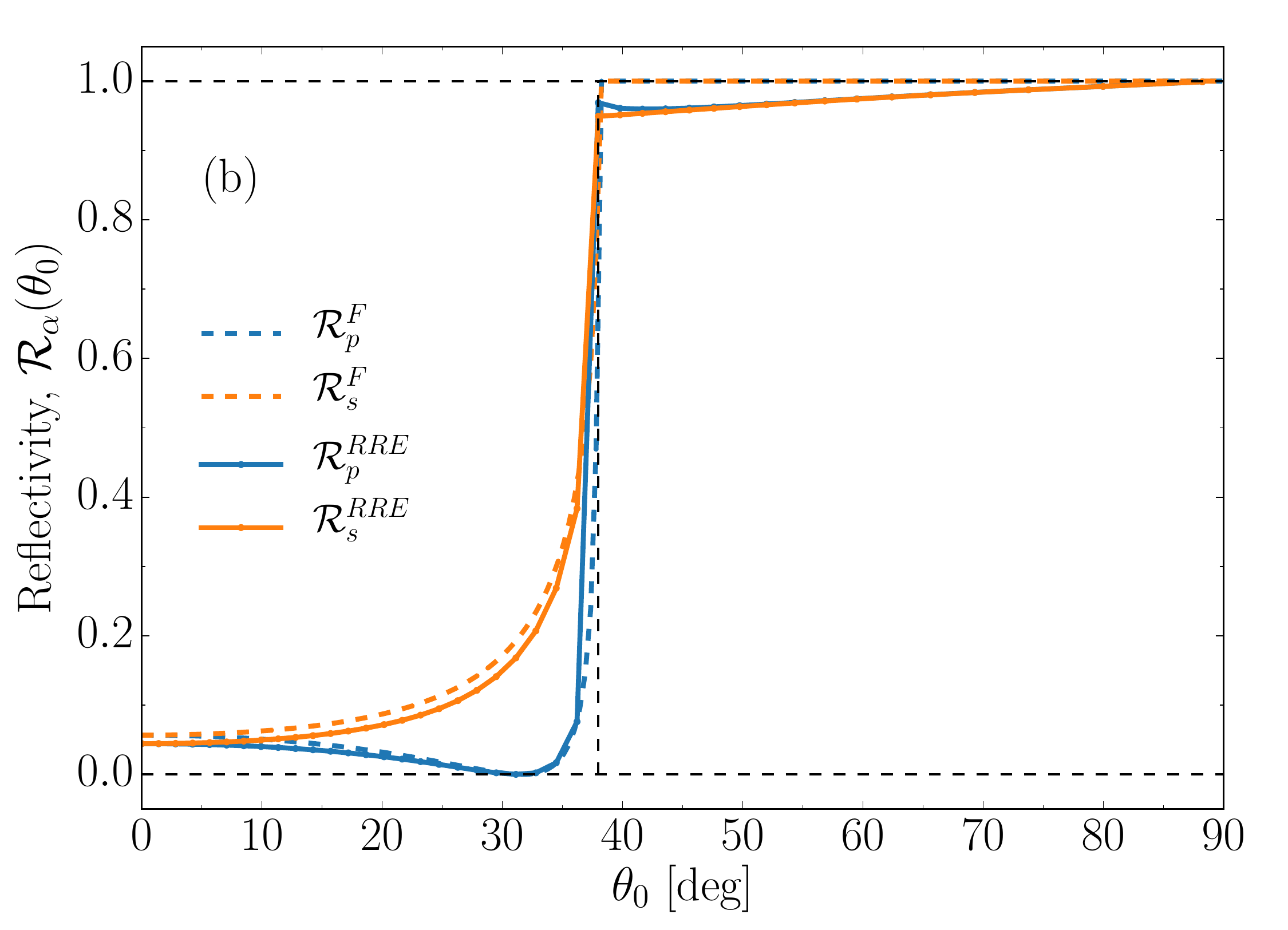}
  \caption{ 
    (a) The reflectivities ${\mathcal R}_\alpha(\theta_0)$ of a two-dimensional randomly rough vacuum-dielectric interface [$\ve_1=1.0$, $\ve_2=2.64$] for p- and s-polarized light as functions of the polar angle of incidence.
    (b) The same as in Fig.~\ref{fig:reflectivity}(a), but for a dielectric-vacuum interface [$\ve_1=2.64$, $\ve_2=1.0$].
    The quantity ${\mathcal R}^F_\alpha(\theta_0)$ indicates the Fresnel reflection coefficient (flat surface reflectivity). 
    The critical angle $\theta_0=\theta_c=\sin^{-1}{\sqrt{\ve_2/\ve_1}}$ for total internal reflection for an equivalent flat-interface system is indicated by a vertical dashed line in Fig.~\ref{fig:reflectivity}(b).
    Several simulations were run with small perturbations in the surface length $L$ in order to obtain reflectivity data with higher angular resolution.
    The roughness parameters assumed in obtaining these results were $\delta=\lambda/40$ and $a=\lambda/4$.
  }
  \label{fig:reflectivity}
\end{figure}

\begin{figure}[tbh]
  \centering
  \includegraphics[width=0.47\columnwidth]{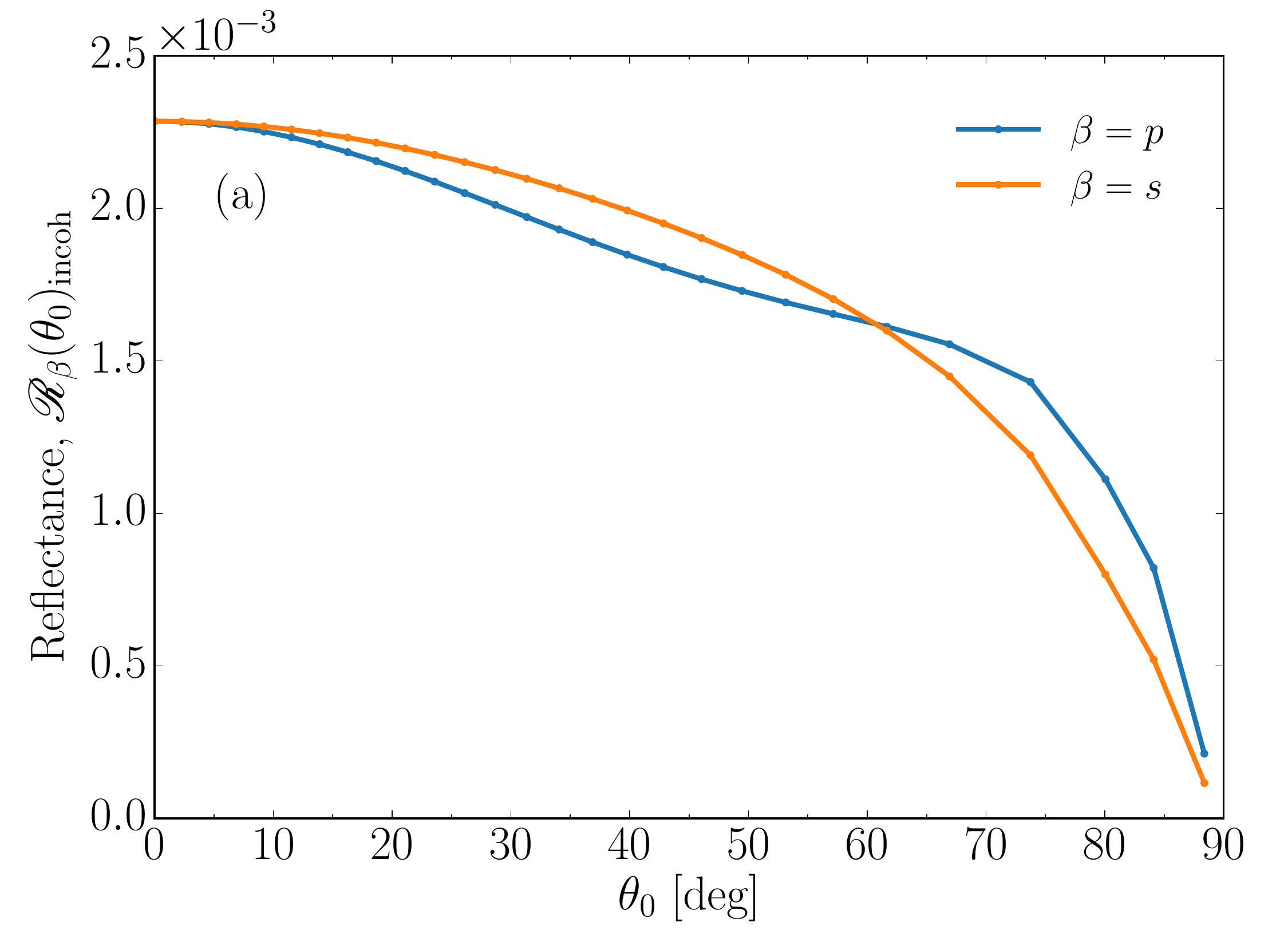}
  \includegraphics[width=0.47\columnwidth]{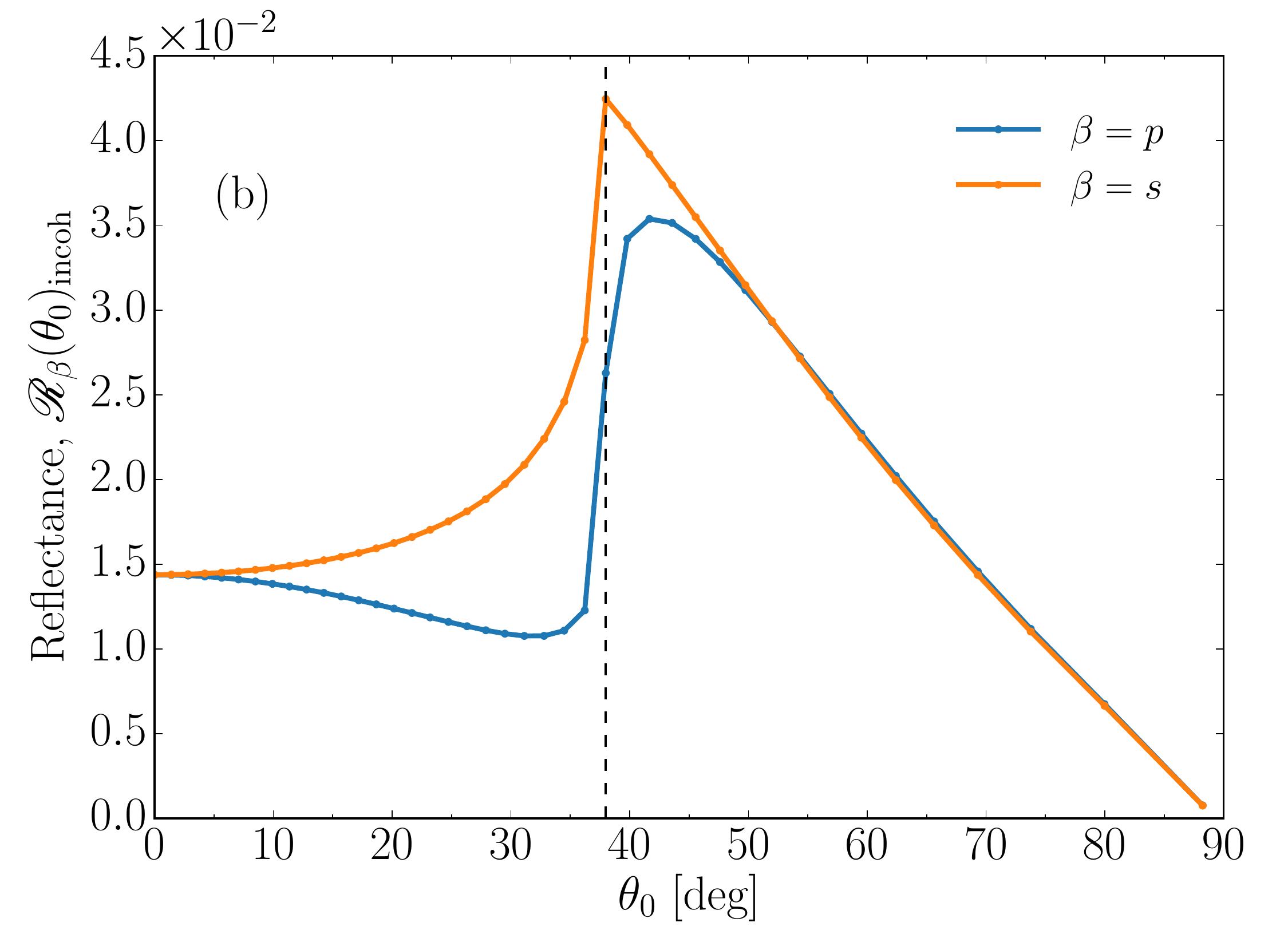}
  \caption{ 
    The $\theta_0$-dependence of the contribution to the reflectance from p- and s-polarized incident light that has been scattered incoherently from a two-dimensional randomly rough surface. This quantity is for $\beta$-polarized incident light defined as ${\mathscr R}_{\beta}(\theta_0)_{\mathrm{incoh}} = {\mathscr R}_{\beta}(\theta_0) - {\mathcal R}_\beta(\theta_0)$.
    (a) The reflectances for a vacuum-dielectric interface [$\ve_1=1.0$, $\ve_2=2.64$] for p- and s-polarized light as functions of the polar angle of incidence. (b) Same as Fig.~\ref{fig:reflectance}(a), but for a dielectric-vacuum interface [$\ve_1=2.64$, $\ve_2=1.0$].
    As in Fig.~\ref{fig:reflectivity}, the critical angle for total internal reflection in a corresponding flat-interface system, $\theta_c$, is indicated by a vertical dashed line in Fig.~\ref{fig:reflectance}(b).
    Several simulations were run with small perturbations in the surface length $L$ in order to obtain reflectance data with higher angular resolution.
    The roughness parameters assumed in obtaining these results were $\delta=\lambda/40$ and $a=\lambda/4$.
  }
  \label{fig:reflectance}
\end{figure}

\subsection{Reflectivity and reflectance}

The reflectivities for the two configurations of media are presented in Fig.~\ref{fig:reflectivity}. Both Figs.~\ref{fig:reflectivity}(a) and \ref{fig:reflectivity}(b) show only small deviations from the Fresnel reflection coefficients for a corresponding flat-surface system, the only notable difference being in Fig.~\ref{fig:reflectivity}(b) where the surface roughness prevents total internal reflection for incoming light with $\theta_0$ larger than $\theta_c=\sin^{-1}{\sqrt{\ve_2/\ve_1}} \approx \ang{38.0}$, the critical angle corresponding to the values of the dielectric constants assumed in these simulations.
The overall reflectivities for both systems are slightly smaller in all cases than the corresponding Fresnel coefficients, which is expected for a rough surface system since some light is scattered diffusely away from the specular direction.
The rough-surface analogues of the Brewster angles for corresponding flat-interface systems, called analogues because the reflectivity does not reach strict zero in the case of surface roughness, are clearly seen for p-polarized light in both figures. 

The differences between the presented results for the reflectivity and the corresponding Fresnel coefficients can be better understood through Fig.~\ref{fig:reflectance}, which presents the contribution to the reflectance from the light that has been reflected incoherently by the interface: ${\mathscr R}_{\beta}(\theta_0)_{\mathrm{incoh}}$ [see Eq.~\eqref{eq:reflectance_sum}]. In both subfigures in Fig.~\ref{fig:reflectance} we see that the amount of diffusely scattered light in general decreases with an increasing angle of incidence, if we ignore the effects of total internal reflection. This is consistent with the general notion that a rough surface is perceived as less rough for large angles of incidence~\cite{Simonsen2010}. 

Figure \ref{fig:reflectance}(a) shows that the incoherent part of the reflectance for the vacuum-dielectric configuration is a monotonically decreasing function of $\theta_0$ for both polarizations, as expected by inspection of Eqs.~\eqref{eq:5.20_R} and \eqref{eq:5.21}, for $\ve_1 < \ve_2$. The functions $|d_p(\kp)|^{-2}$ and $|d_s(\kp)|^{-2}$, and the factor $1/{\cos(\theta_0)}$ are all monotonically increasing functions of $\kp$ (or $\theta_0$), but they do not increase rapidly enough to compensate for the monotonously decreasing factor of $\a_1^2(\kp)$. A closer inspection of the numerical results and a more careful evaluation of the different factors in Eq.~\eqref{eq:5.20_R} have shown that the more rapid initial decrease of ${\mathscr R}_{p}(\theta_0)_{\mathrm{incoh}}$ is due to the contribution from its cross-polarized term, while its co-polarized term is responsible for the eventual less rapid decrease compared to ${\mathscr R}_{s}(\theta_0)_{\mathrm{incoh}}$.

The incoherent part of the reflectance for the dielectric-vacuum configuration is displayed in Fig.~\ref{fig:reflectance}(b). We can find here the explanation for why the curve for p-polarization in Fig.~\ref{fig:reflectivity}(b) showed a stronger peak for $\theta_0$ just beyond the critical angle $\theta_c$ than the curve for s-polarization: less light is scattered incoherently for these angles when the incident light is p-polarized than when it is s-polarized. We can also see that the contribution to the reflectance from the light scattered incoherently increases more than two-fold at the critical angle relative to the contribution at normal incidence. This behaviour can, again, be understood in terms of small amplitude perturbation theory to lowest order in the surface profile function, Eqs.~\eqref{eq:5.20_R} and \eqref{eq:5.21}. The functions  $|d_p(\kp)|^{-2}$ and $|d_s(\kp)|^{-2}$ will both have their maximum at the critical angle $\theta_c$, but while ${\mathscr
R}_{s}(\theta_0)_{\mathrm{incoh}}$ will get monotonously increasing contributions from both its co- and cross-polarized components for $0<\theta_0<\theta_c$, for ${\mathscr R}_{p}(\theta_0)_{\mathrm{incoh}}$ the cross-polarized component will go to zero due to the $\a_2(\kp)$-factor present in Eq.~\eqref{eq:5.20sp}. This dip in $\left\la {\p R_{sp}(\pvec{q} | \pvec{k} )}/{\p\Omega_s}\right\ra_\textrm{incoh}$ is hence the main reason for the differences in the incoherent component of the reflectance for this configuration of media.

\section{Conclusions}
%
We have presented a derivation of the reduced Rayleigh equation (RRE) for the reflection amplitudes of light scattered from a two-dimensional, randomly rough, surface. These equations enable a non-perturbative solution of the scattering problem based on the Rayleigh hypothesis. As an example of its solution by purely numerical means, the full angular distributions for both co- and cross-polarized incoherent components of the mean differential reflection coefficients were reported, for configurations of vacuum and an absorptionless dielectric with a Gaussian surface power spectrum and correlation function. 

It was shown that a configuration of reflection within the optically denser medium leads to Yoneda peaks in the angular distributions of the diffusely scattered light, namely peaks at the critical angle for total internal reflection in the denser medium. The behaviour and development of these peaks for a wide range of angles of incidence and scattering were investigated, and the lack of such peaks for light scattered into p-polarization for polar angles of incidence smaller than the critical angle were explained through small amplitude perturbation theory (SAPT).

Brewster scattering angles, angles where scattering into p-polarization is suppressed to strict zero in SAPT to lowest non-zero order in the surface profile function, were found to explain many of the differences in scattering into s- and p-polarization for the scattering systems investigated in the current work. These angles were first mentioned in the literature by Kawanishi~\etal~in Ref.~\citenum{Kawanishi1997}. Our results are in good agreement with their findings.

Small amplitude perturbation theory, to lowest non-zero order in the surface profile function, was overall shown to reproduce our numerical results qualitatively to a fairly high degree of accuracy, both through analytical arguments and a numerical implementation of that theory. This leads us to believe that the features presented in the results are single-scattering effects.

The scattering of light from a transparent dielectric is well described by solutions obtained by means of small amplitude perturbation theory, including the full angular distribution of the mean DRC for all combinations of the polarizations of the incident and scattered light. The reduced Rayleigh equation is a powerful starting point for studies of higher-order scattering features, such as enhanced backscattering, for example. The results presented here show that for the degree of surface roughness and the values of the dielectric constants assumed in this work no higher-order features are observed. Nevertheless, the RRE still gives more accurate numerical results for the mean DRC than does SAPT to lowest nonzero order in the surface profile function when the surface roughness is increased.

As an investigation of the quality of the results, energy conservation (unitarity) was found to be satisfied within $10^{-4}$ when the total scattered energy from both reflection and transmission was added together, for the roughness parameters and configurations used in this paper. An investigation similar to the present one but for light \textit{transmitted} through the dielectric rough interface will be presented in a separate publication~\cite{Hetland2016b}.

\begin{acknowledgments}
The authors would like to thank J.-P.~Banon for helpful discussions.
The research of \O.S.H. and  I.S. was supported in part by The Research Council of Norway Contract No. 216699.  
This work received support from NTNU and the Norwegian metacenter for High Performance Computing (NOTUR) by an allocation of computer time.
\end{acknowledgments}

\appendix*
\section{Expansion of $R(\pvec{q}|\pvec{k})$ in powers of the surface profile function}
\label{app:SAPT}

In this appendix we outline the derivation of Eq.~\eqref{eq:5.20_R}. 
%
%
To this end, we begin with the expansion
\begin{align}
  \label{app:eq:R-expansion}
  \vec{R}(\pvec{q}|\pvec{k}) &= 
  \sum_{n=0}^\infty \frac{\left(-\imu \right)^n}{n!} \vec{R}^{(n)}(\pvec{q}|\pvec{k}),
\end{align}
where the superscript $n$ denotes the order of the corresponding term in powers of $\zeta(\pvec{x})$. When Eqs.~\eqref{eq:3.13} and \eqref{app:eq:R-expansion} are substituted into Eq.~\eqref{eq:3.24}, the latter becomes
\begin{align}
  \begin{aligned}
  \sum_{m=0}^\infty \sum_{n=0}^m \frac{\left(-\imu \right)^m}{m!} \binom{m}{n} &
  \int \frac{\dint^2q_\parallel}{(2\pi)^2} \left[-\alpha_1(q_\parallel) + \alpha_2(p_\parallel) \right]^{n-1}
  \hz^{(n)}(\pvec{p}-\pvec{q}) \vec{M}^{+}(\pvec{p}|\pvec{q})\, \vec{R}^{(m-n)}(\pvec{q}|\pvec{k})
  \\
  &=
  \sum_{n=0}^m -\frac{\left(-\imu \right)^n}{n!}
  \left[\alpha_1(\kp) + \alpha_2(p_\parallel) \right]^{n-1}
  \hz^{(n)}(\pvec{p}-\pvec{k}) \vec{M}^{-}(\pvec{p}|\pvec{k}).
  \label{app:eq:RRE_sum}
\end{aligned}
\end{align}
When we equate terms of zero order in $\zeta(\pvec{x})$ on both sides of this equation, we obtain 
\begin{align}
  \frac{1}{ -\alpha_1(p_\parallel) + \alpha_2(p_\parallel) }
  \vec{M}^{+}(\pvec{p}|\pvec{p})\, \vec{R}^{(0)}(\pvec{p}|\pvec{k})
  = 
  -(2\pi)^2 \delta\left(\pvec{p}-\pvec{k}\right)
  \frac{1}{\alpha_1(p_\parallel) + \alpha_2(p_\parallel)}
  \vec{M}^{-}(\pvec{p}|\pvec{p}),
  \label{app:eq:zero_order}
\end{align}
which, if we solve for $\vec{R}^{(0)}(\pvec{q}|\pvec{k})$ gives
\begin{align}
  \begin{pmatrix}
    R^{(0)}_{\ppol\ppol}(\pvec{q}|\pvec{k})
    &
    R^{(0)}_{\ppol\spol}(\pvec{q}|\pvec{k})
    \\
    R^{(0)}_{\spol\ppol}(\pvec{q}|\pvec{k})
    &
    R^{(0)}_{\spol\spol}(\pvec{q}|\pvec{k})
  \end{pmatrix}
  &= (2\pi)^2 \delta\left(\pvec{q}-\pvec{k}\right)
  \begin{pmatrix}
      \frac{d^-_p(\kp)}{ d^+_\ppol(\kp) } 
      &
      0
      \\
      0
      &
      \frac{d^-_s(\kp)}{ d^+_\spol(\kp) }
  \end{pmatrix},
  %
  \label{app:eq:A8_R}
\end{align}
where
\begin{subequations}
  \label{app:eq:A11}
\begin{align}
  \label{app:eq:A11a}
  d_\ppol^{\pm}(\kp) &=   \ve_2 \alpha_1(\kp) \pm \ve_1 \alpha_2(\kp)
  \\
  \label{app:eq:A11b}
  d_\spol^{\pm}(\kp) &=        \alpha_1(\kp) \pm      \alpha_2(\kp).
\end{align}
\end{subequations}
Equation~\eqref{app:eq:A8_R} essentially represents the Fresnel coefficients for specular reflection from a flat interface.
%
For $m=1$, Eq.~\eqref{app:eq:RRE_sum} can be simplified to
\begin{align}
  \begin{aligned}
    \frac{1}{ -\alpha_1(p_\parallel) + \alpha_2(p_\parallel) } 
    &
    \vec{M}^{+}(\pvec{p}|\pvec{p}) \vec{R}^{(1)}(\pvec{p}|\pvec{k})
    +
    \int \frac{\dint^2q_\parallel}{(2\pi)^2} 
    \hz^{(1)}(\pvec{p}-\pvec{q}) \vec{M}^{+}(\pvec{p}|\pvec{q})\, \vec{R}^{(0)}(\pvec{q}|\pvec{k})
    \\
    &
    =
    -\hz^{(1)}(\pvec{p}-\pvec{k}) \vec{M}^{-}(\pvec{p}|\pvec{k}).
  \end{aligned}
  \label{app:eq:RRE_m=1}
\end{align}
If  we now use the result that the matrix $\vec{M}^{+}(\pvec{p}|\pvec{p})$ is diagonal and hence readily inverted, and that the matrix $\vec{R}^{(0)}(\pvec{q}|\pvec{k})$ is given by Eq.~\eqref{app:eq:A8_R}, we can simplify Eq.~\eqref{app:eq:RRE_m=1} into
\begin{align}
  \begin{aligned}
    &\vec{R}^{(1)}(\pvec{q}|\pvec{k})
    =
    -\left(\ve_2-\ve_1\right)
    \hz^{(1)}(\pvec{q}-\pvec{k}) 
    \\
    &\times
    \begin{pmatrix}
      \frac{ \sqrt{\ve_1\ve_2} }{ d^+_\ppol(\qp)  d^+_\ppol(\kp)} \left[d^-_p(\kp) M^+_{\ppol\ppol}(\pvec{q}|\pvec{k}) + d^+_\ppol(\kp) M^-_{\ppol\ppol}(\pvec{q}|\pvec{k})\right]
      &
      \frac{ \sqrt{\ve_1\ve_2} }{ d^+_\ppol(\qp)  d^+_\spol(\kp)} M^{\pm}_{\ppol\spol}(\pvec{q}|\pvec{k})\left[ d^-_s(\kp) + d^+_\spol(\kp) \right]
      \\
      \frac{ 1 }{ d^+_\spol(\qp)  d^+_\ppol(\kp)} \left[ d^-_p(\kp) M^+_{\spol\ppol}(\pvec{q}|\pvec{k}) + d^+_\ppol(\kp) M^-_{\spol\ppol}(\pvec{q}|\pvec{k})\right]
      &
      \frac{ 1 }{ d^+_\spol(\qp)  d^+_\spol(\kp)} M^{\pm}_{\spol\spol}(\pvec{q}|\pvec{k})\left[ d^-_s(\kp) + d^+_\spol(\kp) \right]
    \end{pmatrix}
    .
  \end{aligned}
  \label{app:eq:A10_R2}
\end{align}
where the matrix elements $\{ M^{\pm}_{\alpha\beta}(\pvec{q}|\pvec{k})\}$ are given by Eq.~\eqref{eq:3.25a}.
This ultimately gives
\begin{align}
  \begin{aligned}
    \vec{R}^{(1)}&(\pvec{q}|\pvec{k})
    =
    -\left(\ve_2-\ve_1\right)
    \hz^{(1)}(\pvec{q}-\pvec{k})
    \\
    &\times
    \begin{pmatrix}
      \frac{ 1 }{ d^+_\ppol(\qp)  d^+_\ppol(\kp)} 
      \left[ \ve_2\qp\kp - \ve_1\alpha_2(\qp) \left(\pvecUnit{q}\cdot\pvecUnit{k}\right) \alpha_2(\kp)  \right]
      &
      -\frac{ \sqrt{\ve_1} }{ d^+_\ppol(\qp)  d^+_\spol(\kp)} \frac{\w}{c}\alpha_2(\qp )\, \left[\pvecUnit{q} \times \pvecUnit{k}\right]_3
      \\
      -\frac{ \sqrt{\ve_1} }{ d^+_\spol(\qp)  d^+_\ppol(\kp)} \frac{\w}{c} \left[\pvecUnit{q}\times \pvecUnit{k}\right]_3\, \a_2(\kp)
      &
      \frac{ 1 }{ d^+_\spol(\qp)  d^+_\spol(\kp)} \frac{\omega^2}{c^2} \left(\pvecUnit{q}\cdot\pvecUnit{k}\right)
    \end{pmatrix}
    2\alpha_1(\kp)
    .
  \end{aligned}
  \label{app:eq:A103_R}
\end{align}
%
In view of Eq.~\eqref{app:eq:R-expansion} we find that through terms linear in the surface profile function 
\begin{align}
  \begin{aligned}
    \vec{R}(\pvec{q}|\pvec{k})
    =&
  ~\vec{R}^{(0)}(\pvec{q}|\pvec{k})
  %
  - \imu
  ~\vec{R}^{(1)}(\pvec{q}|\pvec{k})
  + \mathcal{O}\left(\zeta^2 \right).
  \end{aligned}
\end{align}
The substitution of these results into Eq.~\eqref{eq:4.14} and use of the result that
$\la\hz(\pvec{Q})\hz(\pvec{Q})^*\ra = S\delta^2g(|\pvec{Q}|)$ yields Eq.~\eqref{eq:5.20_R}.


\begin{thebibliography}{21}%
\makeatletter
\providecommand \@ifxundefined [1]{%
 \@ifx{#1\undefined}
}%
\providecommand \@ifnum [1]{%
 \ifnum #1\expandafter \@firstoftwo
 \else \expandafter \@secondoftwo
 \fi
}%
\providecommand \@ifx [1]{%
 \ifx #1\expandafter \@firstoftwo
 \else \expandafter \@secondoftwo
 \fi
}%
\providecommand \natexlab [1]{#1}%
\providecommand \enquote  [1]{``#1''}%
\providecommand \bibnamefont  [1]{#1}%
\providecommand \bibfnamefont [1]{#1}%
\providecommand \citenamefont [1]{#1}%
\providecommand \href@noop [0]{\@secondoftwo}%
\providecommand \href [0]{\begingroup \@sanitize@url \@href}%
\providecommand \@href[1]{\@@startlink{#1}\@@href}%
\providecommand \@@href[1]{\endgroup#1\@@endlink}%
\providecommand \@sanitize@url [0]{\catcode `\\12\catcode `\$12\catcode
  `\&12\catcode `\#12\catcode `\^12\catcode `\_12\catcode `\%12\relax}%
\providecommand \@@startlink[1]{}%
\providecommand \@@endlink[0]{}%
\providecommand \url  [0]{\begingroup\@sanitize@url \@url }%
\providecommand \@url [1]{\endgroup\@href {#1}{\urlprefix }}%
\providecommand \urlprefix  [0]{URL }%
\providecommand \Eprint [0]{\href }%
\providecommand \doibase [0]{http://dx.doi.org/}%
\providecommand \selectlanguage [0]{\@gobble}%
\providecommand \bibinfo  [0]{\@secondoftwo}%
\providecommand \bibfield  [0]{\@secondoftwo}%
\providecommand \translation [1]{[#1]}%
\providecommand \BibitemOpen [0]{}%
\providecommand \bibitemStop [0]{}%
\providecommand \bibitemNoStop [0]{.\EOS\space}%
\providecommand \EOS [0]{\spacefactor3000\relax}%
\providecommand \BibitemShut  [1]{\csname bibitem#1\endcsname}%
\let\auto@bib@innerbib\@empty
\bibitem [{\citenamefont {Leskova}\ \emph {et~al.}(2011)\citenamefont
  {Leskova}, \citenamefont {Letnes}, \citenamefont {Maradudin}, \citenamefont
  {Nordam},\ and\ \citenamefont {Simonsen}}]{Leskova2011}%
  \BibitemOpen
  \bibfield  {author} {\bibinfo {author} {\bibfnamefont {T.~A.}\ \bibnamefont
  {Leskova}}, \bibinfo {author} {\bibfnamefont {P.~A.}\ \bibnamefont {Letnes}},
  \bibinfo {author} {\bibfnamefont {A.~A.}\ \bibnamefont {Maradudin}}, \bibinfo
  {author} {\bibfnamefont {T.}~\bibnamefont {Nordam}}, \ and\ \bibinfo {author}
  {\bibfnamefont {I.}~\bibnamefont {Simonsen}},\ }\href {\doibase
  10.1117/12.899304} {\bibfield  {journal} {\bibinfo  {journal} {Proc. SPIE}\
  }\textbf {\bibinfo {volume} {8172}},\ \bibinfo {pages} {817209} (\bibinfo
  {year} {2011})}\BibitemShut {NoStop}%
\bibitem [{\citenamefont {Nordam}\ \emph
  {et~al.}(2013{\natexlab{a}})\citenamefont {Nordam}, \citenamefont {Letnes},\
  and\ \citenamefont {Simonsen}}]{Nordam2013a}%
  \BibitemOpen
  \bibfield  {author} {\bibinfo {author} {\bibfnamefont {T.}~\bibnamefont
  {Nordam}}, \bibinfo {author} {\bibfnamefont {P.~A.}\ \bibnamefont {Letnes}},
  \ and\ \bibinfo {author} {\bibfnamefont {I.}~\bibnamefont {Simonsen}},\
  }\href {\doibase 10.3389/fphy.2013.00008} {\bibfield  {journal} {\bibinfo
  {journal} {Front. Phys.}\ }\textbf {\bibinfo {volume} {1}},\ \bibinfo {pages}
  {1} (\bibinfo {year} {2013}{\natexlab{a}})}\BibitemShut {NoStop}%
\bibitem [{\citenamefont {Nakayama}\ \emph {et~al.}(1980)\citenamefont
  {Nakayama}, \citenamefont {Ogura},\ and\ \citenamefont
  {Matsumoto}}]{Nakayama1980}%
  \BibitemOpen
  \bibfield  {author} {\bibinfo {author} {\bibfnamefont {J.}~\bibnamefont
  {Nakayama}}, \bibinfo {author} {\bibfnamefont {H.}~\bibnamefont {Ogura}}, \
  and\ \bibinfo {author} {\bibfnamefont {B.}~\bibnamefont {Matsumoto}},\ }\href
  {http://onlinelibrary.wiley.com/doi/10.1029/RS015i006p01049/full} {\bibfield
  {journal} {\bibinfo  {journal} {Radio Sci.}\ }\textbf {\bibinfo {volume}
  {15}},\ \bibinfo {pages} {1049} (\bibinfo {year} {1980})}\BibitemShut
  {NoStop}%
\bibitem [{\citenamefont {Kawanishi}\ \emph {et~al.}(1997)\citenamefont
  {Kawanishi}, \citenamefont {Ogura},\ and\ \citenamefont
  {Wang}}]{Kawanishi1997}%
  \BibitemOpen
  \bibfield  {author} {\bibinfo {author} {\bibfnamefont {T.}~\bibnamefont
  {Kawanishi}}, \bibinfo {author} {\bibfnamefont {H.}~\bibnamefont {Ogura}}, \
  and\ \bibinfo {author} {\bibfnamefont {Z.~L.}\ \bibnamefont {Wang}},\ }\href
  {\doibase 10.1080/13616679709409805} {\bibfield  {journal} {\bibinfo
  {journal} {Wave. Random Media}\ }\textbf {\bibinfo {volume} {7}},\ \bibinfo
  {pages} {351} (\bibinfo {year} {1997})}\BibitemShut {NoStop}%
\bibitem [{\citenamefont {Yoneda}(1963)}]{Yoneda1963}%
  \BibitemOpen
  \bibfield  {author} {\bibinfo {author} {\bibfnamefont {Y.}~\bibnamefont
  {Yoneda}},\ }\href {\doibase 10.1103/PhysRev.131.2010} {\bibfield  {journal}
  {\bibinfo  {journal} {Phys. Rev.}\ }\textbf {\bibinfo {volume} {131}},\
  \bibinfo {pages} {2010} (\bibinfo {year} {1963})}\BibitemShut {NoStop}%
\bibitem [{\citenamefont {Vineyard}(1982)}]{Vineyard1982}%
  \BibitemOpen
  \bibfield  {author} {\bibinfo {author} {\bibfnamefont {G.~H.}\ \bibnamefont
  {Vineyard}},\ }\href {\doibase 10.1103/PhysRevB.26.4146} {\bibfield
  {journal} {\bibinfo  {journal} {Phys. Rev. B}\ }\textbf {\bibinfo {volume}
  {26}},\ \bibinfo {pages} {4146} (\bibinfo {year} {1982})}\BibitemShut
  {NoStop}%
\bibitem [{\citenamefont {Sinha}\ \emph {et~al.}(1988)\citenamefont {Sinha},
  \citenamefont {Sirota}, \citenamefont {Garoff},\ and\ \citenamefont
  {Stanley}}]{Sinha1988}%
  \BibitemOpen
  \bibfield  {author} {\bibinfo {author} {\bibfnamefont {S.~K.}\ \bibnamefont
  {Sinha}}, \bibinfo {author} {\bibfnamefont {E.~B.}\ \bibnamefont {Sirota}},
  \bibinfo {author} {\bibfnamefont {S.}~\bibnamefont {Garoff}}, \ and\ \bibinfo
  {author} {\bibfnamefont {H.~B.}\ \bibnamefont {Stanley}},\ }\href {\doibase
  10.1103/PhysRevB.38.2297} {\bibfield  {journal} {\bibinfo  {journal} {Phys.
  Rev. B}\ }\textbf {\bibinfo {volume} {38}},\ \bibinfo {pages} {2297}
  (\bibinfo {year} {1988})}\BibitemShut {NoStop}%
\bibitem [{\citenamefont {Leskova}\ and\ \citenamefont
  {Maradudin}(1997)}]{Leskova1997}%
  \BibitemOpen
  \bibfield  {author} {\bibinfo {author} {\bibfnamefont {T.~A.}\ \bibnamefont
  {Leskova}}\ and\ \bibinfo {author} {\bibfnamefont {A.~A.}\ \bibnamefont
  {Maradudin}},\ }\href {\doibase 10.1080/13616679709409807} {\bibfield
  {journal} {\bibinfo  {journal} {Wave. Random Media}\ }\textbf {\bibinfo
  {volume} {7}},\ \bibinfo {pages} {395} (\bibinfo {year} {1997})}\BibitemShut
  {NoStop}%
\bibitem [{\citenamefont {Nieto-Vesperinas}\ and\ \citenamefont
  {S\'{a}nchez-Gil}(1992)}]{Nieto-Vesperinas1992}%
  \BibitemOpen
  \bibfield  {author} {\bibinfo {author} {\bibfnamefont {M.}~\bibnamefont
  {Nieto-Vesperinas}}\ and\ \bibinfo {author} {\bibfnamefont {J.~A.}\
  \bibnamefont {S\'{a}nchez-Gil}},\ }\href {\doibase 10.1364/JOSAA.9.000424}
  {\bibfield  {journal} {\bibinfo  {journal} {J. Opt. Soc. Am. A}\ }\textbf
  {\bibinfo {volume} {9}},\ \bibinfo {pages} {424} (\bibinfo {year}
  {1992})}\BibitemShut {NoStop}%
\bibitem [{\citenamefont {Soubret}\ \emph {et~al.}(2001)\citenamefont
  {Soubret}, \citenamefont {Berginc},\ and\ \citenamefont
  {Bourrely}}]{Soubret2001a}%
  \BibitemOpen
  \bibfield  {author} {\bibinfo {author} {\bibfnamefont {A.}~\bibnamefont
  {Soubret}}, \bibinfo {author} {\bibfnamefont {G.}~\bibnamefont {Berginc}}, \
  and\ \bibinfo {author} {\bibfnamefont {C.}~\bibnamefont {Bourrely}},\ }\href
  {\doibase 10.1103/PhysRevB.63.245411} {\bibfield  {journal} {\bibinfo
  {journal} {Phys. Rev. B}\ }\textbf {\bibinfo {volume} {63}},\ \bibinfo
  {pages} {245411} (\bibinfo {year} {2001})}\BibitemShut {NoStop}%
\bibitem [{\citenamefont {Press}\ \emph {et~al.}(1996)\citenamefont {Press},
  \citenamefont {Teukolsky}, \citenamefont {Vetterling},\ and\ \citenamefont
  {Flannery}}]{Book:Press1996}%
  \BibitemOpen
  \bibfield  {author} {\bibinfo {author} {\bibfnamefont {W.~H.}\ \bibnamefont
  {Press}}, \bibinfo {author} {\bibfnamefont {S.~A.}\ \bibnamefont
  {Teukolsky}}, \bibinfo {author} {\bibfnamefont {W.~T.}\ \bibnamefont
  {Vetterling}}, \ and\ \bibinfo {author} {\bibfnamefont {B.~P.}\ \bibnamefont
  {Flannery}},\ }\href@noop {} {\emph {\bibinfo {title} {Numerical Recipes in
  Fortran 90 : The Art of Parallel Scientific Computing}}}\ (\bibinfo
  {publisher} {Cambridge University Press},\ \bibinfo {address} {Cambridge},\
  \bibinfo {year} {1996})\BibitemShut {NoStop}%
\bibitem [{\citenamefont {Maradudin}\ \emph {et~al.}(1990)\citenamefont
  {Maradudin}, \citenamefont {Michel}, \citenamefont {McGurn},\ and\
  \citenamefont {M\'endez}}]{Maradudin1990}%
  \BibitemOpen
  \bibfield  {author} {\bibinfo {author} {\bibfnamefont {A.~A.}\ \bibnamefont
  {Maradudin}}, \bibinfo {author} {\bibfnamefont {T.}~\bibnamefont {Michel}},
  \bibinfo {author} {\bibfnamefont {A.~R.}\ \bibnamefont {McGurn}}, \ and\
  \bibinfo {author} {\bibfnamefont {E.~R.}\ \bibnamefont {M\'endez}},\ }\href
  {\doibase 10.1016/0003-4916(90)90172-K} {\bibfield  {journal} {\bibinfo
  {journal} {Ann. Phys. (N.Y.)}\ }\textbf {\bibinfo {volume} {203}},\ \bibinfo
  {pages} {255 } (\bibinfo {year} {1990})}\BibitemShut {NoStop}%
\bibitem [{\citenamefont {Simonsen}\ \emph {et~al.}(2011)\citenamefont
  {Simonsen}, \citenamefont {Kryvi}, \citenamefont {Maradudin},\ and\
  \citenamefont {Leskova}}]{Simonsen2011}%
  \BibitemOpen
  \bibfield  {author} {\bibinfo {author} {\bibfnamefont {I.}~\bibnamefont
  {Simonsen}}, \bibinfo {author} {\bibfnamefont {J.~B.}\ \bibnamefont {Kryvi}},
  \bibinfo {author} {\bibfnamefont {A.~A.}\ \bibnamefont {Maradudin}}, \ and\
  \bibinfo {author} {\bibfnamefont {T.~A.}\ \bibnamefont {Leskova}},\ }\href
  {\doibase 10.1016/j.cpc.2011.01.010} {\bibfield  {journal} {\bibinfo
  {journal} {Comput. Phys. Commun.}\ }\textbf {\bibinfo {volume} {182}},\
  \bibinfo {pages} {1904 } (\bibinfo {year} {2011})}\BibitemShut {NoStop}%
\bibitem [{\citenamefont {Simonsen}(2010)}]{Simonsen2010}%
  \BibitemOpen
  \bibfield  {author} {\bibinfo {author} {\bibfnamefont {I.}~\bibnamefont
  {Simonsen}},\ }\href {\doibase 10.1140/epjst/e2010-01221-4} {\bibfield
  {journal} {\bibinfo  {journal} {Eur. Phys. J.-Spec. Top.}\ }\textbf {\bibinfo
  {volume} {181}},\ \bibinfo {pages} {1} (\bibinfo {year} {2010})}\BibitemShut
  {NoStop}%
\bibitem [{\citenamefont {Hetland}\ \emph {et~al.}(2016)\citenamefont
  {Hetland}, \citenamefont {Maradudin}, \citenamefont {Nordam}, \citenamefont
  {Letnes},\ and\ \citenamefont {Simonsen}}]{Hetland2016b}%
  \BibitemOpen
  \bibfield  {author} {\bibinfo {author} {\bibfnamefont {{\O}.~S.}\
  \bibnamefont {Hetland}}, \bibinfo {author} {\bibfnamefont {A.~A.}\
  \bibnamefont {Maradudin}}, \bibinfo {author} {\bibfnamefont {T.}~\bibnamefont
  {Nordam}}, \bibinfo {author} {\bibfnamefont {P.~A.}\ \bibnamefont {Letnes}},
  \ and\ \bibinfo {author} {\bibfnamefont {I.}~\bibnamefont {Simonsen}},\
  }\href@noop {} {\enquote {\bibinfo {title} {Numerical studies of the
  transmission of light through a two-dimensional randomly rough interface},}\
  } (\bibinfo {year} {2016})\BibitemShut {NoStop}%
\bibitem [{\citenamefont {Nordam}\ \emph
  {et~al.}(2013{\natexlab{b}})\citenamefont {Nordam}, \citenamefont {Letnes},\
  and\ \citenamefont {Simonsen}}]{Simonsen2012-04}%
  \BibitemOpen
  \bibfield  {author} {\bibinfo {author} {\bibfnamefont {T.}~\bibnamefont
  {Nordam}}, \bibinfo {author} {\bibfnamefont {P.~A.}\ \bibnamefont {Letnes}},
  \ and\ \bibinfo {author} {\bibfnamefont {I.}~\bibnamefont {Simonsen}},\
  }\href {\doibase 10.1088/1742-6596/454/1/012033} {\bibfield  {journal}
  {\bibinfo  {journal} {J. Phys.: Conf. Ser.}\ }\textbf {\bibinfo {volume}
  {454}},\ \bibinfo {pages} {012033} (\bibinfo {year}
  {2013}{\natexlab{b}})}\BibitemShut {NoStop}%
\bibitem [{\citenamefont {Warren}\ and\ \citenamefont
  {Clarke}(1965)}]{Warren1965}%
  \BibitemOpen
  \bibfield  {author} {\bibinfo {author} {\bibfnamefont {B.~E.}\ \bibnamefont
  {Warren}}\ and\ \bibinfo {author} {\bibfnamefont {J.~S.}\ \bibnamefont
  {Clarke}},\ }\href {\doibase 10.1063/1.1713906} {\bibfield  {journal}
  {\bibinfo  {journal} {J. Appl. Phys.}\ }\textbf {\bibinfo {volume} {36}},\
  \bibinfo {pages} {324} (\bibinfo {year} {1965})}\BibitemShut {NoStop}%
\bibitem [{\citenamefont {Schiff}(1968)}]{Book:Schiff1968}%
  \BibitemOpen
  \bibfield  {author} {\bibinfo {author} {\bibfnamefont {L.~I.}\ \bibnamefont
  {Schiff}},\ }\href@noop {} {\emph {\bibinfo {title} {Quantum Mechanics}}}\
  (\bibinfo  {publisher} {McGraw-Hill},\ \bibinfo {address} {New York},\
  \bibinfo {year} {1968})\BibitemShut {NoStop}%
\bibitem [{\citenamefont {Tamir}(1982)}]{Book:Tamir1982}%
  \BibitemOpen
  \bibfield  {author} {\bibinfo {author} {\bibfnamefont {T.}~\bibnamefont
  {Tamir}},\ }in\ \href {https://books.google.no/books?id=SXJ\_QgAACAAJ} {\emph
  {\bibinfo {booktitle} {Electromagnetic Surface Modes}}},\ \bibinfo {editor}
  {edited by\ \bibinfo {editor} {\bibfnamefont {A.~D.}\ \bibnamefont
  {Boardman}}}\ (\bibinfo  {publisher} {John Wiley \& Sons},\ \bibinfo
  {address} {New York},\ \bibinfo {year} {1982})\ Chap.~\bibinfo {chapter}
  {13}\BibitemShut {NoStop}%
\bibitem [{\citenamefont {Tamir}\ and\ \citenamefont
  {Oliner}(1969)}]{Tamir1969}%
  \BibitemOpen
  \bibfield  {author} {\bibinfo {author} {\bibfnamefont {T.}~\bibnamefont
  {Tamir}}\ and\ \bibinfo {author} {\bibfnamefont {A.~A.}\ \bibnamefont
  {Oliner}},\ }\href {\doibase 10.1364/JOSA.59.000942} {\bibfield  {journal}
  {\bibinfo  {journal} {J. Opt. Soc. Am.}\ }\textbf {\bibinfo {volume} {59}},\
  \bibinfo {pages} {942} (\bibinfo {year} {1969})}\BibitemShut {NoStop}%
\bibitem [{\citenamefont {Nordam}\ \emph {et~al.}(2014)\citenamefont {Nordam},
  \citenamefont {Letnes}, \citenamefont {Simonsen},\ and\ \citenamefont
  {Maradudin}}]{Nordam2014}%
  \BibitemOpen
  \bibfield  {author} {\bibinfo {author} {\bibfnamefont {T.}~\bibnamefont
  {Nordam}}, \bibinfo {author} {\bibfnamefont {P.~A.}\ \bibnamefont {Letnes}},
  \bibinfo {author} {\bibfnamefont {I.}~\bibnamefont {Simonsen}}, \ and\
  \bibinfo {author} {\bibfnamefont {A.~A.}\ \bibnamefont {Maradudin}},\ }\href
  {\doibase 10.1364/JOSAA.31.001126} {\bibfield  {journal} {\bibinfo  {journal}
  {J. Opt. Soc. Am. A}\ }\textbf {\bibinfo {volume} {31}},\ \bibinfo {pages}
  {1126} (\bibinfo {year} {2014})}\BibitemShut {NoStop}%
\end{thebibliography}



%
\end{document}